\documentclass[a4paper,fleqn]{cas-sc}

\usepackage[square,numbers,sort&compress]{natbib}
\usepackage{blkarray}
\usepackage{tikz}
\usepackage[]{sidecap}

\begin{document}
\let\WriteBookmarks\relax
\def\floatpagepagefraction{1}
\def\textpagefraction{.001}
\shorttitle{The quantification of a genuine tetrapartite entanglement}
\shortauthors{H. Vargov\'a, J. Stre\v{c}ka}
%\begin{frontmatter}

\title [mode = title]{The quantification of a genuine tetrapartite entanglement \\in a  mixed spin-(1/2,1) Heisenberg tetramer}

\author[1]{H. Vargov\'a}[
                        orcid=0000-0003-2000-7536]
\cormark[1]
\ead{hcencar@saske.sk}

\credit{Conceptualization, Methodology, Software validation, Formal analysis, Investigation, Resources, Data curation, Writing - Original draft preparation, Visualization }

\affiliation[1]{organization={Institute of Experimental Physics, Slovak Academy of Sciences},
                addressline={Watsonova 47}, 
                city={Ko\v {s}ice},
                postcode={040 01}, 
                state={Slovakia}
                }

\author[2]{J. Stre\v{c}ka}[%
   orcid=0000-0003-1667-6841
   ]

\credit{Conceptualization, Methodology, Software validation, Formal analysis, Investigation, Resources,  Writing - Review and editing, Project administration, Funding acquisition}

\affiliation[2]{organization={Department of Theoretical Physics and Astrophysics, Faculty of Science, Pavol  Jozef \v{S}af\'{a}rik University},
                addressline={Park 
Angelinum 9}, 
                postcode={040 01}, 
                city={Ko\v{s}ice},
                country={Slovakia}}

\cortext[cor1]{Corresponding author}

\begin{abstract}
The genuine tetrapartite entanglement of a quantum mixed spin-(1/2,1) Heisenberg tetramer is quantified according to the three different approaches incorporated all seven global bisections existing within the tetrapartite system. The degree of an entanglement of each bisection is evaluated through the bipartite negativity at zero- and non-zero temperature taking into account ferromagnetic as well as antiferromagnetic type of intra- ($J$) and inter-dimer ($J_1$) exchange coupling inside the square plaquette. Three utilized quantification methods based on   the  generalization of (i) a genuine tripartite negativity, (ii) a Coffman, Kundu and Wootters monogamy relation and (iii) a geometric average of complete trisections, result to the qualitatively and almost quantitatively identical behavior of a genuine tetrapartite negativity. It is shown that the genuine tetrapartite negativity exclusively arises from the antiferromagnetic-inter dimer $J_1>0$ coupling, whereas the character of with respect to $J$ ($J>0$ or $J<0$) determines its zero-temperature magnitude as well as its thermal stability with respect to the magnetic field and temperature. As is demonstrated for $0<J_1/J<1$ the genuine tetrapartite negativity is dramatically reduced due to the preference of magnetic arrangement involving two separable mixed spin-(1/2,1) dimers. In an opposite limit the genuine tetrapartite negativity is significantly stable with a threshold temperature proportional to the strength of an inter-dimer coupling $J_1$. It is found, that all three quantification procedures are insufficient to correctly describe the genuine tetrapartite negativity in a specific part of the parameter space with absence of relevant dimer separable states. Finally, the thermal stability of a genuine tetrapartite negativity is discussed in detail for  selected geometries motivated by the real tetranuclear bimetallic complexes with a Cu$_2$Ni$_2$ magnetic core. 
\end{abstract}


\maketitle

\section{Introduction}

The concept of any hidden global force(s) within a composite system, which prevents its quantum-mechanical description as a simple product of individual subsystem states,  first arose at the beginning of the twentieth century~\cite{EPS}. Thirty years later, J.S. Bell's pivotal work ~\cite{Bell}  established entanglement as a newly accepted inherent feature of many-body quantum systems, rendering it impossible to simulate quantum correlations within any classical formalism. Currently, entanglement lies at the heart of quantum information science~\cite{Nielsen} and is utilized in various groundbreaking applications such as quantum teleportation~\cite{Furusawa}, quantum cryptography~\cite{Deutsch,Bechmann}, and quantum computation~\cite{Loss,Hayashi}. 

From a theoretical perspective,  quantum Heisenberg spin models provide a reasonable theoretical playground for rigorously studying the quantum and thermal entanglement inherent in nature. The focus of theoretical interest lies in exploring various geometries, including dimers~\cite{Arnesen,Lagmago,Terzis,Cenci,Ghannadan,Vargova2021,Vargova2021a,
Vargova2022,Naveena,Wang,Abliz}, trimers~\cite{Bose,Wang2021,Li,Liu,Pal,Cima, Milivojevic,Benabdallah,Han,Najarbashi}, tetramers~\cite{Bose,Cao,Wu2007,Irons,Karlova2023,Szalowski2022, Zad2022,Ghannadan2022,Kuzmak,Vargova2023}, and higher oligomers~\cite{Karlova2018,Szalowski2023,Deb}. Additionally, researchers are investigating the effects of bilinear and biquadratic interactions ~\cite{Wang2005,Wang2007,Bose}, magnetic field inhomogeneity~\cite{Najarbashi,Han,Terzis,Abliz,Wang,Liu,Wu2007,Kuzmak}, exchange interaction anisotropy~\cite{Cenci,Ghannadan,Wang,Lagmago,Abliz,Naveena,Li, Pal, Najarbashi,Benabdallah,Wu2007,Vargova2021a},  the diversity of spin constituents~\cite{Vargova2022,Najarbashi}, etc.

Contemporary quantum studies are heavily focused  on accurately measuring the degree of entanglement.  While  bipartite entanglement is  well understood~\cite{Horodecki2009,Amico},  measuring entanglement in multipartite systems remains an open and intensively investigated task.   This pursuit is driven by the desire to gain deeper insights into the correlation processes between different registers of a quantum computer. One possible approach to determining multipartite entanglement involves a comprehensive analysis of all bipartite  non-local correlations, obtained by reducing the original multipartite system to the bipartite level. Several notable works have explored multipartite entanglement using this method across various Heisenberg models, see for example Refs.~\cite{Ghannadan2022,Szalowski2023,Han,Najarbashi,Bose,Kuzmak,Wu2007, Liu, Pal}. While this procedure provides a qualitative overview of the pairwise distribution of entanglement within the composite system, it does not fully capture the entanglement shared among three or more parties within the multipartite systems.

The measurement of multipartite entanglement, also known as genuine, full, or pure entanglement, represents another alternative approach. It should be emphasized that research in this field is still ongoing, and to date, there is no general quantification procedure applicable to arbitrary systems with various numbers of constituents. For the smallest tripartite systems, multipartite entanglement is typically quantified using the geometric mean of all bisections between one central spin and a spin dimer~\cite{Sabin,Dong2023,Su,Love,Benabdallah,Karlova2023, Milivojevic,Szalowski2022}. Some studies~\cite{Hou,Yu,Pan} prefer the arithmetic mean over the geometric one, however, in such definitions, a separable bisection (with null magnitude) can incorrectly lead to finite global entanglement~\cite{Sabin}.  Genuine tripartite entanglement is also quantified using the Coffman, Kundu, and Wootters (CKW) monogamy inequality~\cite{CKW}, which states that the entanglement between one central spin and the rest of the multipartite system cannot be weaker than the sum of pairwise entanglements. The difference between the left- and right-hand sides of the CKW inequality estimates the residual entanglement not shared in pairwise form, the magnitude of which strongly depends on the choice of a central spin. Subsequently, the combination of all residual entanglements (with respect to the central spin) can offer another quantification of the degree of genuine tripartite entanglement. Recently, a new quantity known as concurrence triangle has been introduced ~\cite{Xie} to measure genuine tripartite entanglement. This quantity corresponds to the area of a triangle with edge lengths matching the square of concurrence between the central spin and spin dimer.

The quantification of genuine entanglement in tetrapartite or higher complex systems is more intricate, and thus far, only the concept based on the CKW inequality has been applied, particularly emphasizing the Heisenberg model~\cite{Zad2022,Li2016,Zepeda,Torres,Oliveira,Dong2020,Dong2019, Dong2020a}. Previous studies have extended the respective concept directly from tripartite to tetrapartite systems, focusing solely on one type of tetrapartite system bisection—specifically, the bisection between the central spin and residual trimer. Notably, the second type of bisection, between two spin dimers, has been entirely overlooked in existing genuine tetrapartite quantification, despite its potential significance. In this paper, we aim to rectify this oversight by introducing three alternative approaches for the accurate quantification of genuine tetrapartite entanglement. The fundamental idea behind all three methods is straightforward: we consider a complete set of all bisections present in the respective multipartite system, as only the combination of all bisections provides comprehensive information about genuine multipartite entanglement. We will introduce (i) an approach defined as the geometric mean of seven non-reduced-system bisections; (ii) an approach derived from an extended CKW relation; and finally, (iii) an approach constructed as the geometric average of all six tripartite entanglement contributions. It is important to note that the number of bisections (trisections) is proportional to the size of elementary magnetic unit of the system.  Subsequently, all three methods will be implemented to study quantum and thermal genuine tetrapartite entanglement in the mixed spin-(1/2,1) Heisenberg tetramer, involving both ferromagnetic and antiferromagnetic exchange interactions under external magnetic fields. Surprisingly, we will demonstrate that each of the introduced approaches leads to qualitatively and almost quantitatively identical predictions, contrary to previous conclusions. Finally, we will discuss in detail the potential thermal stability of genuine entanglement in real tetrapartite bimetallic complexes with a molecular magnetic unit based on  Fe$_2$Ni$_2$~\cite{Wu,Parkin,Rodriguez,Park} and Cu$_2$Ni$_2$~\cite{Lou,Ribas,Osa,Barrios,Nakamura} cores, which exhibit magnetic properties well-suited for the mixed spin-(1/2,1) Heisenberg tetramer model. For completeness, it should be noted that some particular results concerned with the distribution of reduced bipartite~\cite{Vargova2023} and reduced tripartite~\cite{Vargova_arxiv} entanglement have been published in our previous papers. Contrary to this, the global bipartite entanglement as well as the genuine tetrapartite entanglement has not been addressed hitherto and will be the main subject matter of this article.

The structure of the paper is organized as follows. In Section~\ref{model}, we introduce the quantum spin model under investigation and present three different approaches to quantify the degree of genuine entanglement within the tetrapartite system. Section~\ref{results} reports the most interesting results, divided into several subsections. In Section~\ref{Bipartite negativity}, we discuss the most important observations regarding the quantum bipartite negativities of the global system in both bisections. We verify the validity of the monogamy relation and modified monogamy relation in Section~\ref{validity}. The quantification of genuine tetrapartite entanglement for a specific model under investigation through each of the three introduced methods is presented in Section~\ref{Genuine}, with a detailed discussion of behavior at zero and non-zero temperatures. Section~\ref{compare} compares the accuracy of the computational method used thus far, which involves only one type of bisection, with our improved method involving a complete set of bisections. The implications of thermal entanglement for real tetranuclear bimetallic clusters are presented in Section~\ref{exp}. Finally, Section~\ref{conclusion} provides a brief summary of the most important scientific findings. Technical details of calculations are summarized in Appendices~\ref{App A}-\ref{App F}.

\section{\label{model} Model and Method}
\begin{SCfigure}[50][t]
%\begin{center}
{\includegraphics[width=.2\textwidth,trim=1.0cm 4.5cm 16cm 20.5cm, clip]{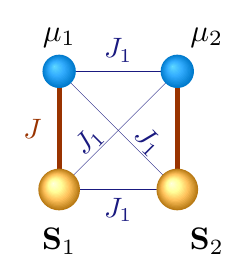}}
%\end{center}
\vspace*{-0.5cm}
\caption{A schematic representation of a mixed spin-(1/2,1) Heisenberg tetramer with a square plaquette. Small and large balls correspond to spins with eigenvalues $S_{1}(S_2)\!=\!1$ and $\mu_{1}(\mu_2)\!=\!1/2$, respectively.  The exchange interaction between the four magnetic ions is represented by the coupling constants  $J$ (thick vertical lines) and $J_1$ (thin horizontal and diagonal lines).}
\label{fig1}
\end{SCfigure}

Let's consider the model Hamiltonian, expressed as the sum of individual cluster Hamiltonians  $\hat{\cal H}\!=\!\sum_{i}\hat{\cal H}_i$, where $\hat{\cal H}_i$ describes the energy of four interacting spins arranged into a square plaquette, as illustrated in the Fig.~\ref{fig1}. 
\allowdisplaybreaks
\begin{align}
\hat{\cal H}_i&=J\left(\hat{\bf S}_{1}\cdot\hat{\boldsymbol\mu}_{1}
\!+\!\hat{\bf S}_{2}\cdot\hat{\boldsymbol\mu}_{2}\right)\!+\! J_1\left(\hat{\bf S}_{1}\!+\!\hat{\boldsymbol\mu}_{1}\right)\cdot\left(\hat{\bf S}_{2}\!+\!\hat{\boldsymbol\mu}_{2}\right)
%\nonumber\\
\!-\!h\left(\hat{S}^z_{1}\!+\!\hat{S}^z_{2}\!+\!\hat{\mu}^z_{1}\!+\!\hat{\mu}^z_{2}\right).
\label{eq1}
\end{align} 
Here,  $\hat{\bf S}_{\alpha}$ and ${\hat{{\boldsymbol\mu}}}_{\alpha}$ ($\alpha=1,2$) are the spin operators with respective spin values of  $S_{\alpha}=1$  and $\mu_{\alpha}=1/2$. Each pair of spins interacts via isotropic Heisenberg interactions $J$ between spins with identical indices forming the mixed spin-(1/2,1) Heisenberg dimer, and $J_1$ between either identical spins or spins with different indices. For a clearer understanding, refer to  Fig.~\ref{fig1}. The influence of an external magnetic field $B$ applied along the $z$ direction is also considered in the final term of Hamiltonian~\eqref{eq1}, where the magnetic field $B$ is incorporated in parameter $h$, expressed  as $h\!=\!g\mu_BB$. Here, $g$ is a  Land\'e $g$-factor, which is identical for both spins, and $\mu_B$ is the Bohr magneton.

To explore the behavior of genuine entanglement in a mixed spin-(1/2,1) tetrapartite system, we employ the concept of bipartite negativity~\cite{Vidal}
\begin{align}
{\cal N}(\rho)=\sum_{\lambda_i<0}\vert \lambda_i\vert.
\label{eq2}
\end{align}
Bipartite negativity quantitatively estimates the Peres and Horodecki separability criterion (PPT)~\cite{Peres,Horodecki} measuring the pairwise entanglement between two selected parts, spins $q_1$ and $q_2$. According to definition~\eqref{eq2}, bipartite negativity can be easily measured as the sum of the absolute values of negative eigenvalues $|\lambda_i|$ of the partially transposed density matrix $\rho^{T_{q_1}}$. The partially transposed density matrix is obtained from the original density matrix $\hat{\rho}$
\begin{align}
\hat{\rho}={\rm e}^{-\hat{\cal H}_i/(k_BT)}/{\cal Z},
\label{eq0}
\end{align}
by transposing only one selected subsystem, e.g., $q_1$. This is indicated by the  index $T_{q_1}$ of the partially transposed density matrix $\rho^{T_{q_1}}$. For completeness, we remark that $k_B$ in Eq.~\eqref{eq0} represents Boltzmann's constant, $T$ is temperature, and ${\cal Z}={\rm Tr}\; {\rm e}^{-\hat{\cal H}_i/(k_BT)}$ is the statistical sum.
 Furthermore, it should be emphasized that the choice of subsystem for partial transposition is irrelevant, resulting in identical magnitudes of negativity despite different eigenvalues of distinct partially transposed density matrices. It is evident from definition (\ref{eq2}) that negativity is a non-negative function. A value of zero (${\cal N}=0$) indicates a separable state, while a positive value (${\cal N}>0$) determines the degree of bipartite entanglement. 

To determine genuine entanglement in a tetrapartite system, we employ three different approaches. 

(i) The first approach is a straightforward extension of the generally accepted procedure used for determining genuine negativity in a tripartite system~\cite{Sabin}. This procedure is based on calculating the geometric average of all different bisections in a tetrapartite system, a concept proposed in a few interesting works~\cite{Love,Ghahi}. However, to the best of our knowledge, it has never been applied to a specific tetrapartite quantum spin model
\begin{align}
\theta_{q_1q_2q_3q_4}&=\left\{ [{\cal N}_{q_1|q_2q_3q_4}{\cal N}_{q_2|q_1q_3q_4}{\cal N}_{q_3|q_1q_2q_4}{\cal N}_{q_4|q_1q_2q_3}]\right.
%\nonumber\\
\left.\times
[{\cal N}_{q_1q_2|q_3q_4} {\cal N}_{q_1q_3|q_2q_4}{\cal N}_{q_1q_4|q_2q_3}
]\right\}^{1/7}.
\label{eq3}
\end{align}
In a tetrapartite system, there are two types of bisections: between a single spin and the spin triplet (e.g., $q_1|q_2q_3q_4$), and between two spin dimers (e.g., $q_1q_2|q_3q_4$). The type of bisection is indicated by the lower indexes on the right-hand side of  Eq.~\eqref{eq3}. A total of seven various bisections have been identified~\cite{Dur}.
Both types of global bipartite negativities are calculated following the generalized definition in Eq.~\eqref{eq2}, wherein the negative eigenvalues  $\lambda_i$ are derived from the respective partially transposed density matrix $\rho^{T_{q_i}}$ or $\rho^{T_{q_iq_j}}$. In our calculations, the choice of a geometric average is preferred over an arithmetic one due to the validity of a fundamental characteristic of global genuine negativity, as deeply discussed in Ref.~\cite{Sabin}. In such a definition, genuine negativity correctly becomes non-zero for fully entangled state, but one must be very careful in cases of zero value. As is generally known, the PPT test is only a necessary condition in dimensions higher than $2\otimes 3$~\cite{Horodecki2009,Horodecki1996}. 

(ii) The second method for estimating genuine tetrapartite entanglement involves the generalization of the CKW monogamy inequality~\cite{CKW} to $n$-partite systems, as extended by Osborne and Verstraete~\cite{Osborne}. They asserted that the entanglement between the spin $q_1$ and the group of other spins in a general pure state  $\vert \psi \rangle$  cannot be weaker than the sum of the individual bipartite entanglements of the reduced system between $q_1$ and each individual spin  $q_i$ 
\begin{align}
\tau^{(1)}_{q_1|q_2\dots q_n}(\vert \psi\rangle)&\geq \tau^{(2)}_{q_1|q_2}(\vert \psi\rangle)+ \tau^{(2)}_{q_1|q_3}(\vert \psi\rangle)+\dots 
%\nonumber\\
+  \tau^{(2)}_{q_1|q_n}(\vert \psi\rangle),
\label{eq4}
\end{align}
expressed through universal functions known as one-tangle $\tau^{(1)}_{q_1|q_2\dots q_n}$ and two-tangles $\tau^{(2)}_{q_i|q_j}$. The difference between the left- and right-hand sides of Eq.~\eqref{eq4} then defines the $n$-tangle or residual entanglement of an $n$-partite system with a central spin $q_1$ in the general pure state $\vert \psi \rangle$
\begin{align}
\delta_{q_1q_2\dots q_n}&(q_1):=\tau^{(1)}_{q_1|q_2\dots q_n}(\vert \psi\rangle)
- \left[\tau^{(2)}_{q_1|q_2}(\vert \psi\rangle)+ \tau^{(2)}_{q_1|q_3}(\vert \psi\rangle)+\dots +  \tau^{(2)}_{q_1|q_n}(\vert \psi\rangle)\right].
\label{eq5}
\end{align}
As clearly stated by Regula et al.~\cite{Regula},  for $n>3$, the difference between the left- and right-hand sides of Eq.~\eqref{eq4} does not express genuine entanglement but rather allows us to roughly estimate the residual entanglement not distributed pairwise.  Considering the hierarchical structure of 4-partite systems ~\cite{Dur}, we propose that for $n=4$, it is desirable to formulate an analogous monogamy relation for a pure state $\vert \psi\rangle$, termed the modified CKW (mCKW) relation. In this relation, the entanglement between pairs of spins, e.g., $q_1q_2$ and  $q_3q_4$, cannot be less than the entanglement between the pair  $q_1q_2$ and the reduced form, i.e., individual spins $q_3$ and $q_4$
\begin{align}
\tau^{(1)}_{q_1q_2|q_3q_4}(\vert \psi\rangle)\geq \tau^{(2)}_{q_1q_2|q_3}(\vert \psi\rangle)+ \tau^{(2)}_{q_1q_2|q_4}(\vert \psi\rangle).
\label{eq6}
\end{align}
The difference between the left- and right-hand sides of  Eq.~\eqref{eq6},
\begin{align}
\pi_{q_1q_2\dots q_n}(q_1q_2)&:=\tau^{(1)}_{q_1q_2|q_3q_4}(\vert \psi\rangle)
- \left[\tau^{(2)}_{q_1q_2|q_3}(\vert \psi\rangle)+ \tau^{(2)}_{q_1q_2|q_4}(\vert \psi\rangle)\right]
\label{eq7}
\end{align}
then provides an estimate of the entanglement not shared between the spin pair and the remaining individual spin. We believe that only the combination of both quantities in Eq.~\eqref{eq5} and Eq.~\eqref{eq7}, considering all possible permutations of a central spin $q_i$ and a spin pair $q_iq_j$, allows us to obtain comprehensive information about the genuine entanglement in a tetrapartite system. Once again, we prefer the definition based on the geometric average. Since the tangle is a universal function, it can be interpreted practically using any appropriate quantity such as negativity~\cite{Vidal}, concurrence~\cite{Wootters}, von Neumann's entropy~\cite{Bennett},  etc.  For the following calculations, we will utilize the square of negativity as an appropriate entanglement monotone~\cite{Ou}, aiming to ensure the continuity of our complex analysis~\cite{Vargova2023,Vargova_arxiv}. Thus, the genuine tetrapartite negativity in the second definition should be mathematically defined as follows
\begin{align}
\nu_{q_1q_2q_3q_4}&=\left\{ \sqrt{\delta_{q_1q_2q_3q_4}(q_1)\delta_{q_1q_2q_3q_4}(q_2)\delta_{q_1q_2q_3q_4}(q_3)\delta_{q_1q_2q_3q_4}(q_4)}\right.
\nonumber\\
&\times
\left.\sqrt{\pi_{q_1q_2q_3q_4}(q_1q_2)\pi_{q_1q_2q_3q_4}(q_1q_3)\pi_{q_1q_2q_3q_4}(q_1q_4)\pi_{q_1q_2q_3q_4}(q_3q_4)
}\right\}^{1/8}.
\label{eq8}
\end{align}
 To the best of our knowledge, all previous theoretical analyses of global entanglement in particular tetrapartite spin systems have been limited to contributions involving individual central spins only ~\cite{Zepeda,Oliveira,Li2016,Torres,Dong2019,Dong2020a,Dong2020, Zad2022}. 
Therefore, our study aims to provide new insights into the quantification of the degree of entanglement in multipartite spin systems with more than three magnetic spin centers. 

(iii) The final approach employed in our study involves constructing a geometric average of all tripartite contributions within a tetrapartite system. This method is based on the principle that if at least one spin can be separated from the others, then the global system cannot be genuinely entangled
\begin{align}
\omega_{\mu_1\mu_2S_1S_2}&=\left\{ {\cal N}_{q_1|q_2|q_3q_4}{\cal N}_{q_1|q_2q_3|q_4}{\cal N}_{q_1|q_2q_4|q_3}\right.
\left.\times
{\cal N}_{q_1q_2|q_3|q_4}{\cal N}_{q_1q_3|q_2|q_4}{\cal N}_{q_1q_4|q_2|q_3} \right\}^{1/6}.
\label{eq9}
\end{align}
In a tetrapartite system, there are six different trisections~\cite{Dur}, which can be easily evaluated using the standard relation~\cite{Sabin}
\begin{align}
{\cal N}_{q_1|q_2|q_3q_4}&=\left\{{\cal N}_{q_1|q_2q_3q_4}{\cal N}_{q_2|q_1q_3q_4}{\cal N}_{q_3q_4|q_1q_2}\right\}^{1/3},
\hspace*{.5cm}
%\nonumber\\
{\cal N}_{q_1|q_2q_3|q_4}=\left\{{\cal N}_{q_1|q_2q_3q_4}{\cal N}_{q_1q_4|q_2q_3}{\cal N}_{q_4|q_1q_2q_3}\right\}^{1/3},
\nonumber\\
{\cal N}_{q_1|q_2q_4|q_3}&=\left\{{\cal N}_{q_1|q_2q_3q_4}{\cal N}_{q_1q_3|q_2q_4}{\cal N}_{q_3|q_1q_2q_4}\right\}^{1/3},
\hspace*{.5cm}
%\nonumber\\
{\cal N}_{q_1q_2|q_3|q_4}=\left\{{\cal N}_{q_1q_2|q_3q_4}{\cal N}_{q_3|q_1q_2q_4}{\cal N}_{q_4|q_1q_2q_3}\right\}^{1/3},
\nonumber\\
{\cal N}_{q_1q_3|q_2|q_4}&=\left\{{\cal N}_{q_1q_3|q_2q_4}{\cal N}_{q_2|q_1q_3q_4}{\cal N}_{q_4|q_1q_2q_3}\right\}^{1/3},
\hspace*{.5cm}
%\nonumber\\
{\cal N}_{q_1q_4|q_2|q_3}=\left\{{\cal N}_{q_1q_4|q_2q_3}{\cal N}_{q_2|q_1q_3q_4}{\cal N}_{q_3|q_1q_2q_4}\right\}^{1/3}.
\label{eq10}
\end{align}
\section{\label{results} Results and discussion}

Before discussing the most interesting results, we would like to recall that an exhaustive analysis of the behavior of bipartite negativity in a mixed spin-(1/2,1) Heisenberg tetramer, particularly when degrees of freedom of remaining two spins are traced out, can be found in our previous papers~\cite{Vargova2023,Vargova_arxiv}. Additionally, a ground-state analysis involving preferred magnetic structures is available in the same papers~\cite{Vargova2023,Vargova_arxiv}. To ensure the present paper is self-contained, we present in Appendix~\ref{App A} the explicit forms of relevant ground-state eigenvectors that contribute to the phase diagram, along with the analytical expressions of their phase boundaries. It is important to note that the finite size of a spin cluster precludes actual finite-temperature phase transitions, and the term 'phase diagram' refers in this context to dissimilar ground states characterized through a set of three quantum numbers $|\sigma_T^z,\sigma_{1},\sigma_{2}\rangle$, where $\sigma^z_T=-\sigma_T,-\sigma_T+1,\dots,\sigma_T$; $\sigma_T\in\langle |\sigma_1-\sigma_2|,\sigma_1+\sigma_2\rangle$ and $\sigma_{\gamma}\in\langle |S_{\gamma}-\mu_{\gamma}|,S_{\gamma}+\mu_{\gamma}\rangle$ with $\gamma=1,2$. 

\begin{figure*}[t!]
\begin{center}
{\includegraphics[width=.38\textwidth,trim=3.6cm 9.8cm 5.8cm 8.8cm, clip]{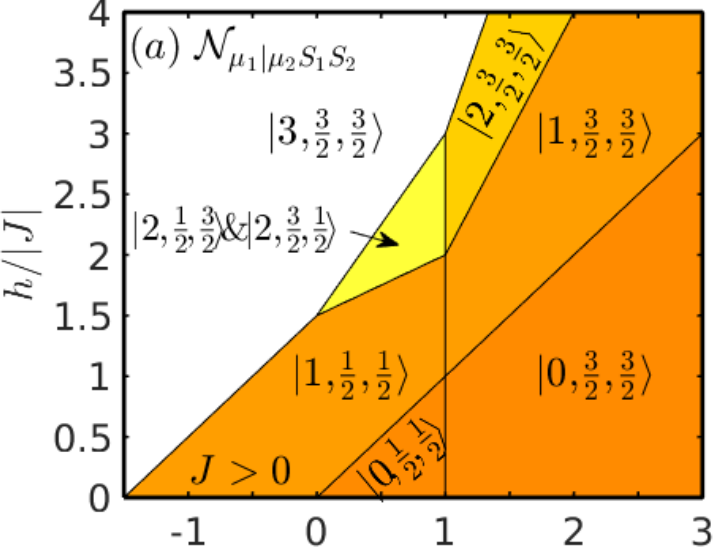}}
{\includegraphics[width=.445\textwidth,trim=4.35cm 9.8cm 3cm 8.8cm, clip]{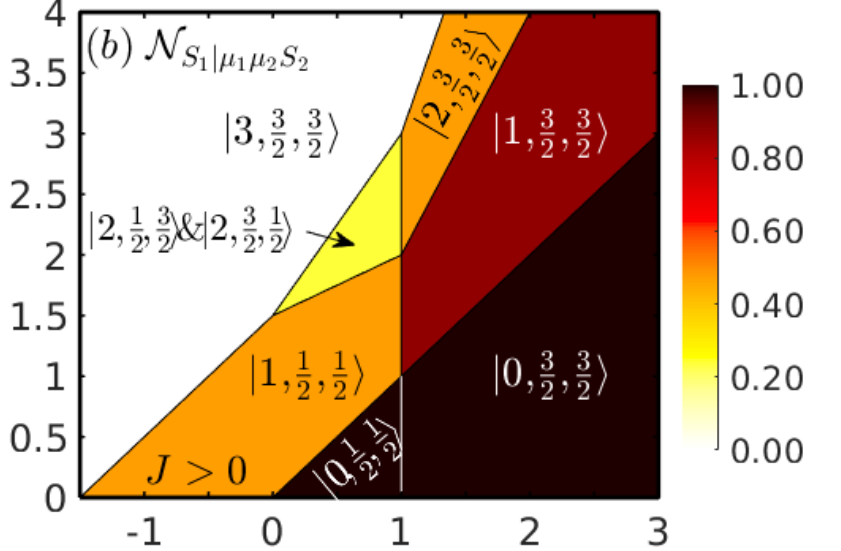}}
{\includegraphics[width=.38\textwidth,trim=3.6cm 8.9cm 5.8cm 8.8cm, clip]{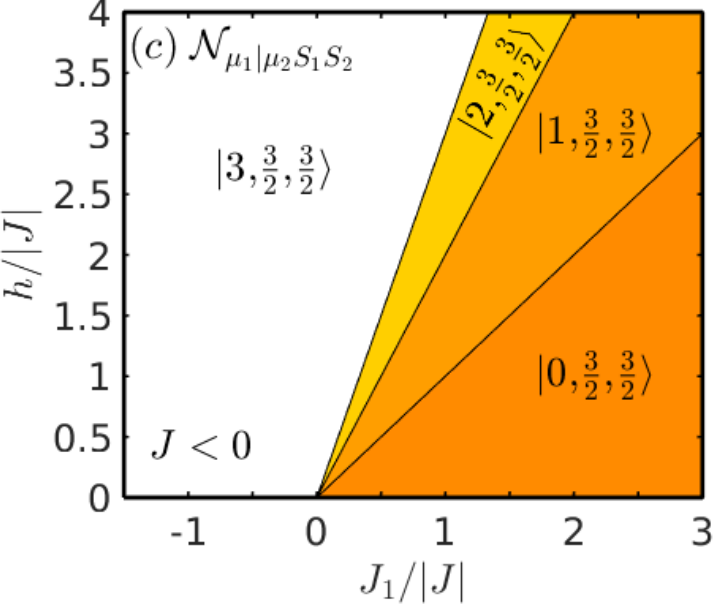}}
{\includegraphics[width=.445\textwidth,trim=4.35cm 8.9cm 3cm 8.8cm, clip]{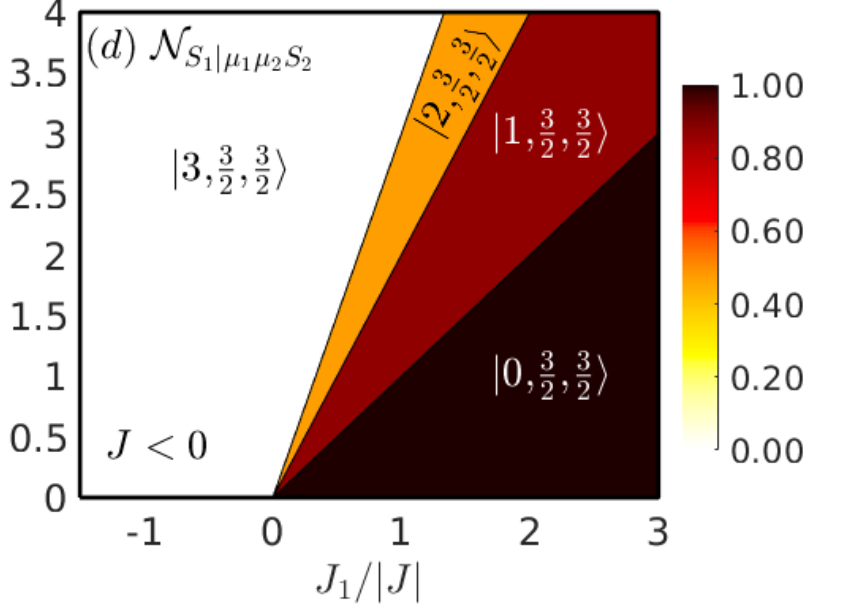}}
\end{center}
\caption{Zero-temperature density plots of global quantum  bipartite negativities  of a mixed-spin Heisenberg tetramer~\eqref{eq1} in the $J_1/|J|-h/|J|$ plane for a bisection with one central spin. Upper (Lower) panels correspond to $J>0$ ($J<0$). Solid black lines visualize borders between different magnetic ground states characterized by a set of three quantum numbers $|\sigma_T^z,\sigma_{1},\sigma_{2}\rangle$. The  explicit value of the global bipartite negativity at each ground state is  presented in Tab.~\ref{tab1}. }
\label{fig2}
\end{figure*}
Furthermore, we provide the density matrix of the entire system, along with the respective partition function, in Appendix~\ref{App A}, which are both necessary quantities for calculating negativity. The partial transposed density matrices ${\rho}^{T_{\mu_1}}_{\mu_1\mu_2S_1S_2}$, $\rho^{T_{S_1}}_{\mu_1\mu_2S_1S_2}$,$\rho^{T_{\mu_1\mu_2}}_{\mu_1\mu_2S_1S_2}$, $\rho^{T_{\mu_1S_1}}_{\mu_1\mu_2S_1S_2}$, and $\rho^{T_{\mu_1S_2}}_{\mu_1\mu_2S_1S_2}$, which are essential prerequisites for calculating the global bipartite negativities  ${\cal N}_{\mu_1|\mu_2S_1S_2}$, ${\cal N}_{S_1|\mu_1\mu_2S_2}$, ${\cal N}_{\mu_1\mu_2|S_1S_2}$, ${\cal N}_{\mu_1S_1|\mu_2S_2}$, and ${\cal N}_{\mu_1S_2|\mu_2S_1}$, are provided in individual Appendices~\ref{App B}-\ref{App F}. 

Lastly, to minimize the vast parametric space, all further dependencies will be normalized with respect to the absolute value of the intra-dimer coupling constant $|J|$, which can take both ferromagnetic ($J<0$) and antiferromagnetic  ($J>0$) values. Similarly, the inter-dimer exchange coupling  $J_1$ can be assumed to be of both types ($J_1>0$ and $J_1<0$). The values of Boltzmann's constant and the Bohr magneton are set to unity.

\subsection{\label{Bipartite negativity} Bipartite negativity ${\cal N}_{q_i|q_jq_kq_l}$ and ${\cal N}_{q_iq_j|q_kq_l}$}

\begin{figure*}[t!]
\begin{center}
{\includegraphics[width=.325\textwidth,trim=3.2cm 9.8cm 5.8cm 8.cm, clip]{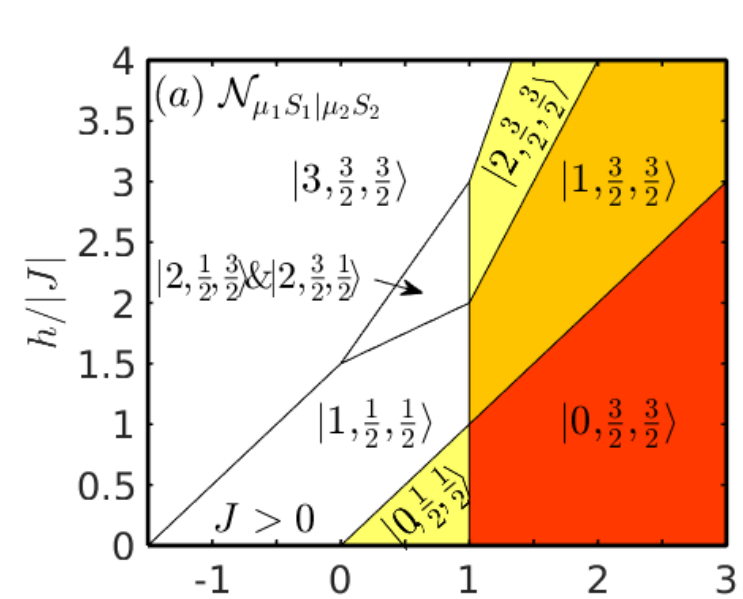}}
{\includegraphics[width=.295\textwidth,trim=4.35cm 9.8cm 5.8cm 8.cm, clip]{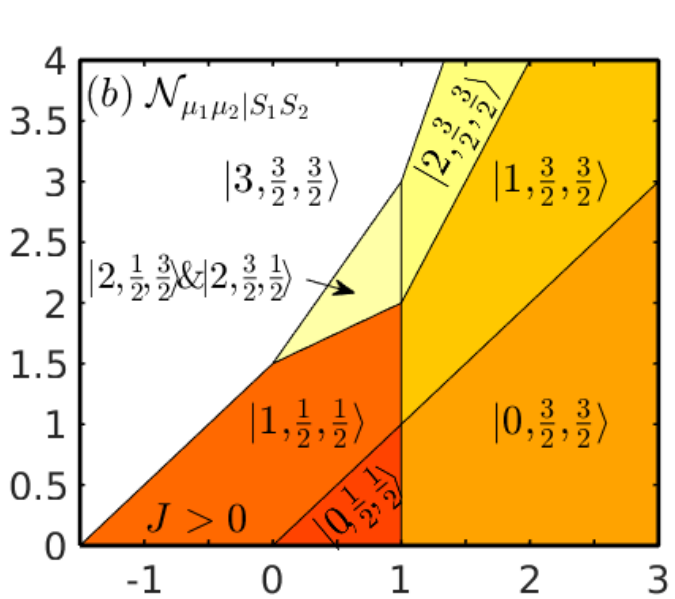}}
{\includegraphics[width=.365\textwidth,trim=4.35cm 9.8cm 3cm 8.cm, clip]{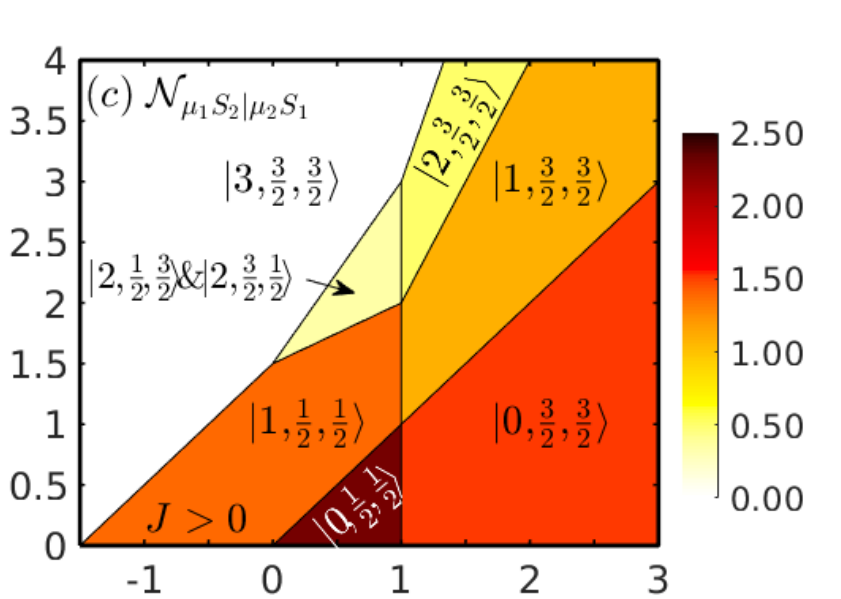}}
{\includegraphics[width=.325\textwidth,trim=3.2cm 8.9cm 5.8cm 8.8cm, clip]{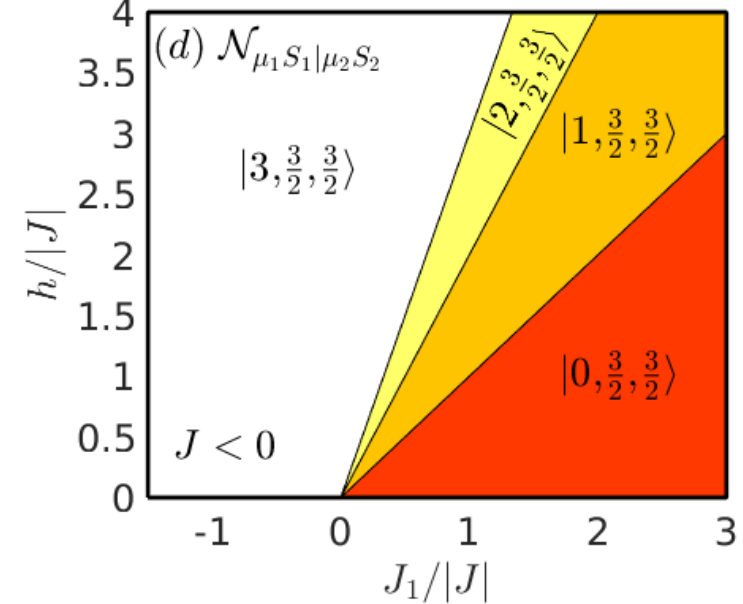}}
{\includegraphics[width=.295\textwidth,trim=4.35cm 8.9cm 5.8cm 8.8cm, clip]{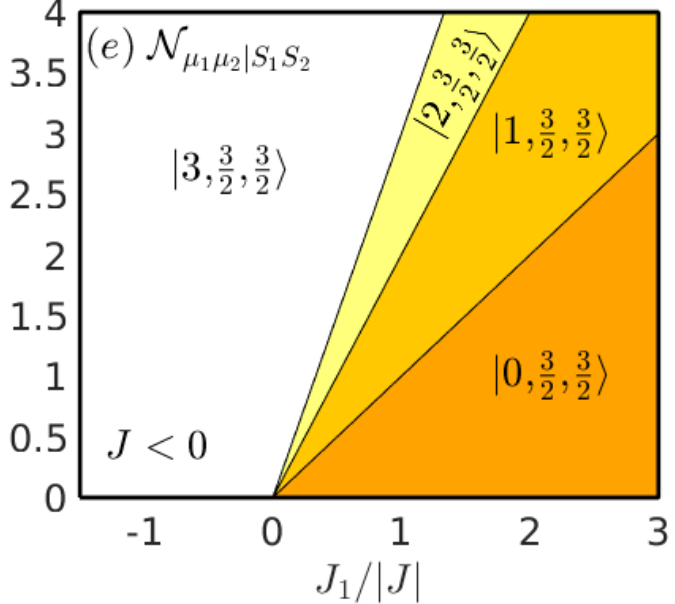}}
{\includegraphics[width=.365\textwidth,trim=4.35cm 8.9cm 3cm 8.8cm, clip]{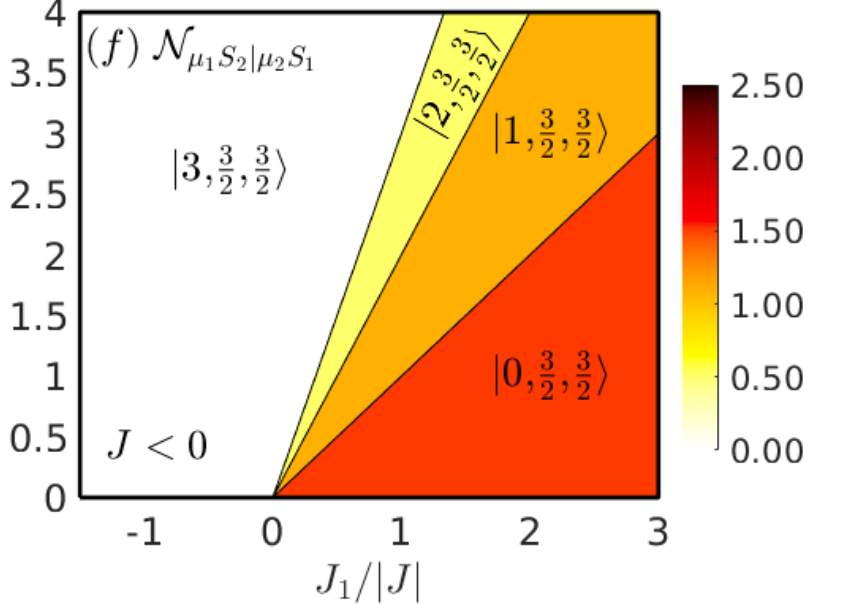}}
\end{center}
\caption{Zero-temperature density plots of a global quantum  bipartite negativity  of a mixed-spin Heisenberg tetramer~\eqref{eq1} in the $J_1/|J|-h/|J|$ plane for a bisection with a central spin dimer.  Upper (Lower) panels correspond to $J>0$ ($J<0$). Solid black lines visualize borders between different magnetic ground states characterized by a set of three quantum numbers $|\sigma_T^z,\sigma_{1},\sigma_{2}\rangle$. The  explicit value of the global bipartite negativity at each ground state is  presented in Tab.~\ref{tab1}. }
\label{fig3}
\end{figure*}

First, let us characterize the behavior of bipartite negativities ${\cal N}_{q_i|q_jq_kq_l}$ and ${\cal N}_{q_iq_j|q_kq_l}$ in a global tetrapartite spin system. The zero-temperature negativities are illustrated in 
Fig.~\ref{fig2} for a bisection with a single central spin and Fig.~\ref{fig3} for a bisection with a central spin dimer, integrated into the framework of the ground-state phase diagram in the $J_1/|J|-h/|J|$ plane.  The explicit magnitude of the global bipartite negativity for each ground state is listed in Tab.~\ref{tab1}.

For a global bipartite negativity with a single central spin  (Fig.~\ref{fig2}),  two different scenarios are identified: for a central spin (i) $\mu_1=\mu_2=1/2$ (Fig.~\ref{fig2}($a$), ($c$)) and (ii) $S_1=S_2=1$ (Fig.~\ref{fig2}($b$), ($d$)), respectively. In both cases, the quantum negativity never exceeds the magnitude of the central spin. The maximum degree of entanglement is achieved exclusively for antiferromagnetic inter-dimer coupling $J_1/|J|>0$, accepting both antiferro- (upper panels) and ferromagnetic (lower panels) intra-dimer interactions $J$. 
Interestingly, the degree of entanglement for the maximally entangled ground states  $\vert 0,1/2,1/2\rangle$ and $\vert 0,3/2,3/2\rangle$  for a selected bisection is identical and independent of the magnitude of the inter-dimer interaction  $J_1/|J|$.  Ferromagnetic inter-dimer coupling ($J_1/|J|<0$) strongly reduces quantum entanglement, with the possibility of detecting the entangled state only for antiferromagnetic intra-dimer coupling $J>0$ (upper panels) in the presence of a weak external magnetic field. As evident from Fig.~\ref{fig2}, the increasing magnetic field  generally gradually reduces the global bipartite quantum negativity  until the fully polarized (separable) state is reached. Comparing both one-central-spin  bisections (left and right panels of Fig.~\ref{fig2}), one can see that the effect of the interaction ratio $J_1/|J|$ on the stability of  quantum negativity is significantly smaller in a bisection $\mu_1|\mu_2S_1S_2$ than in the $S_1|\mu_1\mu_2S_2$ counterpart. This can be explained in connection with the higher effective spin of a tripartite cluster, which reflects in its higher resistance to the influence of magnetic fields in reorienting all spins into a magnetic-field direction and thus minimizing the quantum negativity. 

In the case of a central spin dimer  (Fig.~\ref{fig3}), three different scenarios have been identified, taking into account the symmetry of the analyzed system. In the assumed bisection, two interesting observations have been made in the region of antiferromagnetic intra-dimer interaction $J>0$ (upper panels). The first observation is the separability of both mixed spin-(1/2,1) Heisenberg dimers in the case of  $J_1/|J|<1$ in each ground state with $\sigma_T^z>0$ of the ${\cal N}_{\mu_1S_1|\mu_2S_2}$ bisection (Fig.~\ref{fig3}($a$)). While in the $|1,1/2,1/2\rangle$ phase, separability is a consequence of maximal entanglement within the $\mu_1S_1$ ($\mu_2S_2$) dimer (${\cal N}_{\mu_i|S_i}=\sqrt{2}/3$~\cite{Vargova2023}), the entanglement in the degenerated $|2,1/2,3/2\rangle\&|2,3/2,1/2\rangle$  phase cannot arise due to the separability between the $\mu_1-\mu_2$ (${\cal N}_{\mu_1|\mu_2}=0$), $S_1-S_2$ (${\cal N}_{S_1|S_2}=0$), and $\mu_1-S_2$ (${\cal N}_{\mu_1|S_2}=0$) spins, see Ref.~\cite{Vargova2023}. 
The second interesting observation is a dramatic enhancement of the strength of negativity in a specific antiferromagnetic phase $|0,1/2,1/2\rangle$ in the case of the  ${\cal N}_{\mu_1S_2|\mu_2S_1}$ bisection (Fig.~\ref{fig3}($c$)).  Whereas the quantum negativity in all remaining phases for an arbitrary bisection with a central dimer does not exceed the value ${\cal N}=1.5$, the quantum negativity in the ground state $|0,1/2,1/2\rangle$ for the case of the   ${\cal N}_{\mu_1S_2|\mu_2S_1}$ bisection is around ${\cal N}=2.278$. From the analysis of a reduced system~\cite{Vargova2023,Vargova_arxiv}, we suppose that such unexpectedly high degree of entanglement may originate from the relatively strong entanglement of selected reduced subsystems, for example, between the single spin $S_1$ ($\mu_2$) and spin dimer $\mu_1S_2$ ($\mu_1S_2$) (${\cal N}_{S_1|\mu_1S_2}=0.893$, ${\cal N}_{\mu_2|\mu_1S_2}=0.427$~\cite{Vargova_arxiv}),  as well as between the individual spins $\mu_1-S_1$ ($S_1-S_2$) (${\cal N}_{\mu_1|S_1}=0.333$, ${\cal N}_{S_1|S_2}=0.111$~\cite{Vargova2023}). Finally, it should be noted that the degree of entanglement between two specific spin dimers gradually decreases under the influence of applied external magnetic field, regardless of the (anti)ferromagnetic character of intra-dimer interaction $J$.  

\subsection{\label{validity}Validity of monogamy relation(s)}
With explicit values for all possible bisections, we can directly verify the validity of the CKW monogamy relation Eq.~\eqref{eq4} as well as the mCKW monogamy relation Eq.~\eqref{eq6}. It should be emphasized that determining the validity of monogamy relations is not trivial due to known examples of selected quantum systems with dimensions higher than qubits, such as $3\otimes 3 \otimes 3$ or $3\otimes 2 \otimes 2$ systems~\cite{Ou,Kim}, in which the CKW inequality can be violated. For better visualization, all calculated inequalities are compiled in Tab.~\ref{tab2}. It is worthwhile to recall that the explicit  values of reduced bipartite negativities  ${\cal N}_{q_i|q_j}$ (${\cal N}_{q_iq_j|q_k}\equiv{\cal N}_{q_k|q_iq_j}$) are provided in Ref.~\cite{Vargova2023} (Ref.~\cite{Vargova_arxiv})), and all five types of global bipartite negativities in the entire tetrapartite system are listed in Tab.~\ref{tab1} of Appendix~\ref{App A}. 
\begin{table}[b!]
\caption{The evaluation of the CKW monogamy relation Eq.~\eqref{eq4} (first two rows) and mCKW relation  Eq.~\eqref{eq6} (last four rows) calculated for all available ground states  of a mixed spin-(1/2,1) Heisenberg tetramer~\eqref{eq1}.}
\label{tab2}
\resizebox{1\textwidth}{!}{
\begin{tabular}{l  c c  c  c  c  cc }
\hline\hline
  &    $\vert 0,\frac{1}{2},\frac{1}{2}\rangle$ & 
    $\vert 0,\frac{3}{2},\frac{3}{2}\rangle$ & 
    $\vert 1,\frac{1}{2},\frac{1}{2}\rangle$ & 
    $\vert 1,\frac{3}{2},\frac{3}{2}\rangle$ & 
    $\vert 2,\frac{1}{2},\frac{3}{2}\rangle$ &  
    $\vert 2,\frac{3}{2},\frac{3}{2}\rangle$ & 
    $\vert 3,\frac{3}{2},\frac{3}{2}\rangle$ \\
    &    & 
    & 
  & 
     & 
     $\vert 2,\frac{3}{2},\frac{1}{2}\rangle$ &  
     & 
     \\
\hline
${\cal N}^2_{\mu_1|\mu_2S_1S_2}\geq {\cal N}^2_{\mu_1|\mu_2}+ {\cal N}^2_{\mu_1|S_1}+ {\cal N}^2_{\mu_1|S_2}$&
$0.250\geq0.111$ & $0.250\geq0.077$ & $0.222\geq0.222$ & $0.222\geq0.044$ & $0.056\geq0.009$ & $0.139\geq0.019$ & $0\geq0$
\\
${\cal N}^2_{S_1|\mu_1\mu_2S_2}\geq {\cal N}^2_{S_1|\mu_1}+ {\cal N}^2_{S_1|\mu_1}+ {\cal N}^2_{S_1|S_2}$ & $1.000\geq0.123$ & $1.000\geq0.299$ & $0.222\geq0.222$ & $0.764\geq0.124$ & $0.056\geq0.009$ & $0.222\geq0.059$ & $0\geq0$
\\
${\cal N}^2_{\mu_1\mu_2|S_1S_2}\geq {\cal N}^2_{\mu_1\mu_2|S_1}+ {\cal N}^2_{\mu_1\mu_2|S_2}$& $2.199\geq0.259$ & $1.320\geq0.104$ & $1.924\geq0.444$ & $0.958\geq0.156$ & $0.111\geq0.056$ & $0.222\geq0.085$ & $0\geq0$
\\
${\cal N}^2_{\mu_1S_1|\mu_2S_2}\geq {\cal N}^2_{\mu_1S_1|\mu_2}+ {\cal N}^2_{\mu_1S_1|S_2}$& $0.250\geq0.111$ & $2.250\geq0.507$ & $0\geq0$ & $0.986\geq0.319$ & $0\geq0$ & $0.250\geq0.139$ & $0\geq0$
\\
${\cal N}^2_{\mu_1S_2|\mu_2S_1}\geq {\cal N}^2_{\mu_1S_2|\mu_2}+ {\cal N}^2_{\mu_1S_2|S_1}$& $5.189\geq1.326$ & $2.250\geq0.344$ & $1.924\geq0.444$ & $1.208\geq0.299$ & $0.111\geq0.043$ & $0.250\geq0.139$ & $0\geq0$
\\
${\cal N}^2_{S_1S_2|\mu_1\mu_2}\geq {\cal N}^2_{S_1S_2|\mu_1}+ {\cal N}^2_{S_1S_2|\mu_2}$& $2.199\geq0.484$ & $1.320\geq0.222$ & $1.924\geq0.444$ & $0.958\geq0.185$ & $0.111\geq0.030$ & $0.222\geq0.135$ & $0\geq0$\\
\hline\hline
\end{tabular}
}
\end{table}

Upon inspecting Tab.~\ref{tab2} one can observe that the CKW and mCKW inequalities are valid across the entire parametric space and for all permutations of a central spin (dimer).
One direct consequence of the validity of the aforementioned relations is the feasibility of the suggested concept for estimating genuine multipartite entanglement through the combination of CKW and mCKW monogamy relations.
 Additionally, the CKW monogamy test provides us with another means to identify the separability of states within the regions of $\vert 1,\tfrac{1}{2},\tfrac{1}{2}\rangle$ and $\vert 2,\tfrac{1}{2},\tfrac{3}{2}\rangle\&\vert 2,\tfrac{3}{2},\tfrac{1}{2}\rangle$   concerning the ${\cal N}_{\mu_1S_1|\mu_2S_2}$ bisection. The identity of CKW within these regions suggests that there is no residual entanglement not distributed pairwise. 

\subsection{\label{Genuine}Genuine tetrapartite negativity}
In the following discussion, we will focus solely on the antiferromagnetic intra-dimer interaction $J>0$, as its ferromagnetic counterpart  $J<0$ can be considered as its simpler subgroup. Consequently, all qualitative observations can be generalized to $J<0$ with respect to the structure of the respective ground-state phase diagram. 
\begin{figure}[b!]
\begin{center}
{\includegraphics[width=.43\textwidth,trim=3.6cm 9cm 3cm 8.7cm, clip]{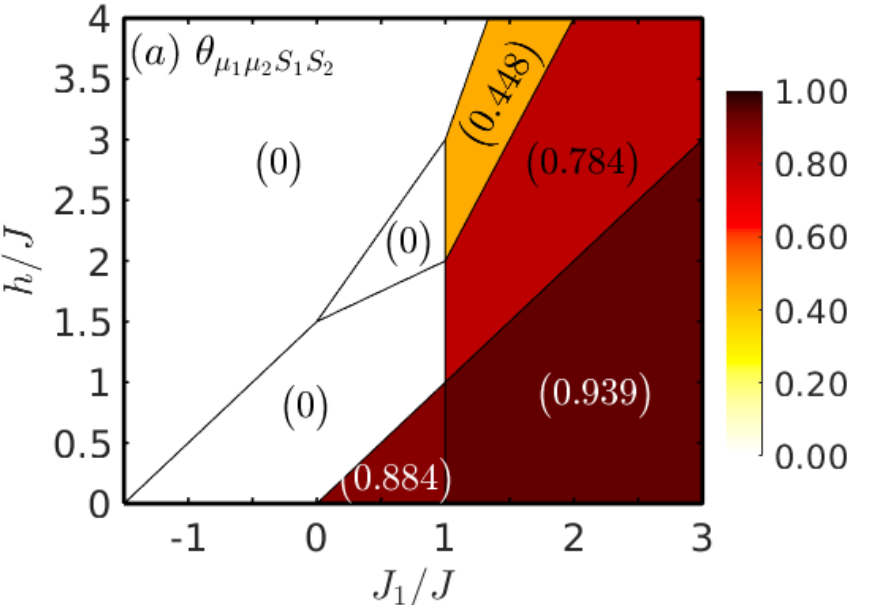}}
{\includegraphics[width=.41\textwidth,trim=4.35cm 9cm 3cm 8.7cm, clip]{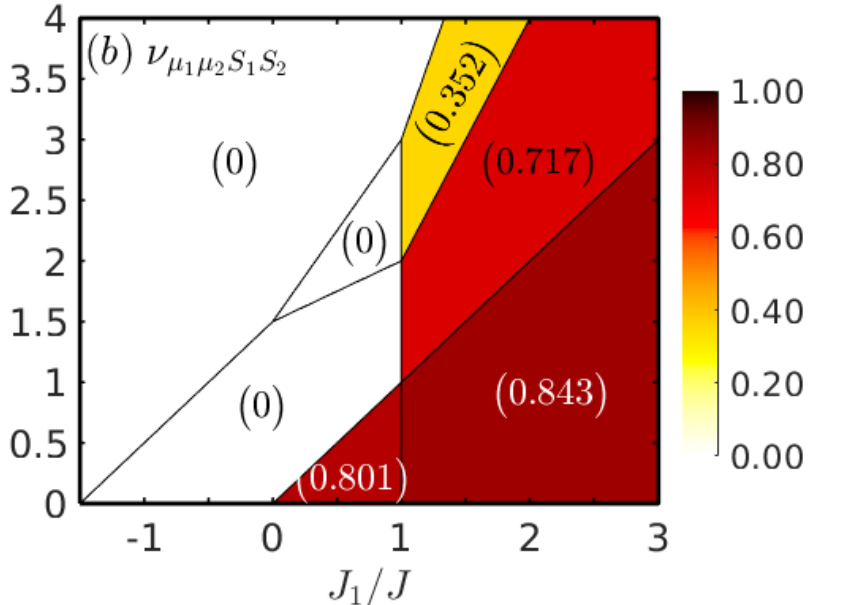}}
{\includegraphics[width=.43\textwidth,trim=3.6cm 9cm 3cm 8.7cm, clip]{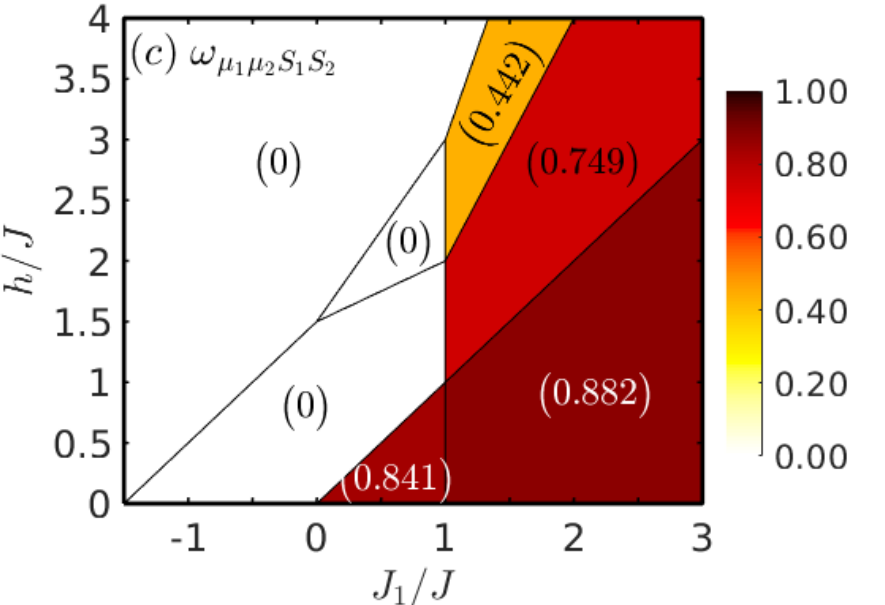}}{\includegraphics[width=.41\textwidth,trim=4.35cm 9cm 3cm 8.7cm, clip]{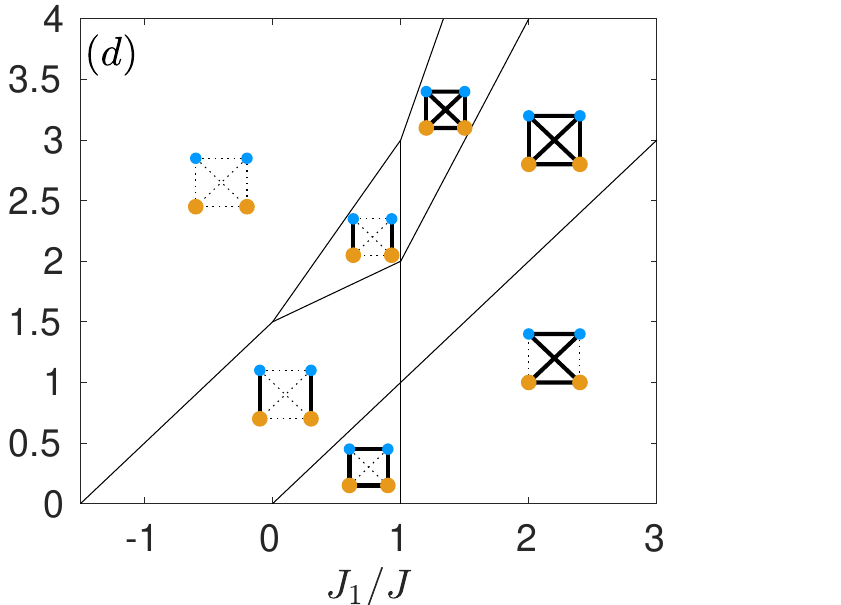}}
\end{center}
\caption{Density plots of quantum genuine   tetrapartite negativities $\theta_{\mu_1\mu_2S_1S_2}$ ($a$), ${\nu}_{\mu_1\mu_2S_1S_2}$ ($b$) and ${\omega}_{\mu_1\mu_2S_1S_2}$ ($c$) of a mixed spin-(1/2,1) Heisenberg tetramer in the $J_1/J-h/J$ plane for $J>0$. Solid black lines visualize borders between different magnetic ground states. The value of the global bipartite negativity within each ground state is indicated in parentheses. ($d$) Graphical visualization of identified classes of genuine tetrapartite entanglement~\cite{Ghahi} in the respective ground states. Small blue (large yellow) dots correspond to spin $\mu_i$ ($S_i$), and the solid thick (dotted thin) lines illustrate  non-zero (zero) specific bipartite entanglement. }
\label{fig5}
\end{figure}

Fig.~\ref{fig5} illustrates the behavior of genuine quantum tetrapartite negativity obtained according to  Eq.~\eqref{eq3} (Fig.~\ref{fig5}($a$)),  Eq.~\eqref{eq8} (Fig.~\ref{fig5}($b$)) and Eq.~\eqref{eq9} (Fig.~\ref{fig5}($c$)), within the context of the ground-state phase diagram for  $J>0$. To maintain clarity, the notation for each respective ground state is omitted, while the explicit magnitude of negativity for each state is indicated in round brackets. To further enhance clarity, panel ($d$) in Fig.~\ref{fig5} visualizes the distribution of reduced bipartite entanglement between specific pairs of spins for each ground state, in alignment with the classification of four-qubit entanglement proposed by Ghahi and Akhtarshenas~\cite{Ghahi}. All three approaches lead to a qualitatively similar picture, in which the variation of the  interaction ratio $J_1/J$ determines three different scenarios. 

In the case of $J_1/J<0$, the mixed spin-(1/2,1) Heisenberg tetramer is always separable. Bearing in mind the previous results on the reduced system~\cite{Vargova2023,Vargova_arxiv},  we identify the biseparable state at the $\vert 1,1/2,1/2\rangle$  phase and a fully separable one at the $\vert 3,3/2,3/2\rangle$ phase. Following the classification of four-qubit entanglement introduced by Ghahi and Akhtarshenas, the aforementioned  {\it biseparable state is the  four vertices with two edges and no circle}  one (\begin{tikzpicture}
\draw[line width=0.2mm, dotted] (0,0) -- (0.3,0);
\draw[line width=0.2mm, dotted] (0,0.3) -- (0.3,0.3);
\draw[line width=0.5mm,] (0,0)  --(0,0.3);
\draw[line width=0.5mm,] (0.3,0)  --(0.3,0.3);
\end{tikzpicture})~\cite{Ghahi}.

In the case of $0<J_1/J<1$, the entangled state can be exclusively realized for the antiferromagnetically arranged plaquette of a mixed spin-(1/2,1) Heisenberg tetramer characterized by a set of quantum spin numbers $\vert0,1/2,1/2\rangle$. The degree of entanglement is relatively high, and the genuine quantum tetrapartite negativity achieves the values $\theta\sim0.884$, $\nu\sim0.801$ and $\omega\sim0.841$, respectively. As identified from the analysis of the reduced bipartite negativity~\cite{Vargova2023}, genuine tetrapartite entanglement can arise  iff each spin of a tetrapartite system is entangled with at least another two spins from the square plaquette.  According to the classification of Ghahi and Akhtarshenas, this entangled state is a  {\it fully inseparable one with  four vertices, four edges and a circle included graph} (\begin{tikzpicture}
\draw[line width=0.5mm] (0,0) -- (0.3,0)--(0.3,0.3)--(0,0.3)--cycle;
\end{tikzpicture})~\cite{Ghahi}. Remaining ground states within this range of  $J_1/J$ interaction are both biseparable ({\it four vertices with two edges and no circle} (\begin{tikzpicture}
\draw[line width=0.2mm, dotted] (0,0) -- (0.3,0);
\draw[line width=0.2mm, dotted] (0,0.3) -- (0.3,0.3);
\draw[line width=0.5mm,] (0,0)  --(0,0.3);
\draw[line width=0.5mm,] (0.3,0)  --(0.3,0.3);
\end{tikzpicture})) with a different degree of entanglement within the mixed-spin dimer $\mu_iS_i$. 

\begin{figure*}[t!]
\begin{center}
{\includegraphics[width=.31\textwidth,trim=4cm 9.8cm 5.6cm 8.8cm, clip]{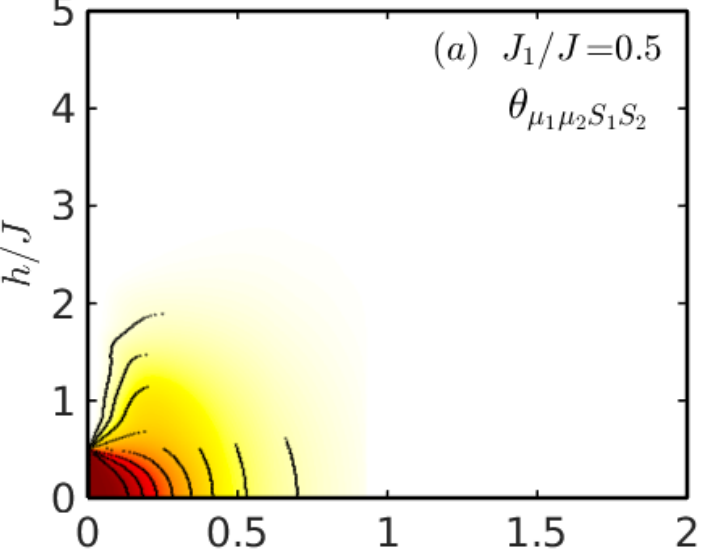}}
{\includegraphics[width=.29\textwidth,trim=4.7cm 9.8cm 5.6cm 8.8cm, clip]{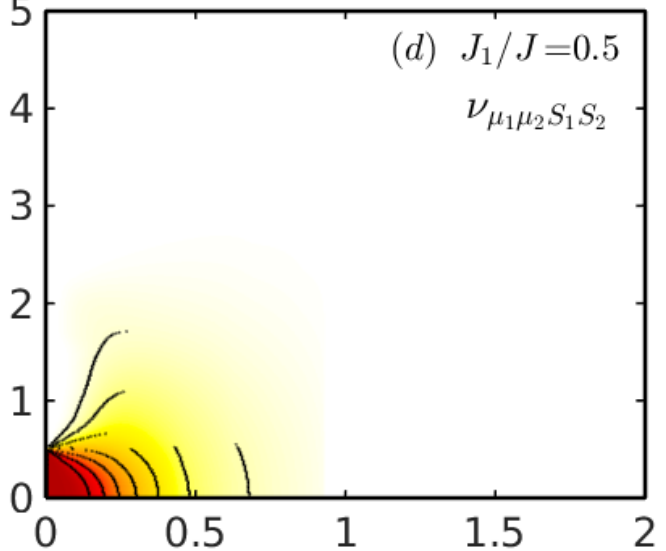}}
{\includegraphics[width=.359\textwidth,trim=4.7cm 9.8cm 3cm 8.cm, clip]{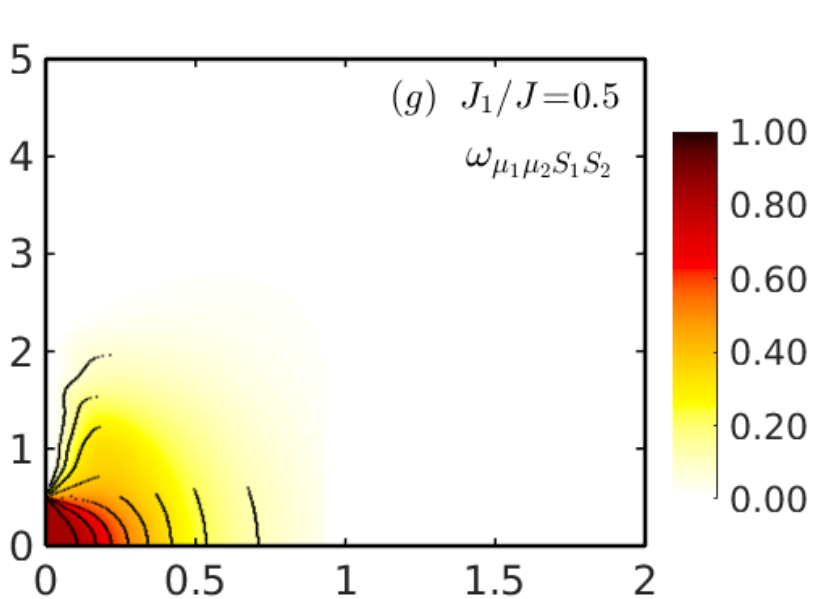}}\\
{\includegraphics[width=.31\textwidth,trim=4cm 9.8cm 5.6cm 8.7cm, clip]{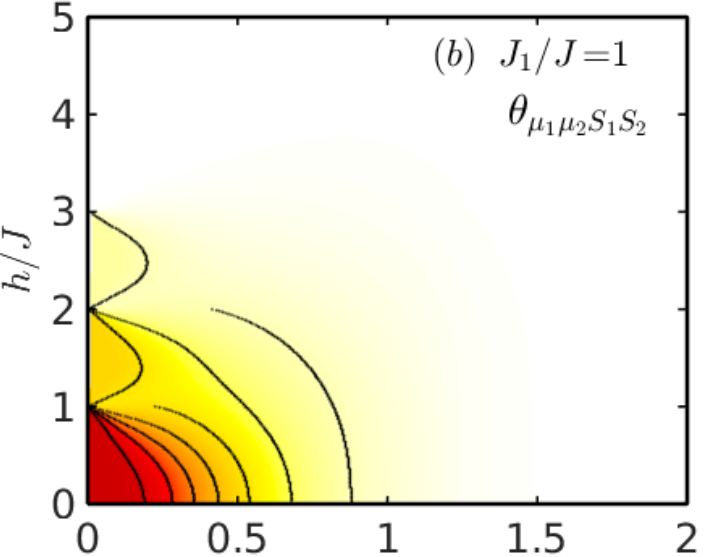}}
{\includegraphics[width=.29\textwidth,trim=4.7cm 9.8cm 5.6cm 8.7cm, clip]{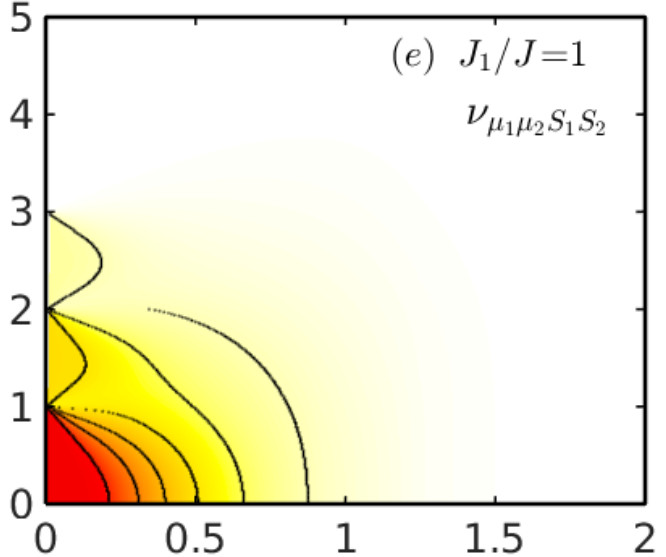}}
{\includegraphics[width=.359\textwidth,trim=4.7cm 9.8cm 3cm 8.7cm, clip]{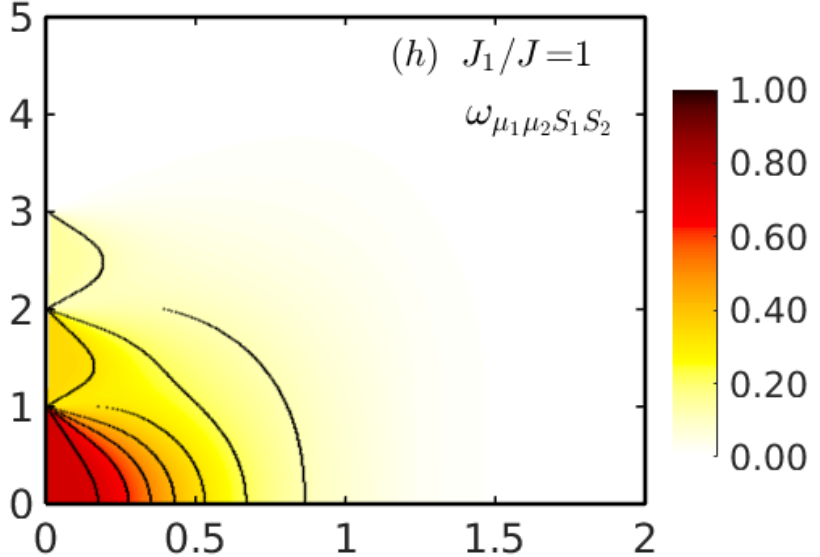}}\\
{\includegraphics[width=.31\textwidth,trim=4cm 8.9cm 5.6cm 8.7cm, clip]{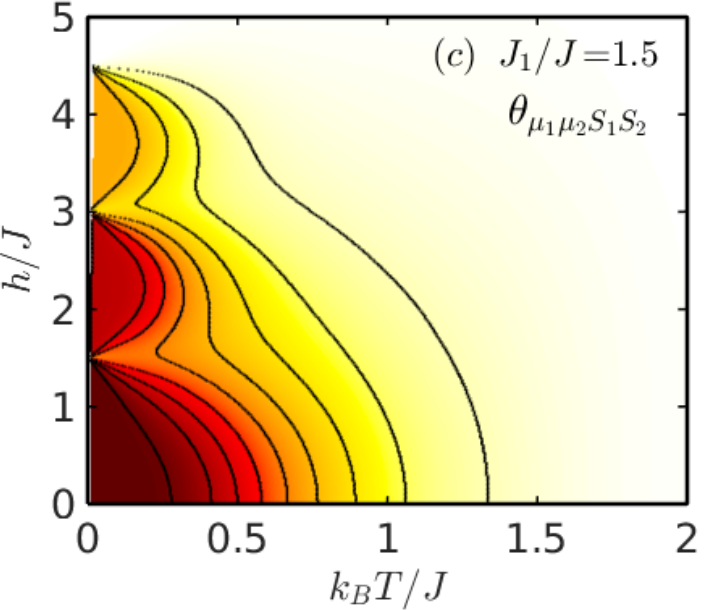}}
{\includegraphics[width=.29\textwidth,trim=4.7cm 8.9cm 5.6cm 8.7cm, clip]{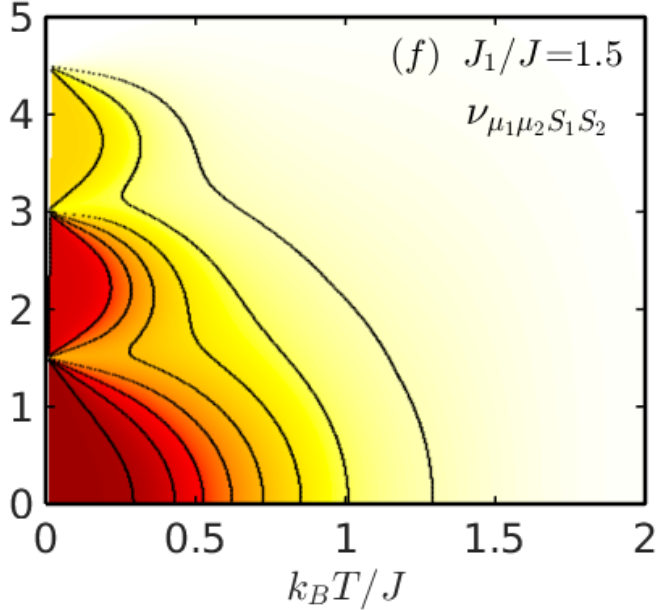}}
{\includegraphics[width=.359\textwidth,trim=4.7cm 8.9cm 3cm 8.7cm, clip]{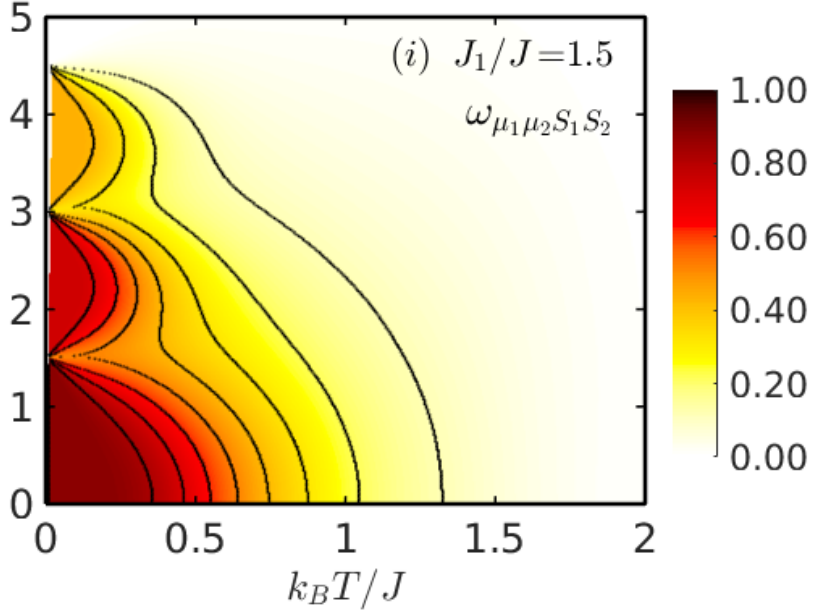}}
\end{center}
\caption{Density plots of a genuine tetrapartite negativities $\theta_{\mu_1\mu_2S_1S_2}$, $\nu_{\mu_1\mu_2S_1S_2}$ and $\omega_{\mu_1\mu_2S_1S_2}$ of a mixed spin-(1/2,1) Heisenberg tetramer~\eqref{eq1} in the $k_BT/J-h/J$ plane for three selected values of  interaction ratio $J_1/J$. The antiferromagnetic intra-dimer interaction is assumed ($J>0$). The black contour  lines are the isovalue curves corresponding to specific values from 1 to 0.1 with a step 0.1 (from left to right).}
\label{fig6}
\end{figure*}

For $J_1/J>1$,  the highest genuine quantum tetrapartite negativity is again detected with an antiferromagnetic spin arrangement, represented by the $\vert0,3/2,3/2\rangle$ ground state. However, the degree of entanglement is slightly higher compared to the previous case, with  $\theta\sim0.939$, $\nu\sim0.843$, and $\omega\sim0.882$. It was identified that this difference directly relates to the increased contribution of three tetrapartite bisections:  ${\cal N}_{\mu_1\mu_2|S_1S_2}$, ${\cal N}_{\mu_1S_1|\mu_2S_2}$, and ${\cal N}_{\mu_1S_2|\mu_2S_1}$, whereas the bisections with a single central spin, ${\cal N}_{\mu_1|\mu_2S_1S_2}$ and ${\cal N}_{S_1|\mu_1\mu_2S_2}$,  remain unchanged upon the enlargement of $J_1/J$. According to the classification of Ghahi and Akhtarshenas, the detected state is again a {\it fully inseparable one with four vertices, four edges and a circle}~\cite{Ghahi}. In contrast to the $\vert 0,1/2,1/2\rangle$ eigenstate, the mixed spin dimer $\mu_iS_i$ is fully separable (\begin{tikzpicture}
\draw[line width=0.5mm] (0,0) -- (0.3,0.3)--(0,0.3)--(0.3,0)--cycle;
\end{tikzpicture}).  The remaining two ground states with $\sigma_T^z>0$ are {\it fully inseparable with four vertices, six edges and a circle} (\begin{tikzpicture}
\draw[line width=0.5mm] (0,0) -- (0.3,0.3)--(0,0.3)--(0.3,0)--(0,0)--(0,0.3)--(0.3,0.3)--(0.3,0)--cycle;
\end{tikzpicture}) when respecting the Ghahi and Akhtarshenas classification. As illustrated in Fig.~\ref{fig5}, the increment of magnetic field in the region of $J_1/J>1$ gradually reduces the degree of genuine  tetrapartite negativity until the fully polarized and fully separable state $\vert 3,3/2,3/2\rangle$ is achieved.

Fig.~\ref{fig6}  visualizes the thermal stability of genuine tetrapartite entanglement through density plots of  $\theta_{\mu_1\mu_2S_1S_2}$ ($\nu_{\mu_1\mu_2S_1S_2}$, $\omega_{\mu_1\mu_2S_1S_2}$) in the $k_BT/J-h/J$ plane at three representative values of $J_1/J>0$. We solely analyze the thermal stability of antiferromagnetic inter-dimer $J_1>0$ coupling because the fully separable and biseparable ground states preclude thermal entanglement in the ferromagnetic $J_1<0$ parametric space. Several interesting observations emerge. It turns out that the qualitative features of the three inspecting approaches are almost indistinguishable, and any of the suggested approaches could be used to describe the qualitatively same behavior of thermal genuine tetrapartite entanglement in the mixed spin-(1/2,1) Heisenberg tetramer~\eqref{eq1}. Additionally, in the case of $0<J_1/J<1$, the genuine tetrapartite negativity $\theta_{\mu_1\mu_2S_1S_2}$ ($\nu_{\mu_1\mu_2S_1S_2}$, $\omega_{\mu_1\mu_2S_1S_2}$) vanishes at relatively low temperature, while thermal fluctuations can induce genuine tetrapartite entanglement sufficiently beyond the threshold magnetic field above which the fully separable ground state is realized. This effect can occur because the competition of subtle thermal fluctuations and a weak magnetic field favors eigenstates reminiscent of the neighboring ground state with a smaller $\sigma_T^z$, whose negativity  is higher than the original one. This phenomenon is typical for arbitrary $0<J_1/J<1$, with a slight increase in the threshold magnetic field and temperature achieved under the enlargement of $J_1/J$ coupling.
At $J_1/J=1$,  a density plot of genuine tetrapartite negativity dramatically changes due to the respective structural changes in the ground state. The threshold magnetic field at zero temperature is considerably enlarged, and as indicated by the isovalue black lines in Fig.~\ref{fig6}, the increasing magnetic field causes a gradual decrement up to the threshold temperature higher than in the case of $J_1/J=0.5$. Our analysis indicates that over $J_1/J=1$, the threshold temperature linearly increases with an increase in $J_1/J$ coupling ratio, and its value is proportional to the strength of $J_1/J$.  This finding is significant for practical applications because it offers a recipe for easily enhancing thermal entanglement in tetranuclear bimetallic complexes by enhancing inter-dimer interaction during chemical synthesis. It should be noted that the non-zero but very small value of negativity can also be measured above the magnetic field where the separable ground state is preferred only, but the determined magnitude is in general less than $1{\rm e}^{-3}$.

\subsection{\label{compare} Genuine tetrapartite negativity with a complete set of bisections vs. genuine tetrapartite negativity involving only the bisection with a one central spin}

We believe that it is very useful to compare our improved approaches with the method utilized previously, in which only the bisection with a single central spin has been taken into account~\cite{Oliveira,Li2016,Torres,Dong2019,Dong2020a,Dong2020,Zad2022}. For this purpose, we will consider a quantity utilized in the mentioned references, which should be perceived as a restriction of our definition~\eqref{eq9}.
Therefore,
 \begin{align}
\nu^*_{q_1q_2q_3q_4}&=\left\{ \sqrt{\delta_{q_1q_2q_3q_4}(q_1)\delta_{q_1q_2q_3q_4}(q_2)\delta_{q_1q_2q_3q_4}(q_3)\delta_{q_1q_2q_3q_4}(q_4)}
\right\}^{1/4}.
\label{eq12}
\end{align}
\begin{figure*}[t!]
\begin{center}
{\includegraphics[width=.322\textwidth,trim=3.6cm 8.9cm 5.8cm 8.6cm, clip]{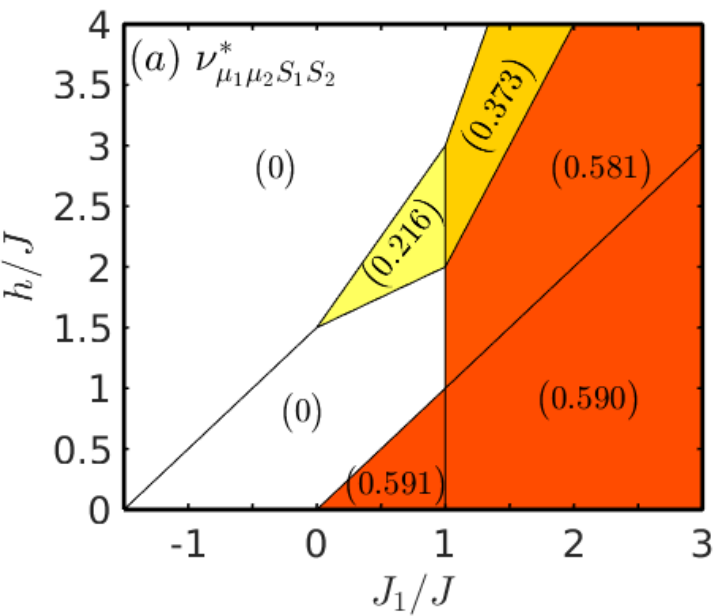}}
{\includegraphics[width=.297\textwidth,trim=4.7cm 8.9cm 5.6cm 8.6cm, clip]{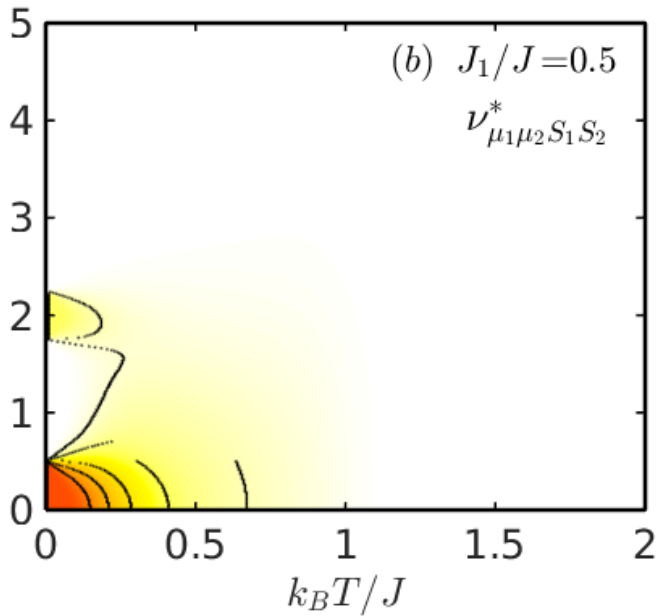}}
{\includegraphics[width=.363\textwidth,trim=4.7cm 8.9cm 3.cm 8.6cm, clip]{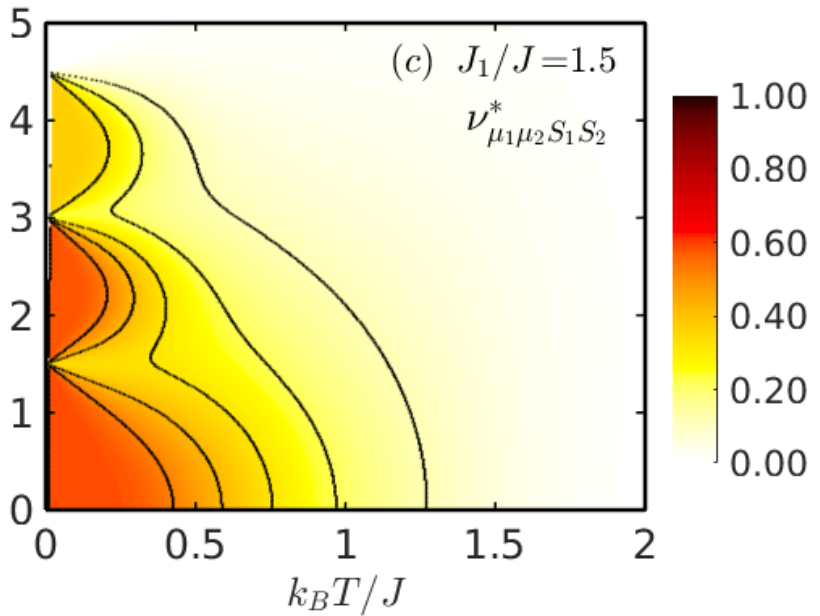}}
\end{center}
\caption{The density plot of a genuine tetrapartite negativity ${\nu}^*_{\mu_1\mu_2S_1S_2}$ of a mixed spin-(1/2,1) Heisenberg tetramer. ($a$) The zero-temperature behavior within the context of ground-state diagram. The explicit magnitude of negativity of each ground state indicted in round brackets. ($b$)-($c$) The thermal behavior for two representative values of $J_1/J$.}
\label{fig8}
\end{figure*}

The functions  $\delta_{q_1q_2q_3q_3}(q_i)$ are identical to the definition~\eqref{eq6}, considering all permutations with a central spin  $q_i$. Because all three proposed approaches lead to almost identical results, for illustration, we choose only one of them, the genuine negativity $\nu_{q_1q_2q_3q_4}$, as defined in Eq.~\eqref{eq6}.
The results obtained at zero temperature are depicted in Fig.~\ref{fig8}($a$). For brevity, we focus solely on the case of antiferromagnetic intra-dimer interaction $J>0$, although the conclusions can be qualitatively extended to the ferromagnetic counterpart, considering the structure of the relevant ground-state phase diagram. 

It is evident that the distribution of genuine zero-temperature negativity differs qualitatively from that involving both types of bisection shown in Fig.~\ref{fig5}($b$). Significant diversity is present within the region of  $J_1/J<1$  for both biseparable states $\vert 1,1/2,1/2\rangle$ and $\vert 2,1/2,3/2\rangle\&\vert 2,3/2,1/2\rangle$. Due to the absence of dimer bisections in the hitherto quantification method, the genuine tetrapartite negativity is incorrectly predicted for the structure where two entangled dimers are separable between themselves. Although the procedure in Eq.~\eqref{eq12} correctly determines the separable state in the phase $\vert 1,1/2,1/2\rangle$ with the maximally entangled spin dimer $\mu_iS_i$, this approach fails if the degree of the spin dimer $\mu_iS_i$ entanglement is reduced (e.g., within the $\vert 2,1/2,3/2\rangle\&\vert 2,3/2,1/2\rangle$ phases). 
Focusing on the thermal stability, as illustrated in Fig.~\ref{fig8}($b$)-($c$), we observe a qualitatively similar behavior as before for $J_1/J\geq 1$, but with a significantly smaller degree of entanglement compared to previous cases. In this interaction limit, two relevant characteristics, namely the threshold temperature and the threshold magnetic field, exhibit almost identical features to those observed for $\nu_{q_1q_2q_3q_4}$.
Conversely, in the opposite interaction limit represented by $J_1/J=0.5$, the thermal stability is entirely different from that of  $\nu_{q_1q_2q_3q_4}$. While the threshold temperature changes almost identically to the case of full bisections with the variation of $J_1/J$, the threshold magnetic field at sufficiently low temperature is shifted to a significantly higher value. Moreover, in the close vicinity of zero temperature and magnetic field values smaller than the threshold magnetic field, a region of full separation is detected in the respective density plot (e.g., Fig.~\ref{fig8}$(b)$).  It has been identified that this area of full separation is dramatically reduced as temperature tends to zero with increasing $J_1/J\to 1$, and relatively strong thermal entanglement is identified even in a region where its existence is unpredicted by the approach with all bisections.

\subsection{\label{exp}Genuine tetrapartite negativity in the heterotetranuclear Cu$_2$Ni$_2$ complex}

 In this section, we aim to apply our improved calculation procedure to assess the thermal stability of genuine tetrapartite entanglement in a real tetranuclear complex with a magnetic structure corresponding to the investigated mixed spin-(1/2,1) Heisenberg tetramer~\eqref{eq1}. There are numerous modern materials, which can be classified as molecular magnets with molecules composed of two magnetic ions with spin-1/2 and two magnetic ions with spin-1, respectively. Examples include heterodinuclear complexes like Fe$_2$Ni$_2$ ($\mu_{\rm Fe(III)}$=1/2, $S_{\rm Ni(II)}$=1)~\cite{Wu,Parkin,Rodriguez,Park} or Cu$_2$Ni$_2$ ($\mu_{\rm Cu(II)}$=1/2, $S_{\rm Ni(II)}$=1)~\cite{Lou,Ribas,Osa,Barrios,Nakamura} magnetic core.  
 
In the former compounds Fe$_2$Ni$_2$, the magnetic ions form an almost perfect orthogonal square plaquette with identical coupling constants around the plaquette. Magnetic data analyses have suggested relatively weak ferromagnetic intramolecular interactions within the Fe$_2$Ni$_2$ complexes~\cite{Wu,Parkin,Rodriguez,Park},  implying the absence of genuine as well as global bipartite entanglement at both zero and non-zero temperatures.
 Conversely, the Cu$_2$Ni$_2$ complexes exhibit a wide variety of configurations depending on the type of nonmagnetic ligands present, which determine both the planar geometry of the magnetic unit and the nature of intramolecular exchange interactions. For instance, the magnetic ions in Cu$_2$Ni$_2$ complexes may adopt a square planar coordination geometry with two identical antiferromagnetic couplings between different magnetic ions ($J/k_B$=$J_1/k_B$=45.46 K)~\cite{Osa}.  Alternatively, two strongly antiferromagnetically coupled (CuNi) dimers may weakly interact ferromagnetically, with estimated coupling strengths of  $J/k_B$=165.42 K (intra-dimer) and $J_1/k_B$=-0.04 K (inter-dimer)~\cite{Ribas}.
Additionally,  the magnetic unit could consist of two separate (CuNi) dimers localized in a tetranuclear linear array, for which DFT calculations predict a strong antiferromagnetic intra-dimer coupling ($J/k_B$=142.69K) and a weak antiferromagnetic inter-dimer coupling ($J_1/k_B$=0.12K)~\cite{Barrios}.  Furthermore,   Cu$_2$Ni$_2$ complexes may also exhibit a defect double-cubane structure, wherein weak antiferromagnetic intra-dimer (CuNi) interactions ($J/k_B$=10.07 K) and weak ferromagnetic inter-dimer interactions ($J_1/k_B$=-5.75 K) have been estimated~\cite{Nakamura}.
The presence of two antiferromagnetic couplings in Cu$_2$Ni$_2$ complexes provides most favorable condition for the existence of an entanglement.
\begin{figure*}[t!]
\begin{center}
{\includegraphics[width=.32\textwidth,trim=3.2cm 8.5cm 5.6cm 8.cm, clip]{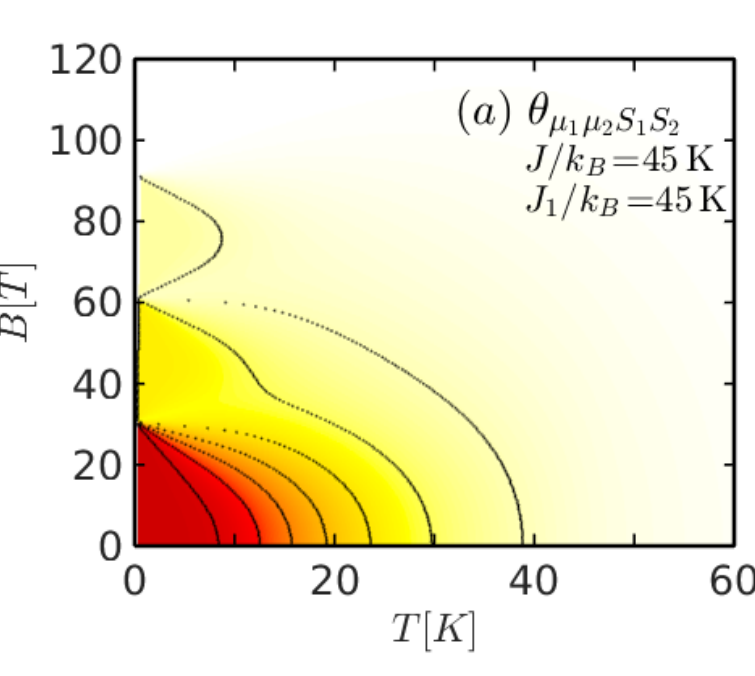}}
{\includegraphics[width=.29\textwidth,trim=4.3cm 8.5cm 5.6cm 8.cm, clip]{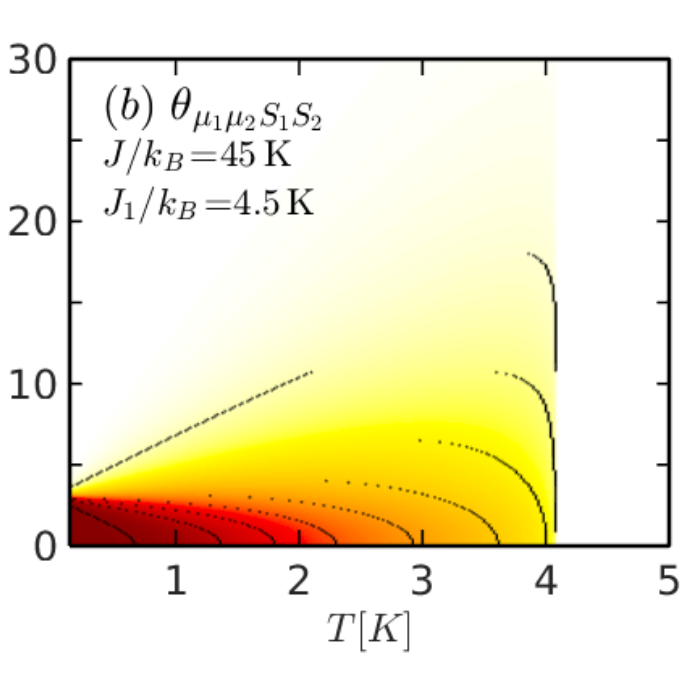}}
{\includegraphics[width=.355\textwidth,trim=4.2cm 8.5cm 3.cm 8.cm, clip]{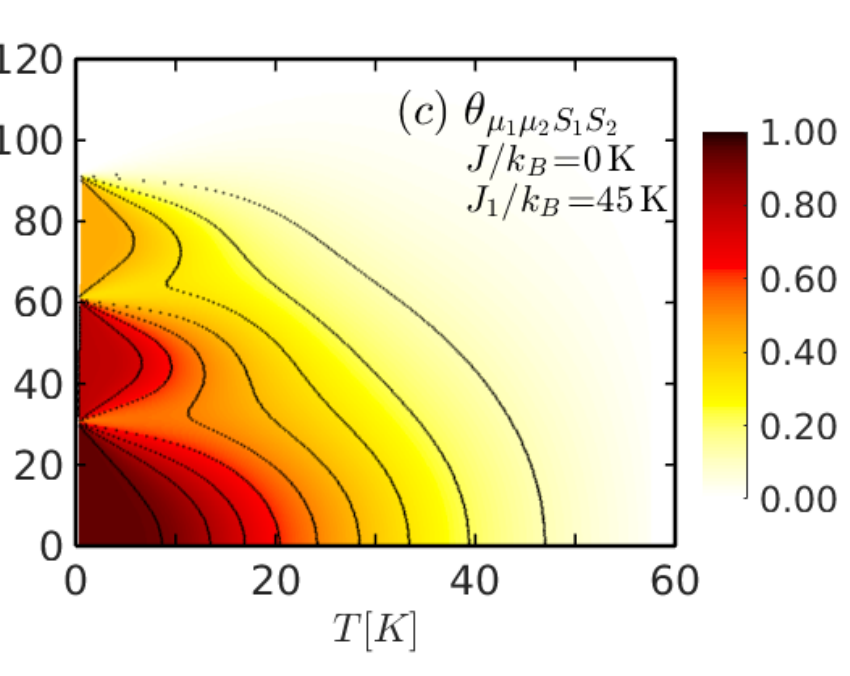}}
\begin{tikzpicture}[overlay]
 \node  at (-13.8, 3.5) {{\includegraphics[width=.09\textwidth,trim=0.2cm 5cm 15.5cm 19.5cm, clip]{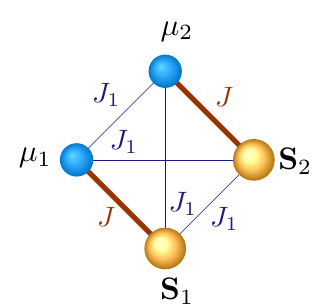}}};
\node  at (-7.2, 3.55) {{\includegraphics[width=.09\textwidth,trim=0.2cm 5cm 14.5cm 19.5cm, clip]{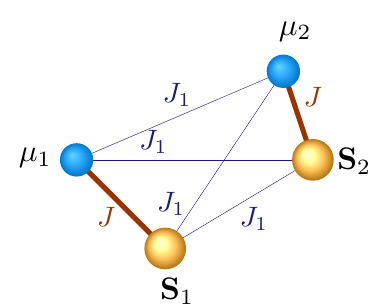}}};
\node  at (-3.7, 3.5) {{\includegraphics[width=.07\textwidth,trim=1cm 5cm 16cm 21cm, clip]{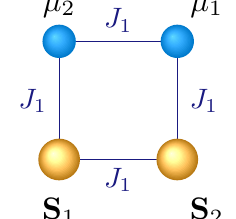}}};
\end{tikzpicture}
\end{center}
\caption{Density plots of genuine  tetrapartite negativity $\theta_{\mu_1\mu_2S_1S_2}$ of a mixed spin-(1/2,1) Heisenberg tetramer in the $k_BT/J-h/J$ plane calculated for three set of model parameters represent ($a$) an ideal tetrahedral geometry, ($b$) a distorted tetrahedral geometry and ($c$) a square planar geometry. Land\'e $g$-factor is identical for both magnetic ions, $g=2.2$. The black contour  lines correspond to values from 0.1 to 1 with a step 0.1 in order (from left to right).}
\label{fig11}
\end{figure*}

For further analysis, we choose three sets of exchange interactions, $J/k_B, J_1/k_B$, to approximate real materials: ($a$) an ideal case of perfect tetrahedral geometry with  $J/k_B=$45 K and $J_1/k_B=$45 K;  ($b$) a distorted tetrahedral geometry with $J/k_B=$45 K and $J_1/k_B=$4.5 K, very close to the complexes analyzed in Ref.~\cite{Barrios,Ribas}, in which the intra-dimer interaction $J/k_B$  is significantly higher than the inter-dimer one $J_1/k_B$, although some interactions in real complexes may be absent; ($c$) a square planar geometry with  $J/k_B=$0 K and $J_1/k_B=$45 K, emulating the situation in a tetranuclear bimetallic complex discussed in Ref.~\cite{Osa}.  An identical value of the Land\'e g-factor $g=2.2$  for each magnetic ion is implemented in our calculations, where the $\theta_{q_1q_2q_3q_4}$  approach has been selected.

 The obtained results are depicted in  Fig.~\ref{fig11}.  Interestingly, we find that in cases ($a$) and ($c$), the tetranuclear bimetallic complexes could exhibit thermal entanglement up to relatively high temperatures, around $T\sim 55$ K. This temperature range is comparable to the relevant inter-dimer exchange coupling  $J_1/k_B$,  suggesting the possibility of optimizing the synthesis of novel heteronuclear bimetallic complexes as better candidates for technical applications at room temperature. In such geometries, thermal entanglement persists up to relatively high threshold magnetic fields  $B\sim 100$ T, which gradually decrease with increasing thermal fluctuations. In case ($a$), the decrease of the threshold magnetic field is visibly faster compared to case ($c$) due to weaker zero-temperature entanglement caused by ground-state degeneracy at the special case $J_1/J=1$.
 As illustrated in Fig.~\ref{fig11}($b$), thermal entanglement is also possible in the case of strong nonequivalence between both intramolecular couplings, see the case ($b$). Nevertheless, thermal negativity is possible only at very low temperatures ($T \lesssim 4$ K) and is rapidly destroyed by a relatively weak magnetic field of $\sim 10$ T.

Considering the obtained dependencies, one can theoretically predict that heteronuclear bimetallic complexes with a Cu$_2$Ni$_2$ magnetic  core and a relatively strong antiferromagnetic  coupling within two CuNi dimers should be appropriate materials for modern smart devices utilizing the entanglement property.

\section{\label{conclusion}Conclusion}
In this study, we rigorously analyzed the genuine tetrapartite negativity within  the mixed spin-(1/2,1) Heisenberg model on a square plaquette. By employing the method of full exact diagonalization, we determined all global bipartite negativities with respect to all permutations of a central spin and a central spin dimer. An in-depth understanding of these negativities is crucial for determining the global tetrapartite negativity.  Focusing on the global bipartite negativity at zero temperature, we showed that the simultaneous consideration of  ferromagnetic intra-dimer ($J<0$) and inter-dimer ($J_1<0$) couplings favors the fully polarized separable state, in which quantum entanglement is strictly prohibited in this parameter space. The antiferromagnetic character of the intra-dimer ($J>0$) interaction allows  bipartite entanglement, except in the case of two $\mu_iS_i$ dimers, which exhibit a very narrow region of stability with respect to the applied magnetic field. For an antiferromagnetic inter-dimer coupling ($J_1>0$), we identified that the zero-temperature global bipartite negativity with a single central spin never exceeds the magnitude of the respective central spin and is gradually reduced by an increasing external magnetic field due to magnetic-field-driven structural changes. The same observation can also be made for a central spin dimer, except in the case of an antiferromagnetic intra-dimer coupling $0<J_1/J<1$. In this limit, the investigated mixed spin-(1/2,1) Heisenberg tetramer model~\eqref{eq1} is biseparable for two ground states  ($\vert 1,1/2,1/2\rangle$ and $\vert 2,1/2,3/2\rangle\&\vert 2,3/2,1/2\rangle$) and additionally exhibits an unusually high degree of entanglement at the $\vert 0,1/2,1/2\rangle$ phase due to the relatively high contribution of a few reduced negativities.
 
The most interesting and novel observations pertain to the implementation of three proposed calculation procedures for quantifying genuine entanglement in a tetrapartite system. The first approach, inspired by the works of Love~\cite{Love} and Ghahi~\cite{Ghahi}, suggests that genuine tetrapartite entanglement can be quantified as a geometric mean of all bipartite contributions involving a central spin as well as a central spin dimer. While this procedure was discussed conceptually, it had not been applied to investigate specific quantum spin systems before. The second approach is an improved quantification method~\cite{Oliveira,Li2016,Torres,Dong2019,Dong2020a,Dong2020,Zad2022} 
based on the CKW monogamy relation~\cite{CKW}, which additionally includes bisections with a central spin-dimer. The last approach defines a geometric mean of all tripartite contributions evaluated through the standard definition~\cite{Sabin}. It was illustrated that all three approaches lead to qualitatively and almost quantitatively identical results, indicating that genuine tetrapartite negativity can only exist for antiferromagnetic inter-dimer coupling, with both ferromagnetic and antiferromagnetic intra-dimer couplings accepted. Furthermore, it was shown that entangled states with $\sigma_1=\sigma_2=3/2$ are thermally stable over a wide range of temperature and magnetic fields, with the threshold temperature linearly increasing with an increment of inter-dimer interaction. Conversely, the thermal stability of entangled states with $\sigma_1,\sigma_2\neq3/2$  is  confined to sufficiently low temperatures and magnetic fields due to spontaneous induction at very small magnetic fields. A comparison of our method with previous ones~\cite{Oliveira,Li2016,Torres,Dong2019,Dong2020a,Dong2020,Zad2022} 
demonstrated the deficiency of the previous procedure, which led to the incorrect detection of genuine entanglement in states where two weakly entangled mixed spin-(1/2,1) dimers are fully separable.

The investigation of this system was motivated by the  tetranuclear bimetallic complexes with various magnetic cores, consisting of a molecular unit comprising two magnetic ions with a spin-1/2 and two magnetic ions with a spin-1. Examples of such complexes include   Fe$_2$Ni$_2$~\cite{Wu,Parkin,Rodriguez,Park} and Cu$_2$Ni$_2$~\cite{Lou,Ribas,Osa,Barrios,Nakamura} molecular complexes, which are characterized by  diverse  intramolecular exchange interactions. By employing the estimated values of both couplings within  three designed lattice geometries, we have demonstrated that genuine tetrapartite entanglement can indeed occur in real bimetallic complexes. Furthermore, we discussed how the thermal stability of genuine tetrapartite entanglement is strongly influenced by the strength of the inter-dimer coupling. Specifically, we highlighted that for optimal entanglement across a wide range of temperature and magnetic field conditions, the strength of the inter-dimer coupling  should be at least proportional to the intra-dimer coupling. This insight underscores the importance of understanding the interplay between different coupling parameters in determining the entanglement properties of these systems.
\\\\
{\bf Acknowledgments}\\
This work was financially supported by the grant of the Slovak Research and Development Agency provided under the contract No. APVV-20-0150 and by the grant of  Sciences and The Ministry of Education, Science, Research, and Sport of the Slovak Republic provided under the contract No. VEGA 1/0695/23. 
\newline
\\
\appendix
{\bf Appendix}\\
\section{\label{App A} }

\setcounter{equation}{0}
\setcounter{table}{0}
\renewcommand{\theequation}{\thesection.\arabic{equation}}
\renewcommand{\thetable}{\thesection.\arabic{table}}

The ground-state eigenvectors $\vert \sigma_T^z,\sigma_1,\sigma_2\rangle$  of the mixed spin-(1/2,1) Heisenberg tetramer~\eqref{eq1} for non-zero $h/J$ are defined in terms of a linear combination of standard basis vectors $\vert \mu_1^z,S_1^z,\mu_2^z,S_2^z\rangle$~\cite{Vargova2023}
\begin{align}
\vert 3,\tfrac{3}{2},\tfrac{3}{2}\rangle&=\vert\tfrac{1}{2},1,\tfrac{1}{2},1\rangle;\nonumber\\
\vert 2,\tfrac{3}{2},\tfrac{3}{2}\rangle&=\frac{1}{\sqrt{6}}\left( 
\sqrt{2}\vert\tfrac{1}{2},1,\tfrac{1}{2},0\rangle+\vert\tfrac{1}{2},1,-\tfrac{1}{2},1\rangle
-\sqrt{2}\vert\tfrac{1}{2},0,\tfrac{1}{2},1\rangle-\vert-\tfrac{1}{2},1,\tfrac{1}{2},1\rangle
\right);\nonumber\\
\vert 2,\tfrac{1}{2},\tfrac{3}{2}\rangle&=\frac{1}{\sqrt{3}}\left( 
\vert\tfrac{1}{2},1,\tfrac{1}{2},0\rangle-\sqrt{2}\vert\tfrac{1}{2},1,-\tfrac{1}{2},1\rangle\right)
;\nonumber\\
\vert2,\tfrac{3}{2},\tfrac{1}{2}\rangle&=
\frac{1}{\sqrt{3}}\left( 
\vert\tfrac{1}{2},0,\tfrac{1}{2},1\rangle-\sqrt{2}\vert-\tfrac{1}{2},1,\tfrac{1}{2},1\rangle\right)
;\nonumber\\
\vert 1,\tfrac{3}{2},\tfrac{3}{2}\rangle&=\frac{1}{\sqrt{10}}\left( 
\vert\tfrac{1}{2},1,\tfrac{1}{2},-1\rangle+\sqrt{2}\vert\tfrac{1}{2},1,-\tfrac{1}{2},0\rangle
-\frac{4}{3}\vert\tfrac{1}{2},0,\tfrac{1}{2},0\rangle-\frac{2\sqrt{2}}{3}\vert\tfrac{1}{2},0,-\tfrac{1}{2},1\rangle\right.
\nonumber\\
&\left.\hspace*{1.1cm}-\frac{2\sqrt{2}}{3}\vert-\tfrac{1}{2},1,\tfrac{1}{2},0\rangle-\frac{2}{3}\vert-\tfrac{1}{2},1,-\tfrac{1}{2},1\rangle
+\vert\tfrac{1}{2},-1,\tfrac{1}{2},1\rangle+\sqrt{2}\vert-\tfrac{1}{2},0,\tfrac{1}{2},1\rangle
\right);\nonumber\\
\vert 1,\tfrac{1}{2},\tfrac{1}{2}\rangle&=\frac{1}{3}\left( 
\vert\tfrac{1}{2},0,\tfrac{1}{2},0\rangle-\sqrt{2}\vert\tfrac{1}{2},0,-\tfrac{1}{2},1\rangle-\sqrt{2}\vert-\tfrac{1}{2},1,\tfrac{1}{2},0\rangle+2\vert-\tfrac{1}{2},1,-\tfrac{1}{2},1\rangle
\right);\nonumber\\
\vert 1,\tfrac{1}{2},\tfrac{3}{2}\rangle&=\frac{1}{6}\left( 
\sqrt{2}\vert\tfrac{1}{2},0,\tfrac{1}{2},0\rangle+\vert\tfrac{1}{2},0,-\tfrac{1}{2},1\rangle
-2\vert-\tfrac{1}{2},1,\tfrac{1}{2},0\rangle
-\sqrt{2}\vert-\tfrac{1}{2},1,-\tfrac{1}{2},1\rangle\right.
\nonumber\\
&\left.\hspace*{0.9cm}-3\sqrt{2}\vert\tfrac{1}{2},-1,\tfrac{1}{2},1\rangle
+3\vert-\tfrac{1}{2},0,\tfrac{1}{2},1\rangle
\right);\nonumber\\
\vert 1,\tfrac{3}{2},\tfrac{1}{2}\rangle&=-\frac{1}{6}\left( 
\sqrt{2}\vert\tfrac{1}{2},0,\tfrac{1}{2},0\rangle+\vert-\tfrac{1}{2},1,\tfrac{1}{2},0\rangle
-2\vert\tfrac{1}{2},0,-\tfrac{1}{2},1\rangle
-\sqrt{2}\vert-\tfrac{1}{2},1,-\tfrac{1}{2},1\rangle\right.
\nonumber\\
&\left.\hspace*{0.9cm}-3\sqrt{2}\vert\tfrac{1}{2},1,\tfrac{1}{2},-1\rangle
+3\vert\tfrac{1}{2},1,-\tfrac{1}{2},0\rangle
\right);\nonumber\\
\vert 0,\tfrac{3}{2},\tfrac{3}{2}\rangle&=\frac{1}{6}\left(
3\vert\tfrac{1}{2},1,-\tfrac{1}{2},-1\rangle-\sqrt{2}\vert\tfrac{1}{2},0,\tfrac{1}{2},-1\rangle-2\vert\tfrac{1}{2},0,-\tfrac{1}{2},0\rangle-\vert-\tfrac{1}{2},1,\tfrac{1}{2},-1\rangle-\sqrt{2}\vert-\tfrac{1}{2},1,-\tfrac{1}{2},0\rangle\right.\nonumber\\
&\left.\hspace*{0.6cm}-3\vert-\tfrac{1}{2},-1,\tfrac{1}{2},1\rangle
+\sqrt{2}\vert-\tfrac{1}{2},0,-\tfrac{1}{2},1\rangle+2\vert-\tfrac{1}{2},0,\tfrac{1}{2},0\rangle+\vert\tfrac{1}{2},-1,-\tfrac{1}{2},1\rangle+\sqrt{2}\vert\tfrac{1}{2},-1,\tfrac{1}{2},0\rangle
\right);\nonumber\\
\vert 0,\tfrac{1}{2},\tfrac{1}{2}\rangle&=\frac{1}{3\sqrt{2}}\left(
\sqrt{2}\vert\tfrac{1}{2},0,\tfrac{1}{2},-1\rangle-\vert\tfrac{1}{2},0,-\tfrac{1}{2},0\rangle-2\vert-\tfrac{1}{2},1,\tfrac{1}{2},-1\rangle+\sqrt{2}\vert-\tfrac{1}{2},1,-\tfrac{1}{2},0\rangle\right.\nonumber\\
&\left.\hspace*{1.1cm}-\sqrt{2}\vert-\tfrac{1}{2},0,-\tfrac{1}{2},1\rangle+\vert-\tfrac{1}{2},0,\tfrac{1}{2},0\rangle+2\vert\tfrac{1}{2},-1,-\tfrac{1}{2},1\rangle-\sqrt{2}\vert\tfrac{1}{2},-1,\tfrac{1}{2},0\rangle
\right).
\label{a00}
\end{align}
The analytical expressions of the phase boundaries emerging in the ground-state phase diagrams depicted in Fig.~\ref{fig2} and Fig.~\ref{fig3} are given by
\begin{align}
\vert 0,\tfrac{3}{2},\tfrac{3}{2}\rangle-\vert 1,\tfrac{3}{2},\tfrac{3}{2}\rangle:&\; h=J_1;\hspace*{2cm}
\vert 0,\tfrac{1}{2},\tfrac{1}{2}\rangle-\vert 1,\tfrac{1}{2},\tfrac{1}{2}\rangle:\; h=J_1;
\nonumber\\
\vert 1,\tfrac{3}{2},\tfrac{3}{2}\rangle-\vert 2,\tfrac{3}{2},\tfrac{3}{2}\rangle:&\; h=2J_1;\hspace*{1.8cm}
\vert 1,\tfrac{1}{2},\tfrac{1}{2}\rangle-\vert 2,\tfrac{1}{2},\tfrac{3}{2}\rangle\&\vert 2,\tfrac{3}{2},\tfrac{1}{2}\rangle:\; h=\frac{3}{2}J+\frac{J_1}{2};
\nonumber\\
\vert 2,\tfrac{3}{2},\tfrac{3}{2}\rangle-\vert 3,\tfrac{3}{2},\tfrac{3}{2}\rangle:&\; h=3J_1;\hspace*{1.8cm}
\vert 2,\tfrac{1}{2},\tfrac{3}{2}\rangle\&\vert 2,\tfrac{3}{2},\tfrac{1}{2}\rangle-\vert 3,\tfrac{3}{2},\tfrac{3}{2}\rangle:\; h=\frac{3}{2}J+\frac{3}{2}J_1;
\nonumber\\
\vert 1,\tfrac{1}{2},\tfrac{1}{2}\rangle-\vert 3,\tfrac{3}{2},\tfrac{3}{2}\rangle:&\; h=\frac{3}{2}J+J_1;\hspace*{1.05cm}
\vert 0,\tfrac{1}{2},\tfrac{1}{2}\rangle-\vert 0,\tfrac{3}{2},\tfrac{3}{2}\rangle:\; J=J_1;
\nonumber\\
\vert 1,\tfrac{1}{2},\tfrac{1}{2}\rangle-\vert 1,\tfrac{3}{2},\tfrac{3}{2}\rangle:&\; J=J_1;\hspace*{1.95cm}
\vert 2,\tfrac{1}{2},\tfrac{3}{2}\rangle\&\vert 2,\tfrac{3}{2},\tfrac{1}{2}\rangle-\vert 2,\tfrac{3}{2},\tfrac{3}{2}\rangle:\; J=J_1.
\label{a0}
\end{align}
The  density operator $\hat{\rho}_{\mu_{1}S_1\mu_{2}S_{2}}$ is defined in a standard form 
\begin{align}
\hat{\rho}_{\mu_{1}S_1\mu_{2}S_{2}}&\!=\!\frac{1}{\cal Z}\sum_{k=1}^{36}  {\rm e}^{-\beta \varepsilon_k}|\psi_k\rangle\langle\psi_k|,
\label{a1}
\end{align}
where ${\cal Z}$ is a partition function given by ${\cal Z}=\sum_{k=1}^{36}{\rm e}^{-\beta \varepsilon_k}$ with $\beta=1/k_BT$ representing the reciprocal temperature. Here, $k_B$ is  Boltzmann's constant, and $\varepsilon_k$ ($\psi_k$) denotes the k-th eigenvalue (respective eigenvector) of the Hamiltonian~\eqref{eq1}. The explicit form of the partition function ${\cal Z}$, as well as the eigenvalues and eigenvectors, have been derived in our previous study ~\cite{Vargova2023}.
The  corresponding  density matrix $\hat{\rho}_{\mu_{1}S_1\mu_{2}S_{2}}$  in the basis of $\vert \mu^z_{1},S^z_1,\mu_{2}^z,S^z_{2}\rangle$ is a square matrix of size $36\times 36$ 
\begin{align}
\allowdisplaybreaks
{\hat{\rho}_{\mu_{1}S_1\mu_{2}S_{2}}}\!=\!\left(
\begin{array}{cc}
A(h) & B(h)\\
B^T(h)& A^{\tau}(-h)\\
\end{array}\right).
\label{a2}
\end{align}
In this notation, $A(h)$ and $B(h)$ represent two $18\times 18$ matrices with the following explicit forms
\begin{align}
\allowdisplaybreaks
A(h)\!=\!
\resizebox{0.92\textwidth}{!}{$
\begin{blockarray}{r cccccc cccccc cccccc }
 & \vert \varphi_1^+\rangle & \vert \varphi_2^+\rangle &\vert  \varphi_6^+\rangle  
 & \vert  \varphi_3^+\rangle & \vert  \varphi_{7}^+\rangle &\vert  \varphi_{14}^-\rangle
 & \vert  \varphi_4^+\rangle & \vert  \varphi_8^+\rangle &\vert  \varphi_{15}^-\rangle & \vert  \varphi_{9}^+\rangle & \vert  \varphi_{16}^-\rangle & \vert  \varphi_{13}^-\rangle
 & \vert  \varphi_{12}^+\rangle & \vert  \varphi_{18}^+\rangle &\vert  \varphi_{11}^-\rangle & \vert  \varphi_{17}^+\rangle & \vert  \varphi_{10}^-\rangle & \vert  \varphi_{5}^-\rangle
 \\
\begin{block}{r(cccccc cccccc cccccc)}
\vert \varphi_1^+\rangle  \;\;\;&\rho_{1,1}(h) & 0 & 0 & 0 & 0 & 0& 0 & 0 & 0& 0 & 0 & 0 & 0 & 0 & 0& 0 & 0 & 0\\
\vert \varphi_2^+\rangle  \;\;\;&0 & \rho_{2,2}(h) & 0 & \rho_{2,4}(h) & 0 & 0 & \rho_{2,7}(h) & 0 & 0 & 0 & 0 & 0& 0 & 0 & 0& 0 & 0 & 0\\
\vert \varphi_6^+\rangle \;\;\; & 0 & 0 & \rho_{3,3}(h) &  0 &\rho_{3,5}(h)& 0&  0 &\rho_{3,8}(h)& 0& \rho_{3,10}(h) & 0 & 0 & \rho_{3,13}(h) & 0 & 0& 0 & 0 & 0\\
 \vert \varphi_3^+\rangle \; \;\;& 0& \rho_{4,2}(h) &0 & \rho_{4,4}(h) & 0 & 0 &\rho_{4,7}(h) & 0 & 0 & 0 & 0 & 0& 0 & 0 & 0 & 0 & 0&0\\
\vert \varphi_7^+\rangle \;\;\;&0 &0 & \rho_{5,3}(h) &0  &  \rho_{5,5}(h) & 0 &  0 &\rho_{5,8}(h)& 0& \rho_{5,10}(h) & 0  & 0 & \rho_{5,13}(h) & 0 & 0& 0 & 0 & 0\\
 \vert \varphi_{14}^-\rangle \;\;\;& 0 & 0 & 0 & 0 & 0 & \rho_{6,6}(h)& 0 & 0 & \rho_{6,9}(h) & 0 & \rho_{6,11}(h)& 0 &0&\rho_{6,14}(h)& 0 &\rho_{6,16}(h)& 0 &0\\
 \vert \varphi_4^+\rangle  \;\;\;&0 & \rho_{7,2}(h) & 0 & \rho_{7,4}(h) & 0 & 0 & \rho_{7,7}(h) & 0 & 0 & 0 & 0 & 0& 0 & 0 & 0& 0 & 0 & 0
 \\
 \vert \varphi_8^+\rangle  \;\;\;& 0 & 0 & \rho_{8,3}(h) &  0 &\rho_{8,5}(h)& 0&  0 &\rho_{8,8}(h)& 0& \rho_{8,10}(h) & 0 & 0& \rho_{8,13}(h)  & 0 & 0& 0 & 0 & 0
 \\
 \vert \varphi_{15}^-\rangle  \;\;\;&0 & 0 & 0 & 0 & 0 & \rho_{9,6}(h)& 0 & 0 & \rho_{9,9}(h) & 0 & \rho_{9,11}(h)& 0& 0 & \rho_{9,14}(h)  & 0& \rho_{9,16}(h)  & 0 & 0
 \\
  \vert \varphi_9^+\rangle \;\;\;& 0 & 0 & \rho_{10,3}(h) &  0 &\rho_{10,5}(h)& 0&  0 &\rho_{10,8}(h)& 0& \rho_{10,10}(h) & 0 & 0& \rho_{10,13}(h)  & 0 & 0& 0 & 0 & 0
 \\
 \vert \varphi_{16}^-\rangle  \;\;\;&0 & 0 & 0 & 0 & 0 & \rho_{11,6}(h)& 0 & 0 & \rho_{11,9}(h) & 0 & \rho_{11,11}(h)& 0& 0 & \rho_{11,14}(h)  & 0& \rho_{11,16}(h)  & 0 & 0
 \\
 \vert \varphi_{13}^-\rangle  \;\;\;&0 & 0 & 0 & 0 & 0& 0 & 0 & 0& 0 & 0 & 0 & \rho_{12,12}(h)& 0 & 0 & \rho_{12,15}(h) & 0 & \rho_{12,17}(h)  & 0
 \\
  \vert \varphi_{12}^+\rangle  \;\;\;&0 & 0 &  \rho_{13,3} (h)& 0 &  \rho_{13,5}(h)& 0 & 0 &  \rho_{13,8}(h)& 0 &  \rho_{13,10}(h) & 0 & 0& \rho_{13,13}(h) & 0 & 0 & 0 & 0  & 0
 \\
  \vert \varphi_{18}^+\rangle  \;\;\;&0 & 0 &  0 & 0 &  0& \rho_{14,6}(h) & 0 & 0& \rho_{14,9}(h)& 0 &  \rho_{14,11}(h) & 0 & 0& \rho_{14,14}(h) & 0 & \rho_{14,16}(h) & 0  & 0
 \\
  \vert \varphi_{11}^-\rangle  \;\;\;&0 & 0 &  0 & 0 &  0& 0 & 0 & 0& 0&0& 0 &  \rho_{15,12}(h) & 0 & 0& \rho_{15,15}(h) & 0 & \rho_{15,17}(h)  & 0
 \\
  \vert \varphi_{17}^+\rangle  \;\;\;&0 & 0 &  0 & 0 &  0& \rho_{16,6}(h) & 0 & 0& \rho_{16,9}(h)& 0 &  \rho_{16,11}(h) & 0 & 0& \rho_{16,14}(h) & 0 & \rho_{16,16}(h)& 0  & 0
 \\
 \vert \varphi_{10}^-\rangle  \;\;\;&0 & 0 &  0 & 0 &  0& 0 & 0 & 0& 0&0& 0 &  \rho_{17,12}(h) & 0 & 0& \rho_{17,15}(h) & 0 & \rho_{17,17}(h)  & 0
 \\
 \vert \varphi_{5}^-\rangle  \;\;\;&0 & 0 &  0 & 0 &  0& 0 & 0 & 0& 0& 0 &  0 & 0 & 0& 0 & 0 & 0  & 0 & \rho_{18,18}(h)
 \\
\end{block}
\end{blockarray}\;\;,$}
\label{b3}
\end{align}
\begin{align}
\allowdisplaybreaks
B(h)\!=\!
\resizebox{0.92\textwidth}{!}{$
\begin{blockarray}{r cccccc cccccc cccccc }
 & \vert \varphi_5^+\rangle & \vert \varphi_{10}^+\rangle &\vert  \varphi_{17}^-\rangle  
 & \vert  \varphi_{11}^+\rangle & \vert  \varphi_{18}^-\rangle &\vert  \varphi_{12}^-\rangle
 & \vert  \varphi_{13}^+\rangle & \vert  \varphi_{16}^+\rangle &\vert  \varphi_{9}^-\rangle & \vert  \varphi_{15}^+\rangle & \vert  \varphi_{8}^-\rangle & \vert  \varphi_{4}^-\rangle
 & \vert  \varphi_{14}^+\rangle & \vert  \varphi_{7}^-\rangle &\vert  \varphi_{3}^-\rangle & \vert  \varphi_{6}^-\rangle & \vert  \varphi_{2}^-\rangle & \vert  \varphi_{1}^-\rangle
 \\
\begin{block}{r(cccccc cccccc cccccc)}
\vert \varphi_1^+\rangle  \;\;\;&0 & 0 & 0 & 0 & 0 & 0& 0 & 0 & 0& 0 & 0 & 0 & 0 & 0 & 0& 0 & 0 & 0\\
\vert \varphi_2^+\rangle  \;\;\;& \rho_{2,19}(h) &0 & 0 & 0 & 0 & 0 & 0 & 0 & 0 & 0 & 0 & 0& 0 & 0 & 0& 0 & 0 & 0\\
\vert \varphi_6^+\rangle \;\;\; & 0 & \rho_{3,20}(h) &  0 &\rho_{3,22}(h)& 0&  0 &\rho_{3,25}(h)& 0& 0 & 0 & 0 & 0&0 & 0 & 0& 0 & 0 & 0\\
 \vert \varphi_3^+\rangle \; \;\;& \rho_{4,19}(h)& 0 &0 & 0 & 0 & 0 &0 & 0 & 0 & 0 & 0 & 0& 0 & 0 & 0 & 0 & 0&0\\
\vert \varphi_7^+\rangle \;\;\;& 0 & \rho_{5,20}(h) &  0 &\rho_{5,22}(h)& 0&  0 &\rho_{5,25}(h)& 0& 0 & 0 & 0 & 0&0 & 0 & 0& 0 & 0 & 0\\
 \vert \varphi_{14}^-\rangle \;\;\;& 0 & 0 & \rho_{6,21}(h) & 0 & \rho_{6,23}(h) & 0&0&\rho_{6,26}(h)& 0 & \rho_{6,28}(h) & 0&0 & \rho_{6,31}(h)& 0 &0& 0 & 0 &0\\
 \vert \varphi_4^+\rangle  \;\;\;& \rho_{7,19}(h)&0  & 0 & 0 & 0 & 0 & 0 & 0 & 0 & 0 & 0 & 0& 0 & 0 & 0& 0 & 0 & 0\\
 \vert \varphi_8^+\rangle  \;\;\;& 0 & \rho_{8,20}(h) &  0 &\rho_{8,22}(h)& 0&  0 &\rho_{8,25}(h)& 0& 0 & 0 & 0 & 0&0 & 0 & 0& 0 & 0 & 0\\
 \vert \varphi_{15}^-\rangle  \;\;\;& 0 & 0 & \rho_{9,21}(h) & 0 & \rho_{9,23}(h) & 0&0&\rho_{9,26}(h)& 0 & \rho_{9,28}(h) & 0&0 & \rho_{9,31}(h)& 0 &0& 0 & 0 &0\\
  \vert \varphi_9^+\rangle \;\;\;& 0 & \rho_{10,20}(h) &  0 &\rho_{10,22}(h)& 0&  0 &\rho_{10,25}(h)& 0& 0 & 0 & 0 & 0&0 & 0 & 0& 0 & 0 & 0\\
 \vert \varphi_{16}^-\rangle  \;\;\;& 0 & 0 & \rho_{11,21}(h) & 0 & \rho_{11,23}(h) & 0&0&\rho_{11,26}(h)& 0 & \rho_{11,28}(h) & 0&0 & \rho_{11,31}(h)& 0 &0& 0 & 0 &0\\
 \vert \varphi_{13}^-\rangle  \;\;\;&0 & 0 & 0 & 0 & 0& \rho_{12,24}(h) & 0 & 0& \rho_{12,27}(h) & 0 & \rho_{12,29}(h)& 0 & 0 & \rho_{12,32}(h) & 0 & \rho_{12,34}(h)&0  & 0 \\
  \vert \varphi_{12}^+\rangle  \;\;\;& 0 & \rho_{13,20}(h) &  0 &\rho_{13,22}(h)& 0&  0 &\rho_{13,25}(h)& 0& 0 & 0 & 0 & 0&0 & 0 & 0& 0 & 0 & 0\\
  \vert \varphi_{18}^+\rangle  \;\;\;& 0 & 0 & \rho_{14,21}(h)& 0 & \rho_{14,23}(h) & 0&0&\rho_{14,26}(h)& 0 & \rho_{14,28}(h)& 0&0 & \rho_{14,31}(h)& 0 &0& 0 & 0 &0\\
  \vert \varphi_{11}^-\rangle  \;\;\;&0 & 0 & 0 & 0 & 0& \rho_{15,24}(h) & 0 & 0& \rho_{15,27}(h) & 0 & \rho_{15,29}(h)& 0 & 0 & \rho_{15,32}(h) & 0 & \rho_{15,34}(h)&0  & 0 \\
  \vert \varphi_{17}^+\rangle  \;\;\;& 0 & 0 & \rho_{16,21}(h) & 0 & \rho_{16,23}(h) & 0&0&\rho_{16,26}(h)& 0 & \rho_{16,28}(h) & 0&0 & \rho_{16,31}(h)& 0 &0& 0 & 0 &0\\
   \vert \varphi_{10}^-\rangle  \;\;\;&0 & 0 & 0 & 0 & 0& \rho_{17,24}(h) & 0 & 0& \rho_{17,27}(h) & 0 & \rho_{17,29}(h)& 0 & 0 & \rho_{17,32}(h) & 0 & \rho_{17,34}(h)&0  & 0 \\
 \vert \varphi_{5}^-\rangle  \;\;\;&0 & 0 &  0 & 0 &  0& 0 & 0 & 0& 0& 0 &  0 & \rho_{18,30}(h)& 0 & 0 &  \rho_{18,33}(h)  & 0 & \rho_{18,35}(h) & 0
 \\
\end{block}
\end{blockarray}\;\;,$}
\label{b4}
\end{align}
The symbol $B^T(h)$ used in Eq.~\eqref{a2} denotes the transposed  matrix $B(h)$, while $A^{\tau}(-h)$ is a square matrix obtained from the original matrix $A(h)$ by transposing it  over the anti-diagonal with a substitution $h=-h$. Specifically,
\begin{align}
A^{\tau}(-h)=
J_{18}
A^T(-h)J_{18},
\label{b5}
\end{align}
 where $J_{18}$ is an 18$\times$18 reversal matrix with non-zero unit elements residing on the anti-diagonal.
Each non-zero density matrix element $\rho_{i,j}(h)$ has a specific form given by the following equations. 
\begin{flalign}
\rho_{1,1}(h)\!&=\!\frac{{\rm e}^{ \frac{\beta }{4}J_1}}{{\cal Z}}
{\rm e}^{3\beta h}{\rm e}^{-\beta(J_x+\frac{5}{2}J_1)},
\label{b6}\\
\rho_{2,2}(h)\!&=\!\frac{{\rm e}^{ \frac{\beta }{4}J_1}}{3{\cal Z}}
{\rm e}^{2\beta h}\left[{\rm e}^{-\beta(J_x-\frac{J_1}{2})}\!+\!{\rm e}^{-\beta(J_x+\frac{5}{2}J_1)}\!+\!{\rm e}^{-\beta(-\frac{J_x}{2}+J_1)} \right],
\label{b7}\\
\rho_{3,3}(h)\!&=\!\frac{{\rm e}^{ \frac{\beta }{4}J_1}}{30{\cal Z}}
{\rm e}^{\beta h}\left[5{\rm e}^{-\beta(J_x-\frac{J_1}{2})}\!+\!2{\rm e}^{-\beta(J_x+\frac{5}{2}J_1)}\!+\!3{\rm e}^{-\beta(J_x-\frac{5}{2}J_1)}\!+\!5{\rm e}^{-\beta(-\frac{J_x}{2}+J_1)} \!+\!15{\rm e}^{-\beta(-\frac{J_x}{2}-J_1)} \right],
\label{b8}\\
\rho_{4,4}(h)\!&=\!\frac{{\rm e}^{ \frac{\beta }{4}J_1}}{6{\cal Z}}
{\rm e}^{2\beta h}\left[{\rm e}^{-\beta(J_x-\frac{J_1}{2})}\!+\!{\rm e}^{-\beta(J_x+\frac{5}{2}J_1)}\!+\!4{\rm e}^{-\beta(-\frac{J_x}{2}+J_1)} \right],
\label{b9}\\
\rho_{5,5}(h)\!&=\!\frac{{\rm e}^{ \frac{\beta }{4}J_1}}{60{\cal Z}}
{\rm e}^{\beta h}\left[20{\rm e}^{-\beta(J_x-\frac{J_1}{2})}\!+\!8{\rm e}^{-\beta(J_x+\frac{5}{2}J_1)}\!+\!12{\rm e}^{-\beta(J_x-\frac{5}{2}J_1)}\!+\!5{\rm e}^{-\beta(-\frac{J_x}{2}+J_1)} \!+\!15{\rm e}^{-\beta(-\frac{J_x}{2}-J_1)} \right],
\label{b10}\\
\rho_{6,6}(h)\!&=\!\frac{{\rm e}^{ \frac{\beta }{4}J_1}}{20{\cal Z}}
\left[5{\rm e}^{-\beta(J_x-\frac{J_1}{2})}\!+\!{\rm e}^{-\beta(J_x+\frac{5}{2}J_1)}\!+\!9{\rm e}^{-\beta(J_x-\frac{5}{2}J_1)}\!+\!5{\rm e}^{-\beta(J_x-\frac{7}{2}J_1)} \right],
\label{b11}\\
\rho_{7,7}(h)\!&=\!\rho_{2,2}(h),
\label{b12}\\
\rho_{8,8}(h)\!&=\!\frac{{\rm e}^{ \frac{\beta }{4}J_1}}{45{\cal Z}}
{\rm e}^{\beta h}\left[12{\rm e}^{-\beta(J_x+\frac{5}{2}J_1)}\!+\!8{\rm e}^{-\beta(J_x-\frac{5}{2}J_1)}\!+\!5{\rm e}^{-\beta(-2J_x+\frac{J_1}{2})}\!+\!15{\rm e}^{-\beta(-\frac{J_x}{2}+J_1)} \!+\!5{\rm e}^{-\beta(-\frac{J_x}{2}-J_1)} \right],
\label{b13}\\
\rho_{9,9}(h)\!&=\!\frac{{\rm e}^{ \frac{\beta }{4}J_1}}{90{\cal Z}}
\left[5{\rm e}^{-\beta(J_x-\frac{J_1}{2})}\!+\!9{\rm e}^{-\beta(J_x+\frac{5}{2}J_1)}\!+\!{\rm e}^{-\beta(J_x-\frac{5}{2}J_1)}\!+\!5{\rm e}^{-\beta(J_x-\frac{7}{2}J_1)}\!+\!10{\rm e}^{-\beta(-2J_x+\frac{J_1}{2})} \!+\!10{\rm e}^{-\beta(-2J_x-\frac{J_1}{2})}\right.
\nonumber\\
\!&+\!\left.25{\rm e}^{-\beta(-\frac{J_x}{2}+J_1)}\!+\!25{\rm e}^{-\beta(-\frac{J_x}{2}-J_1)}\right],
\label{b14}\\
\rho_{10,10}(h)\!&=\!\frac{{\rm e}^{ \frac{\beta }{4}J_1}}{180{\cal Z}}
{\rm e}^{\beta h}\left[24{\rm e}^{-\beta(J_x+\frac{5}{2}J_1)}\!+\!16{\rm e}^{-\beta(J_x-\frac{5}{2}J_1)}\!+\!40{\rm e}^{-\beta(-2J_x+\frac{J_1}{2})}\!+\!75{\rm e}^{-\beta(-\frac{J_x}{2}+J_1)} \!+\!25{\rm e}^{-\beta(-\frac{J_x}{2}-J_1)} \right],
\label{b15}\\
\rho_{11,11}(h)\!&=\!\frac{{\rm e}^{ \frac{\beta }{4}J_1}}{90{\cal Z}}
\left[10{\rm e}^{-\beta(J_x-\frac{J_1}{2})}\!+\!18{\rm e}^{-\beta(J_x+\frac{5}{2}J_1)}\!+\!2{\rm e}^{-\beta(J_x-\frac{5}{2}J_1)}\!+\!10{\rm e}^{-\beta(J_x-\frac{7}{2}J_1)}\!+\!5{\rm e}^{-\beta(-2J_x+\frac{J_1}{2})} \!+\!5{\rm e}^{-\beta(-2J_x-\frac{J_1}{2})}\right.
\nonumber\\
\!&+\!\left.20{\rm e}^{-\beta(-\frac{J_x}{2}+J_1)}\!+\!20{\rm e}^{-\beta(-\frac{J_x}{2}-J_1)}\right],
\label{b16}\\
\rho_{12,12}(h)\!&=\!\rho_{25,25}(-h),
\label{b17}\\
\rho_{13,13}(h)\!&=\!\rho_{3,3}(h),
\label{b18}\\
\rho_{14,14}(h)\!&=\!\rho_{9,9}(h),
\label{b19}\\
\rho_{15,15}(h)\!&=\!\rho_{22,22}(-h),
\label{b20}\\
\rho_{16,16}(h)\!&=\!\rho_{21,21}(-h),
\label{b21}\\
\rho_{17,17}(h)\!&=\!\rho_{20,20}(-h),
\label{b22}\\
\rho_{18,18}(h)\!&=\!\rho_{19,19}(-h),
\label{b23}\\
\rho_{19,19}(h)\!&=\!\rho_{4,4}(h),
\label{b24}\\
\rho_{20,20}(h)\!&=\!\rho_{10,10}(h),
\label{b25}\\
\rho_{21,21}(h)\!&=\!\frac{{\rm e}^{ \frac{\beta }{4}J_1}}{180{\cal Z}}
\left[5{\rm e}^{-\beta(J_x-\frac{J_1}{2})}\!+\!9{\rm e}^{-\beta(J_x+\frac{5}{2}J_1)}\!+\!{\rm e}^{-\beta(J_x-\frac{5}{2}J_1)}\!+\!5{\rm e}^{-\beta(J_x-\frac{7}{2}J_1)}\!+\!40{\rm e}^{-\beta(-2J_x+\frac{J_1}{2})} \!+\!40{\rm e}^{-\beta(-2J_x-\frac{J_1}{2})}\right.
\nonumber\\
\!&+\!\left.40{\rm e}^{-\beta(-\frac{J_x}{2}+J_1)}\!+\!40{\rm e}^{-\beta(-\frac{J_x}{2}-J_1)}\right],
\label{b26}\\
\rho_{22,22}(h)\!&=\!\frac{{\rm e}^{ \frac{\beta }{4}J_1}}{90{\cal Z}}
{\rm e}^{\beta h}\left[6{\rm e}^{-\beta(J_x+\frac{5}{2}J_1)}\!+\!4{\rm e}^{-\beta(J_x-\frac{5}{2}J_1)}\!+\!40{\rm e}^{-\beta(-2J_x+\frac{J_1}{2})}\!+\!30{\rm e}^{-\beta(-\frac{J_x}{2}+J_1)} \!+\!10{\rm e}^{-\beta(-\frac{J_x}{2}-J_1)} \right],
\label{b27}\\
\rho_{23,23}(h)\!&=\!\rho_{14,14}(-h),
\label{b28}\\
\rho_{24,24}(h)\!&=\!\rho_{13,13}(-h),
\label{b29}\\
\rho_{25,25}(h)\!&=\!\rho_{5,5}(h),
\label{b30}\\
\rho_{26,26}(h)\!&=\!\rho_{11,11}(-h),
\label{b31}\\
\rho_{27,27}(h)\!&=\!\rho_{10,10}(-h),
\label{b32}\\
\rho_{28,28}(h)\!&=\!\rho_{9,9}(-h),
\label{b33}\\
\rho_{29,29}(h)\!&=\!\rho_{8,8}(-h),
\label{b34}\\
\rho_{30,30}(h)\!&=\!\rho_{7,7}(-h),
\label{b35}\\
\rho_{31,31}(h)\!&=\!\rho_{6,6}(-h),
\label{b36}\\
\rho_{32,32}(h)\!&=\!\rho_{5,5}(-h),
\label{b37}\\
\rho_{33,33}(h)\!&=\!\rho_{4,4}(-h),
\label{b38}\\
\rho_{34,34}(h)\!&=\!\rho_{3,3}(-h),
\label{b39}\\
\rho_{35,35}(h)\!&=\!\rho_{2,2}(-h),
\label{b40}\\
\rho_{36,36}(h)\!&=\!\rho_{1,1}(-h),
\label{b41}\\
\rho_{2,4}(h)\!&=\!\rho_{4,2}(h)\!=\!\sqrt{2}\frac{{\rm e}^{ \frac{\beta }{4}J_1}}{6{\cal Z}}
{\rm e}^{2\beta h}\left[{\rm e}^{-\beta(J_x-\frac{J_1}{2})}\!+\!{\rm e}^{-\beta(J_x+\frac{5}{2}J_1)} \!-\!2{\rm e}^{-\beta(-\frac{J_x}{2}+J_1)} \right],
\label{b42}\\
\rho_{2,7}(h)\!&=\!\rho_{7,2}(h)\!=\!-\frac{{\rm e}^{ \frac{\beta }{4}J_1}}{3{\cal Z}}
{\rm e}^{2\beta h}\left[{\rm e}^{-\beta(J_x-\frac{J_1}{2})}\!-\!{\rm e}^{-\beta(J_x+\frac{5}{2}J_1)} \right],
\label{b43}\\
\rho_{2,19}(h)\!&=\!\rho_{19,2}(h)\!=\!\frac{\sqrt{2}}{2}\rho_{2,7}(h),
\label{b44}\\
\rho_{3,5}(h)\!&=\!\rho_{5,3}(h)\!=\!\sqrt{2}\frac{{\rm e}^{ \frac{\beta }{4}J_1}}{60{\cal Z}}
{\rm e}^{\beta h}\left[10{\rm e}^{-\beta(J_x-\frac{J_1}{2})}\!+\!4{\rm e}^{-\beta(J_x+\frac{5}{2}J_1)}\!+\!6{\rm e}^{-\beta(J_x-\frac{5}{2}J_1)}\!-\!5{\rm e}^{-\beta(-\frac{J_x}{2}+J_1)} \!-\!15{\rm e}^{-\beta(-\frac{J_x}{2}-J_1)} \right],
\label{b45}\\
\rho_{3,8}(h)\!&=\!\rho_{8,3}(h)\!=\!\frac{{\rm e}^{ \frac{\beta }{4}J_1}}{30{\cal Z}}
{\rm e}^{\beta h}\left[4{\rm e}^{-\beta(J_x+\frac{5}{2}J_1)}\!-\!4{\rm e}^{-\beta(J_x-\frac{5}{2}J_1)}\!+\!5{\rm e}^{-\beta(-\frac{J_x}{2}+J_1)} \!-\!5{\rm e}^{-\beta(-\frac{J_x}{2}-J_1)} \right],
\label{b46}\\
\rho_{3,10}(h)\!&=\!\rho_{10,3}(h)\!=\!\sqrt{2}\frac{{\rm e}^{ \frac{\beta }{4}J_1}}{30{\cal Z}}
{\rm e}^{\beta h}\left[2{\rm e}^{-\beta(J_x+\frac{5}{2}J_1)}\!-\!2{\rm e}^{-\beta(J_x-\frac{5}{2}J_1)}\!-\!5{\rm e}^{-\beta(-\frac{J_x}{2}+J_1)} \!+\!5{\rm e}^{-\beta(-\frac{J_x}{2}-J_1)} \right],
\label{b47}\\
\rho_{3,13}(h)\!&=\!\rho_{13,3}(h)\!=\!-\frac{{\rm e}^{ \frac{\beta }{4}J_1}}{30{\cal Z}}
{\rm e}^{\beta h}\left[5{\rm e}^{-\beta(J_x-\frac{J_1}{2})}\!-\!2{\rm e}^{-\beta(J_x+\frac{5}{2}J_1)}\!-\!3{\rm e}^{-\beta(J_x-\frac{5}{2}J_1)}\right],
\label{b48}\\
\rho_{3,20}(h)\!&=\!\rho_{20,3}(h)\!=\!\frac{\sqrt{2}}{2}\rho_{3,8}(h),
\label{b49}\\
\rho_{3,22}(h)\!&=\!\rho_{22,3}(h)\!=\!\frac{\sqrt{2}}{2}\rho_{3,10}(h),
\label{b50}\\
\rho_{3,25}(h)\!&=\!\rho_{25,3}(h)\!=\!\sqrt{2}\rho_{3,13}(h),
\label{b51}\\
\rho_{4,7}(h)\!&=\!\rho_{7,4}(h)\!=\!\rho_{2,19}(h),
\label{b52}\\
\rho_{4,19}(h)\!&=\!\rho_{19,4}(h)\!=\!\frac{1}{2}\rho_{2,7}(h),
\label{b53}\\
\rho_{5,8}(h)\!&=\!\rho_{8,5}(h)\!=\!\sqrt{2}\frac{{\rm e}^{ \frac{\beta }{4}J_1}}{60{\cal Z}}
{\rm e}^{\beta h}\left[8{\rm e}^{-\beta(J_x+\frac{5}{2}J_1)}\!-\!8{\rm e}^{-\beta(J_x-\frac{5}{2}J_1)}\!-\!5{\rm e}^{-\beta(-\frac{J_x}{2}+J_1)} \!+\!5{\rm e}^{-\beta(-\frac{J_x}{2}-J_1)}\right],
\label{b54}\\
\rho_{5,10}(h)\!&=\!\rho_{10,5}(h)\!=\!\rho_{3,8}(h),
\label{b55}\\
\rho_{5,13}(h)\!&=\!\rho_{13,5}(h)\!=\!\rho_{3,25}(h),
\label{b56}\\
\rho_{5,20}(h)\!&=\!\rho_{20,5}(h)\!=\!\frac{\sqrt{2}}{2}\rho_{5,8}(h),
\label{b57}\\
\rho_{5,22}(h)\!&=\!\rho_{22,5}(h)\!=\!\rho_{3,20}(h),
\label{b58}\\
\rho_{5,25}(h)\!&=\!\rho_{25,5}(h)\!=\!2\rho_{3,13}(h),
\label{b59}\\
\rho_{6,9}(h)\!&=\!\rho_{9,6}(h)\!=\!\sqrt{2}\frac{{\rm e}^{ \frac{\beta }{4}J_1}}{60{\cal Z}}
\left[5{\rm e}^{-\beta(J_x-\frac{J_1}{2})}\!+\!3{\rm e}^{-\beta(J_x+\frac{5}{2}J_1)}\!-\!3{\rm e}^{-\beta(J_x-\frac{5}{2}J_1)}\!-\!5{\rm e}^{-\beta(J_x-\frac{7}{2}J_1)}\right],
\label{b60}\\
\rho_{6,11}(h)\!&=\!\rho_{11,6}(h)\!=\!\sqrt{2}\rho_{6,9}(h),
\label{b61}\\
\rho_{6,14}(h)\!&=\!\rho_{14,6}(h)\!=\!-\sqrt{2}\frac{{\rm e}^{ \frac{\beta }{4}J_1}}{60{\cal Z}}
\left[5{\rm e}^{-\beta(J_x-\frac{J_1}{2})}\!-\!3{\rm e}^{-\beta(J_x+\frac{5}{2}J_1)}\!+\!3{\rm e}^{-\beta(J_x-\frac{5}{2}J_1)}\!-\!5{\rm e}^{-\beta(J_x-\frac{7}{2}J_1)}\right],
\label{b62}\\
\rho_{6,16}(h)\!&=\!\rho_{16,6}(h)\!=\!\frac{\sqrt{2}}{2}\rho_{6,14}(h),
\label{b63}\\
\rho_{6,21}(h)\!&=\!\rho_{21,6}(h)\!=\!\frac{\sqrt{2}}{2}\rho_{6,9}(h),
\label{b64}\\
\rho_{6,23}(h)\!&=\!\rho_{23,6}(h)\!=\!\rho_{6,9}(h),
\label{b65}\\
\rho_{6,26}(h)\!&=\!\rho_{26,6}(h)\!=\!\frac{2}{\sqrt{2}}\rho_{6,14}(h),
\label{b66}\\
\rho_{6,28}(h)\!&=\!\rho_{28,6}(h)\!=\!\rho_{6,14}(h),
\label{b67}\\
\rho_{6,31}(h)\!&=\!\rho_{31,6}(h)\!=\!-\frac{{\rm e}^{ \frac{\beta }{4}J_1}}{20{\cal Z}}
\left[5{\rm e}^{-\beta(J_x-\frac{J_1}{2})}\!-\!{\rm e}^{-\beta(J_x+\frac{5}{2}J_1)}\!-\!9{\rm e}^{-\beta(J_x-\frac{5}{2}J_1)}\!+\!5{\rm e}^{-\beta(J_x-\frac{7}{2}J_1)}\right],
\label{b68}\\
\rho_{7,19}(h)\!&=\!\rho_{19,7}(h)\!=\!\rho_{2,4}(h),
\label{b69}\\
\rho_{8,10}(h)\!&=\!\rho_{10,8}(h)\!=\!\sqrt{2}\frac{{\rm e}^{ \frac{\beta }{4}J_1}}{180{\cal Z}}
{\rm e}^{\beta h}\left[24{\rm e}^{-\beta(J_x+\frac{5}{2}J_1)}\!+\!16{\rm e}^{-\beta(J_x-\frac{5}{2}J_1)}\!-\!20{\rm e}^{-\beta(-2J_x+\frac{J_1}{2})}\!-\!15{\rm e}^{-\beta(-\frac{J_x}{2}+J_1)}\right.
\nonumber\\
 \!&-\!\left.5{\rm e}^{-\beta(-\frac{J_x}{2}-J_1)} \right],
\label{b70}\\
\rho_{8,13}(h)\!&=\!\rho_{13,8}(h)\!=\!\rho_{3,8}(h),
\label{b71}\\
\rho_{8,20}(h)\!&=\!\rho_{20,8}(h)\!=\!\rho_{8,10}(h),
\label{b72}\\
\rho_{8,22}(h)\!&=\!\rho_{22,8}(h)\!=\!\frac{{\rm e}^{ \frac{\beta }{4}J_1}}{90{\cal Z}}
{\rm e}^{\beta h}\left[12{\rm e}^{-\beta(J_x+\frac{5}{2}J_1)}\!+\!8{\rm e}^{-\beta(J_x-\frac{5}{2}J_1)}\!+\!20{\rm e}^{-\beta(-2J_x+\frac{J_1}{2})}\!-\!30{\rm e}^{-\beta(-\frac{J_x}{2}+J_1)} \!-\!10{\rm e}^{-\beta(-\frac{J_x}{2}-J_1)} \right],
\label{b73}\\
\rho_{8,25}(h)\!&=\!\rho_{25,8}(h)\!=\!\rho_{5,8}(h),
\label{b74}\\
\rho_{9,11}(h)\!&=\!\rho_{11,9}(h)\!=\!\sqrt{2}\frac{{\rm e}^{ \frac{\beta }{4}J_1}}{90{\cal Z}}
\left[5{\rm e}^{-\beta(J_x-\frac{J_1}{2})}\!+\!9{\rm e}^{-\beta(J_x+\frac{5}{2}J_1)}\!+\!{\rm e}^{-\beta(J_x-\frac{5}{2}J_1)}\!+\!5{\rm e}^{-\beta(J_x-\frac{7}{2}J_1)}\right.
\nonumber\\
\!&-\!5{\rm e}^{-\beta(-2J_x+\frac{J_1}{2})} \!-\!5{\rm e}^{-\beta(-2J_x-\frac{J_1}{2})}\!-\!\left.5{\rm e}^{-\beta(-\frac{J_x}{2}+J_1)}\!-\!5{\rm e}^{-\beta(-\frac{J_x}{2}-J_1)}\right],
\label{b75}\\
\rho_{9,14}(h)\!&=\!\rho_{14,9}(h)\!=\!-\frac{{\rm e}^{ \frac{\beta }{4}J_1}}{90{\cal Z}}
\left[5{\rm e}^{-\beta(J_x-\frac{J_1}{2})}\!-\!9{\rm e}^{-\beta(J_x+\frac{5}{2}J_1)}\!-\!{\rm e}^{-\beta(J_x-\frac{5}{2}J_1)}\!+\!5{\rm e}^{-\beta(J_x-\frac{7}{2}J_1)}\!-\!10{\rm e}^{-\beta(-2J_x+\frac{J_1}{2})}\right.
\nonumber\\
 \!&+\!10{\rm e}^{-\beta(-2J_x-\frac{J_1}{2})}\!-\!\left.20{\rm e}^{-\beta(-\frac{J_x}{2}+J_1)}\!+\!20{\rm e}^{-\beta(-\frac{J_x}{2}-J_1)}\right],
\label{b76}\\
\rho_{9,16}(h)\!&=\!\rho_{16,9}(h)\!=\!-\sqrt{2}\frac{{\rm e}^{ \frac{\beta }{4}J_1}}{180{\cal Z}}
\left[5{\rm e}^{-\beta(J_x-\frac{J_1}{2})}\!-\!9{\rm e}^{-\beta(J_x+\frac{5}{2}J_1)}\!-\!{\rm e}^{-\beta(J_x-\frac{5}{2}J_1)}\!+\!5{\rm e}^{-\beta(J_x-\frac{7}{2}J_1)}\right.
\nonumber\\
\!&+\!20{\rm e}^{-\beta(-2J_x+\frac{J_1}{2})} \!-\!20{\rm e}^{-\beta(-2J_x-\frac{J_1}{2})}\!+\!\left.10{\rm e}^{-\beta(-\frac{J_x}{2}+J_1)}\!-\!10{\rm e}^{-\beta(-\frac{J_x}{2}-J_1)}\right],
\label{b77}\\
\rho_{9,21}(h)\!&=\!\rho_{21,9}(h)\!=\!\sqrt{2}\frac{{\rm e}^{ \frac{\beta }{4}J_1}}{180{\cal Z}}
\left[5{\rm e}^{-\beta(J_x-\frac{J_1}{2})}\!+\!9{\rm e}^{-\beta(J_x+\frac{5}{2}J_1)}\!+\!{\rm e}^{-\beta(J_x-\frac{5}{2}J_1)}\!+\!5{\rm e}^{-\beta(J_x-\frac{7}{2}J_1)}\!-\!20{\rm e}^{-\beta(-2J_x+\frac{J_1}{2})}\right.
\nonumber\\
 \!&-\!20{\rm e}^{-\beta(-2J_x-\frac{J_1}{2})}\!+\!\left.10{\rm e}^{-\beta(-\frac{J_x}{2}+J_1)}\!+\!10{\rm e}^{-\beta(-\frac{J_x}{2}-J_1)}\right],
\label{b78}\\
\rho_{9,23}(h)\!&=\!\rho_{23,9}(h)\!=\!\frac{{\rm e}^{ \frac{\beta }{4}J_1}}{90{\cal Z}}
\left[5{\rm e}^{-\beta(J_x-\frac{J_1}{2})}\!+\!9{\rm e}^{-\beta(J_x+\frac{5}{2}J_1)}\!+\!{\rm e}^{-\beta(J_x-\frac{5}{2}J_1)}\!+\!5{\rm e}^{-\beta(J_x-\frac{7}{2}J_1)}\!+\!10{\rm e}^{-\beta(-2J_x+\frac{J_1}{2})} \right.
\nonumber\\
\!&+\!10{\rm e}^{-\beta(-2J_x-\frac{J_1}{2})}\!-\!\left.20{\rm e}^{-\beta(-\frac{J_x}{2}+J_1)}\!-\!20{\rm e}^{-\beta(-\frac{J_x}{2}-J_1)}\right],
\label{b79}\\
\rho_{9,26}(h)\!&=\!\rho_{26,9}(h)\!=\!-\sqrt{2}\frac{{\rm e}^{ \frac{\beta }{4}J_1}}{90{\cal Z}}
\left[5{\rm e}^{-\beta(J_x-\frac{J_1}{2})}\!-\!9{\rm e}^{-\beta(J_x+\frac{5}{2}J_1)}\!-\!{\rm e}^{-\beta(J_x-\frac{5}{2}J_1)}\!+\!5{\rm e}^{-\beta(J_x-\frac{7}{2}J_1)}\!+\!5{\rm e}^{-\beta(-2J_x+\frac{J_1}{2})} \right.
\nonumber\\
\!&-\!5{\rm e}^{-\beta(-2J_x-\frac{J_1}{2})}\!-\!\left.5{\rm e}^{-\beta(-\frac{J_x}{2}+J_1)}\!+\!5{\rm e}^{-\beta(-\frac{J_x}{2}-J_1)}\right],
\label{b80}\\
\rho_{9,28}(h)\!&=\!\rho_{28,9}(h)\!=\!-\frac{{\rm e}^{ \frac{\beta }{4}J_1}}{90{\cal Z}}
\left[5{\rm e}^{-\beta(J_x-\frac{J_1}{2})}\!-\!9{\rm e}^{-\beta(J_x+\frac{5}{2}J_1)}\!-\!{\rm e}^{-\beta(J_x-\frac{5}{2}J_1)}\!+\!5{\rm e}^{-\beta(J_x-\frac{7}{2}J_1)}\!-\!10{\rm e}^{-\beta(-2J_x+\frac{J_1}{2})} \right.
\nonumber\\
\!&+\!10{\rm e}^{-\beta(-2J_x-\frac{J_1}{2})}\!+\!\left.25{\rm e}^{-\beta(-\frac{J_x}{2}+J_1)}\!-\!25{\rm e}^{-\beta(-\frac{J_x}{2}-J_1)}\right],
\label{b81}\\
\rho_{9,31}(h)\!&=\!\rho_{31,9}(h)\!=\!\rho_{6,28}(-h),
\label{b82}\\
\rho_{10,13}(h)\!&=\!\rho_{13,10}(h)\!=\!\rho_{3,20}(h);
\label{b83}\\
\rho_{10,20}(h)\!&=\!\rho_{20,10}(h)\!=\!\rho_{8,22}(h);
\label{b84}\\
\rho_{10,22}(h)\!&=\!\rho_{22,10}(h)\!=\!\sqrt{2}\frac{{\rm e}^{ \frac{\beta }{4}J_1}}{180{\cal Z}}
{\rm e}^{\beta h}\left[12{\rm e}^{-\beta(J_x+\frac{5}{2}J_1)}\!+\!8{\rm e}^{-\beta(J_x-\frac{5}{2}J_1)}\!-\!40{\rm e}^{-\beta(-2J_x+\frac{J_1}{2})}\!+\!15{\rm e}^{-\beta(-\frac{J_x}{2}+J_1)} \right.
\nonumber\\
\!&+\!\left.5{\rm e}^{-\beta(-\frac{J_x}{2}-J_1)} \right],
\label{b85}\\
\rho_{10,25}(h)\!&=\!\rho_{25,10}(h)\!=\!\rho_{5,20}(h);
\label{b91}\\
\rho_{11,14}(h)\!&=\!\rho_{14,11}(h)\!=\!\rho_{9,26}(h);
\label{b86}\\
\rho_{11,16}(h)\!&=\!\rho_{16,11}(h)\!=\!\rho_{9,14}(h);
\label{b87}\\
\rho_{11,21}(h)\!&=\!\rho_{21,11}(h)\!=\!\rho_{9,23}(h);
\label{b88}\\
\rho_{11,23}(h)\!&=\!\rho_{23,11}(h)\!=\!\rho_{9,11}(h);
\label{b89}\\
\rho_{11,26}(h)\!&=\!\rho_{26,11}(h)\!=\!-\frac{{\rm e}^{ \frac{\beta }{4}J_1}}{90{\cal Z}}
\left[10{\rm e}^{-\beta(J_x-\frac{J_1}{2})}\!-\!18{\rm e}^{-\beta(J_x+\frac{5}{2}J_1)}\!-\!2{\rm e}^{-\beta(J_x-\frac{5}{2}J_1)}\!+\!10{\rm e}^{-\beta(J_x-\frac{7}{2}J_1)}\!-\!5{\rm e}^{-\beta(-2J_x+\frac{J_1}{2})} \right.
\nonumber\\
\!&+\!5{\rm e}^{-\beta(-2J_x-\frac{J_1}{2})}\!+\!\left.20{\rm e}^{-\beta(-\frac{J_x}{2}+J_1)}\!-\!20{\rm e}^{-\beta(-\frac{J_x}{2}-J_1)}\right],
\label{b90}\\
\rho_{11,28}(h)\!&=\!\rho_{28,11}(h)\!=\!\rho_{9,26}(-h),
\label{b92}\\
\rho_{11,31}(h)\!&=\!\rho_{31,11}(h)\!=\!\rho_{6,26}(-h),
\label{b93}\\
\rho_{12,15}(h)\!&=\!\rho_{15,12}(h)\!=\!\rho_{22,25}(-h),
\label{b94}\\
\rho_{12,17}(h)\!&=\!\rho_{17,12}(h)\!=\!\rho_{20,25}(-h),
\label{b95}\\
\rho_{12,24}(h)\!&=\!\rho_{24,12}(h)\!=\!\rho_{13,25}(-h),
\label{b96}\\
\rho_{12,27}(h)\!&=\!\rho_{27,12}(h)\!=\!\rho_{10,25}(-h),
\label{b97}\\
\rho_{12,29}(h)\!&=\!\rho_{29,12}(h)\!=\!\rho_{8,25}(-h),
\label{b98}\\
\rho_{12,32}(h)\!&=\!\rho_{32,12}(h)\!=\!\rho_{5,25}(-h),
\label{b99}\\
\rho_{12,34}(h)\!&=\!\rho_{34,12}(h)\!=\!\rho_{3,25}(-h),
\label{b100}\\
\rho_{13,20}(h)\!&=\!\rho_{20,13}(h)\!=\!\rho_{3,10}(h),
\label{b101}\\
\rho_{13,22}(h)\!&=\!\rho_{22,13}(h)\!=\!\rho_{3,22}(h),
\label{b102}\\
\rho_{13,25}(h)\!&=\!\rho_{25,13}(h)\!=\!\rho_{3,5}(h),
\label{b103}\\
\rho_{14,16}(h)\!&=\!\rho_{16,14}(h)\!=\!\rho_{9,21}(h),
\label{b104}\\
\rho_{14,21}(h)\!&=\!\rho_{21,14}(h)\!=\!\rho_{9,16}(h),
\label{b105}\\
\rho_{14,23}(h)\!&=\!\rho_{23,14}(h)\!=\!\rho_{9,28}(h),
\label{b106}\\
\rho_{14,26}(h)\!&=\!\rho_{26,14}(h)\!=\!\rho_{11,23}(-h),
\label{b107}\\
\rho_{14,28}(h)\!&=\!\rho_{28,14}(h)\!=\!\rho_{9,23}(-h),
\label{b108}\\
\rho_{14,31}(h)\!&=\!\rho_{31,14}(h)\!=\!\rho_{6,23}(-h),
\label{b109}\\
\rho_{15,17}(h)\!&=\!\rho_{17,15}(h)\!=\!\rho_{20,22}(-h),
\label{b110}\\
\rho_{15,24}(h)\!&=\!\rho_{24,15}(h)\!=\!\rho_{13,22}(-h),
\label{b111}\\
\rho_{15,27}(h)\!&=\!\rho_{27,15}(h)\!=\!\rho_{10,22}(-h),
\label{b112}\\
\rho_{15,29}(h)\!&=\!\rho_{29,15}(h)\!=\!\rho_{8,22}(-h),
\label{b113}\\
\rho_{15,32}(h)\!&=\!\rho_{32,15}(h)\!=\!\rho_{5,22}(-h),
\label{b114}\\
\rho_{15,34}(h)\!&=\!\rho_{34,15}(h)\!=\!\rho_{3,22}(-h),
\label{b115}\\
\rho_{16,21}(h)\!&=\!\rho_{21,16}(h)\!=\!-\frac{{\rm e}^{ \frac{\beta }{4}J_1}}{180{\cal Z}}
\left[5{\rm e}^{-\beta(J_x-\frac{J_1}{2})}\!-\!9{\rm e}^{-\beta(J_x+\frac{5}{2}J_1)}\!-\!{\rm e}^{-\beta(J_x-\frac{5}{2}J_1)}\!+\!5{\rm e}^{-\beta(J_x-\frac{7}{2}J_1)}\!-\!40{\rm e}^{-\beta(-2J_x+\frac{J_1}{2})} \right.
\nonumber\\
\!&+\!40{\rm e}^{-\beta(-2J_x-\frac{J_1}{2})}\!+\!\left.40{\rm e}^{-\beta(-\frac{J_x}{2}+J_1)}\!-\!40{\rm e}^{-\beta(-\frac{J_x}{2}-J_1)}\right],
\label{b116}\\
\rho_{16,23}(h)\!&=\!\rho_{23,16}(h)\!=\!\rho_{14,21}(-h),
\label{b117}\\
\rho_{16,26}(h)\!&=\!\rho_{26,16}(h)\!=\!\rho_{11,21}(-h),
\label{b118}\\
\rho_{16,28}(h)\!&=\!\rho_{28,16}(h)\!=\!\rho_{9,21}(-h),
\label{b119}\\
\rho_{16,31}(h)\!&=\!\rho_{31,16}(h)\!=\!\rho_{6,21}(-h),
\label{b120}\\
\rho_{17,24}(h)\!&=\!\rho_{24,17}(h)\!=\!\rho_{13,20}(-h),
\label{b121}\\
\rho_{17,27}(h)\!&=\!\rho_{27,17}(h)\!=\!\rho_{10,20}(-h),
\label{b122}\\
\rho_{17,29}(h)\!&=\!\rho_{29,17}(h)\!=\!\rho_{8,20}(-h),
\label{b123}\\
\rho_{17,32}(h)\!&=\!\rho_{32,17}(h)\!=\!\rho_{5,20}(-h),
\label{b124}\\
\rho_{17,34}(h)\!&=\!\rho_{34,17}(h)\!=\!\rho_{3,20}(-h),
\label{b125}\\
\rho_{18,30}(h)\!&=\!\rho_{30,18}(h)\!=\!\rho_{7,19}(-h),
\label{b126}\\
\rho_{18,33}(h)\!&=\!\rho_{33,18}(h)\!=\!\rho_{4,19}(-h),
\label{b127}\\
\rho_{18,35}(h)\!&=\!\rho_{35,18}(h)\!=\!\rho_{2,19}(-h),
\label{b128}\\
\rho_{20,22}(h)\!&=\!\rho_{22,20}(h)\!=\!\rho_{10,22}(h);
\label{b129}\\
\rho_{20,25}(h)\!&=\!\rho_{25,20}(h)\!=\!\rho_{3,8}(h);
\label{b130}\\
\rho_{21,23}(h)\!&=\!\rho_{23,21}(h)\!=\!\rho_{14,16}(-h),
\label{b131}\\
\rho_{21,26}(h)\!&=\!\rho_{26,21}(h)\!=\!\rho_{11,16}(-h),
\label{b132}\\
\rho_{21,28}(h)\!&=\!\rho_{28,21}(h)\!=\!\rho_{9,16}(-h),
\label{b133}\\
\rho_{21,31}(h)\!&=\!\rho_{31,21}(h)\!=\!\rho_{6,16}(-h),
\label{b134}\\
\rho_{22,25}(h)\!&=\!\rho_{25,22}(h)\!=\!\rho_{3,20}(h);
\label{b135}\\
\rho_{23,26}(h)\!&=\!\rho_{26,23}(h)\!=\!\rho_{11,14}(-h),
\label{b136}\\
\rho_{23,28}(h)\!&=\!\rho_{28,23}(h)\!=\!\rho_{9,14}(-h),
\label{b137}\\
\rho_{23,31}(h)\!&=\!\rho_{31,23}(h)\!=\!\rho_{6,14}(-h),
\label{b138}\\
\rho_{24,27}(h)\!&=\!\rho_{27,24}(h)\!=\!\rho_{10,13}(-h),
\label{b139}\\
\rho_{24,29}(h)\!&=\!\rho_{29,24}(h)\!=\!\rho_{8,13}(-h),
\label{b140}\\
\rho_{24,32}(h)\!&=\!\rho_{32,24}(h)\!=\!\rho_{5,13}(-h),
\label{b141}\\
\rho_{24,34}(h)\!&=\!\rho_{34,24}(h)\!=\!\rho_{3,13}(-h),
\label{b142}\\
\rho_{26,28}(h)\!&=\!\rho_{28,26}(h)\!=\!\rho_{9,11}(-h),
\label{b143}\\
\rho_{26,31}(h)\!&=\!\rho_{31,26}(h)\!=\!\rho_{6,11}(-h),
\label{b144}\\
\rho_{27,29}(h)\!&=\!\rho_{29,27}(h)\!=\!\rho_{8,10}(-h),
\label{b145}\\
\rho_{27,32}(h)\!&=\!\rho_{32,27}(h)\!=\!\rho_{5,10}(-h),
\label{b146}\\
\rho_{27,34}(h)\!&=\!\rho_{34,27}(h)\!=\!\rho_{3,10}(-h),
\label{b147}\\
\rho_{28,31}(h)\!&=\!\rho_{31,28}(h)\!=\!\rho_{6,9}(-h),
\label{b148}\\
\rho_{29,32}(h)\!&=\!\rho_{32,29}(h)\!=\!\rho_{5,8}(-h),
\label{b149}\\
\rho_{29,34}(h)\!&=\!\rho_{34,29}(h)\!=\!\rho_{3,8}(-h),
\label{b150}\\
\rho_{30,33}(h)\!&=\!\rho_{33,30}(h)\!=\!\rho_{4,7}(-h),
\label{b151}\\
\rho_{30,35}(h)\!&=\!\rho_{35,30}(h)\!=\!\rho_{2,7}(-h),
\label{b152}\\
\rho_{32,34}(h)\!&=\!\rho_{34,32}(h)\!=\!\rho_{3,5}(-h),
\label{b153}\\
\rho_{33,35}(h)\!&=\!\rho_{35,33}(h)\!=\!\rho_{2,4}(-h).
\label{b154}
\end{flalign}
\begin{table}[h!]
\caption{The explicit values of five global quantum  bipartite   negativities (${\cal N}_{\mu_1|\mu_2S_1S_2}$, ${\cal N}_{S_1|\mu_1\mu_2S_2}$, ${\cal N}_{\mu_1\mu_2|S_1S_2}$, ${\cal N}_{\mu_1S_1|\mu_2S_2}$, and ${\cal N}_{\mu_1S_2|\mu_2S_1}$)  calculated for all available ground states  (GS) of a mixed spin-(1/2,1) Heisenberg tetramer~\eqref{eq1}, respecting the ground-state phase diagram for $J>0$ and $J<0$.}
\label{tab1}
\resizebox{0.98\textwidth}{!}{
\begin{tabular}{l  l  l  l  l  l l }
\hline\hline
   &GS & ${\cal N}_{\mu_{1}|\mu_2S_{1}S_{2}}$ & ${\cal N}_{S_{1}|\mu_{1}\mu_2S_{2}}$ & ${\cal N}_{\mu_1\mu_{2}|S_{1}S_{2}}$ & ${\cal N}_{\mu_{1}S_1|\mu_{2}S_{2}}$ & ${\cal N}_{\mu_{1}S_2|\mu_{2}S_{1}}$\\\hline
   $J_1/|J|<1$ & $\vert 0,\frac{1}{2},\frac{1}{2}\rangle$ & $1/2= 0.500$ &1& $(2+\sqrt{6})/3\approx 1.483$ & 1/2=0.500 &41/18$\approx 2.278$\\
         &&  & &  & &\\
         & $\vert 1,\frac{1}{2},\frac{1}{2}\rangle$ &$\sqrt{2}/3\approx 0.471$ & $\sqrt{2}/3\approx 0.471$  & $2(3\sqrt{2}+2)/9\approx 1.387$ & 0&$2(3\sqrt{2}+2)/9\approx 1.387$\\
          &&  & &  & &\\
       & $\vert 2,\frac{1}{2},\frac{3}{2}\!\rangle$, &  $1/\sqrt{18}\approx 0.236$ & $1/\sqrt{18}\approx 0.236$ & $1/3\approx 0.333$ & 0&$1/3\approx 0.333$ \\
       & $\vert 2,\frac{3}{2},\frac{1}{2}\rangle$ &  & &   &&\\ 
           &&  & &  & &\\
   & $\vert 3,\frac{3}{2},\frac{3}{2}\rangle$  & 0 & 0 & 0 & 0&0\\
           &&  & &  & \\
\hline
%%%%%%%%%%%%%%%%%%%%%%%%
   $J_1/J=1$ & $\vert 0,\frac{1}{2},\frac{1}{2}\rangle$, &1/2=0.500 &1/3$\approx 0.333$  &1/2=0.500   &  1&1\\
&$\vert 0,\frac{3}{2},\frac{3}{2}\rangle$ &  & &  & &\\
         &&  & &  & &\\
         & $\vert 1,\frac{1}{2},\frac{1}{2}\rangle$,   &$\sqrt{3}/8\approx 0.217$  &$(1\!+\!\sqrt{3}\!+\!\sqrt{6})/40\approx 0.355$ & $(5\!+\!\sqrt{3}\!+\!\sqrt{6}\!+\!3\sqrt{2})/40\approx 0.336$ &$3(1\!+\!3\sqrt{3})/40\approx 0.465$ &$3(1\!+\!3\sqrt{3})/40\approx 0.465$\\
         &$\vert 1,\frac{3}{2},\frac{3}{2}\rangle$, &  & &  & \\
        &$\vert 1,\frac{1}{2},\frac{3}{2}\rangle$, &    & &  & &\\
        &$\vert 1,\frac{3}{2},\frac{1}{2}\rangle$  &  & &  & &\\
         &&  & &   &&\\
        & $\vert 2,\frac{1}{2},\frac{3}{2}\!\rangle$,  &$\sqrt{5}/18\approx 0.124$  &$\sqrt{2}/9\approx 0.157$  &$\sqrt{2}/9\approx 0.157$  &$1/6\approx 0.167$ & $1/6\approx 0.167$\\
& $\vert 2,\frac{3}{2},\frac{1}{2}\rangle$, &  & &   &&\\ 
          & $\vert 2,\frac{3}{2},\frac{3}{2}\rangle$ &  & &   &&\\ 
          &&  & &  & &\\
    & $\vert 3,\frac{3}{2},\frac{3}{2}\rangle$  & 0 & 0 & 0 & 0&0 \\
            &&  & &  &&\\
	      \hline	   
	      %%%%%%%%%%%%%%%%%%%%%%
	  $J_1/|J|>1$ & $\vert 0,\frac{3}{2},\frac{3}{2}\rangle$  & 1/2=0.500 & 1 & $(1+\sqrt{6})/3\approx 1.149$ & 3/2=1.500&3/2=1.500 \\
	         &&  & &  & &\\
	         & $\vert 1,\frac{3}{2},\frac{3}{2}\rangle$   &$\sqrt{2}/3\approx 0.471$ &A$\approx 0.874$&$(6\sqrt{2}+6\sqrt{17}+5+\sqrt{34})/45\approx 0.979$ &$(4\sqrt{3}+3)/10\approx 0.993$& $B\approx 1.099$\\
           &&  & &  & &\\
               & $\vert 2,\frac{3}{2},\frac{3}{2}\rangle$ & $ \sqrt{5}/6\approx 0.373$ &$\sqrt{2}/3\approx 0.471$ & $\sqrt{2}/3\approx 0.471$ & 1/2=0.500&1/2=0.500\\ 
             & & &  & & \\
  & $\vert 3,\frac{3}{2},\frac{3}{2}\rangle$ & 0 & 0 & 0 & 0&0\\
          &&  & &  & & \\
             \hline
                   %-----------------------
               \multicolumn{7}{l}{$A\!=\!\tfrac{1}{45}|7\!+\!\sqrt{313}\cos(\frac{\phi_1}{3}\!+\!\frac{6\pi}{3})|\!+\!\sqrt{\frac{7}{150}}\!+\!\frac{1}{3}\sqrt{\frac{91}{150}}; 
             \hspace*{2cm}\phi_1\!=\!\arctan\left( \frac{\sqrt{p_1^3\!-\!q_1^2}}{q_1}\right), p_1\!=\!\frac{313}{(90)^2}, q_1\!=\!-\frac{2170}{(90)^3}$} \\
                \multicolumn{7}{l}{$B\!=\!\frac{2}{45}\left(\sqrt{119+\frac{51}{2}\sqrt{19}}+\sqrt{119-\frac{51}{2}\sqrt{19}}\right)+\frac{1}{9}+\frac{17}{90}$;} \\
                   %-----------------------
                      \hline\hline
\end{tabular}
}

\end{table}

\section{\label{App B}Global bipartite entanglement ${\cal N}_{\mu_{1}|\mu_{2}S_1S_{2}}$ with a central spin $\mu_1$ }

Before proceeding with the  calculation details, we would like to emphasize that the global bipartite negativity ${\cal N}_{\mu_1|\mu_2S_1S_2}$ is identical to the global bipartite negativity ${\cal N}_{\mu_2|\mu_1S_1S_2}$ due to  system symmetry. Consequently, we will discuss the former case only in the following. The global bipartite negativity 
${\cal N}_{\mu_1|\mu_2S_1S_2}$ is defined according to definition~\eqref{eq2}, where $\lambda_i$  is a negative eigenvalue of the partially transposed global density matrix $\rho^{T_{\mu_1}}_{\mu_{1}S_1\mu_{2}S_{2}}$. The partial transposition of the original global density matrix $\hat{\rho}_{\mu_{1}S_1\mu_{2}S_{2}}$, Eq.~\eqref{a2}, is performed over the $\mu_{1}$ spin subsystem, indicated by the upper index  $T_{\mu_{1}}$.
The obtained transposed density matrix   $\hat{\rho}_{\mu_{1}S_1\mu_{2}S_{2}}^{T_{\mu_{1}}}$ has a block-diagonal structure that includes two $1\times 1$ matrices $\mathbf{Q}_1^{\mu_{1}}(\pm h)$, two $4\times 4$ matrices $\mathbf{Q}_2^{\mu_1}(\pm h)$, two $8\times 8$ matrices $\mathbf{Q}_3^{\mu_1}(\pm h)$, and one $10\times 10$ matrix $\mathbf{Q}_4^{\mu_{1}}(h)$ with the following elements:
\begin{align}
\allowdisplaybreaks
&\mathbf{Q}_1^{\mu_{1}}(\pm h)\!=\!\left(
\begin{array}{c}
 \rho_{18,18}(\pm h)
\end{array}\right),\;\;
%\label{f24}
\mathbf{Q}_2^{\mu_{1}}(\pm h)\!=\!\left(
\begin{array}{cccc}
 \rho_{1,1}(\pm h) & \rho_{2,19}(\pm h) & \rho_{4,19}(\pm h)& \rho_{7,19}(\pm h)\\
 \rho_{2,19} (\pm h)& \rho_{20,20}(\pm h) & \rho_{20,22}(\pm h)& \rho_{20,25}(\pm h)\\
 \rho_{4,19}(\pm h) & \rho_{20,22}(\pm h) & \rho_{22,22}(\pm h)& \rho_{22,25}(\pm h)\\
 \rho_{7,19}(\pm h) & \rho_{20,25}(\pm h) & \rho_{22,25}(\pm h)& \rho_{25,25}(\pm h)\\
\end{array}\right),\;\;
\nonumber\\
&\mathbf{Q}_3^{\mu_{1}}(\pm h)\!=\!\left(
\begin{array}{cccccccc}
 \rho_{2,2}(\pm h) & \rho_{2,4}(\pm h) & \rho_{2,7}(\pm h)& \rho_{3,20}(\pm h) &\rho_{5,20}(\pm h) & \rho_{8,20}(\pm h) & \rho_{10,20}(\pm h)& \rho_{13,20}(\pm h)\\
\rho_{2,4}(\pm h) & \rho_{4,4}(\pm h) & \rho_{4,7}(\pm h)& \rho_{3,22}(\pm h)&\rho_{5,22}(\pm h) & \rho_{8,22}(\pm h) & \rho_{10,22}(\pm h)& \rho_{13,22}(\pm h)\\
\rho_{2,7}(\pm h) & \rho_{4,7}(\pm h) & \rho_{7,7}(\pm h)& \rho_{3,25}(\pm h) &\rho_{5,25}(\pm h) & \rho_{8,25}(\pm h) & \rho_{10,25}(\pm h)& \rho_{13,25}(\pm h)\\
\rho_{3,20}(\pm h) & \rho_{3,22}(\pm h) & \rho_{3,25}(\pm h)& \rho_{21,21}(\pm h) &\rho_{21,23}(\pm h) & \rho_{21,26}(\pm h) & \rho_{21,28}(\pm h)& \rho_{21,31}(\pm h)\\
\rho_{5,20}(\pm h) & \rho_{5,22}(\pm h) & \rho_{5,25}(\pm h)& \rho_{21,23}(\pm h) &\rho_{23,23}(\pm h) & \rho_{23,26}(\pm h) & \rho_{23,28}(\pm h)& \rho_{23,31}(\pm h)\\
\rho_{8,20}(\pm h) & \rho_{8,22}(\pm h) & \rho_{8,25}(\pm h)& \rho_{21,26}(\pm h) &\rho_{23,26}(\pm h) & \rho_{26,26}(\pm h) & \rho_{26,28}(\pm h)& \rho_{26,31}(\pm h)\\
\rho_{10,20}(\pm h) & \rho_{10,22}(\pm h) & \rho_{10,25}(\pm h)& \rho_{21,28}(\pm h) &\rho_{23,28}(\pm h) & \rho_{26,28}(\pm h) & \rho_{28,28}(\pm h)& \rho_{28,31}(\pm h)\\
\rho_{13,20}(\pm h) & \rho_{13,22}(\pm h) & \rho_{13,25}(\pm h)& \rho_{21,31}(\pm h) &\rho_{23,31}(\pm h) & \rho_{26,31}(\pm h) & \rho_{28,31}(\pm h)& \rho_{31,31}(\pm h)\\
\end{array}\right),
\nonumber\\
&\resizebox{1\textwidth}{!}{$
\mathbf{Q}_4^{\mu_{1}}(h)\!=\!\left(
\begin{array}{cccccccccc}
 \rho_{3,3}(-h) & \rho_{3,5}(-h) & \rho_{3,8}(-h)& \rho_{3,10}(-h) &\rho_{3,13}(-h) & \rho_{6,21}(-h) & \rho_{9,21}(-h)& \rho_{11,21}(-h)& \rho_{14,21}(-h)& \rho_{16,21}(h)\\
 \rho_{3,5}(-h) & \rho_{5,5}(-h) & \rho_{5,8}(-h)& \rho_{5,10}(-h) &\rho_{5,13}(-h) & \rho_{6,23}(-h) & \rho_{9,23}(-h)& \rho_{11,23}(-h)& \rho_{14,23}(h)& \rho_{14,21}(h)\\
  \rho_{3,8}(-h) & \rho_{5,8}(-h) & \rho_{8,8}(-h)& \rho_{8,10}(-h)& \rho_{8,13}(-h)& \rho_{6,26}(-h) & \rho_{9,26}(-h)& \rho_{11,26}(h)& \rho_{11,23}(h)& \rho_{11,21}(h)\\
   \rho_{3,10}(-h) & \rho_{5,10}(-h) & \rho_{8,10}(-h)& \rho_{10,10}(-h)& \rho_{10,13}(-h) & \rho_{6,28}(-h) & \rho_{9,28}(h)& \rho_{9,26}(h)& \rho_{9,23}(h)& \rho_{9,21}(h)\\
   \rho_{3,13}(-h) & \rho_{5,13}(-h) & \rho_{8,13}(-h)& \rho_{10,13}(-h)& \rho_{13,13}(-h) & \rho_{6,31}(h) & \rho_{6,28}(h)& \rho_{6,26}(h)& \rho_{6,23}(h)& \rho_{6,21}(h)\\
    \rho_{6,21}(-h) & \rho_{6,23}(-h) & \rho_{6,26}(-h)& \rho_{6,28}(-h)& \rho_{6,31}(h) & \rho_{13,13}(h) & \rho_{10,13}(h)& \rho_{8,13}(h)& \rho_{5,13}(h)& \rho_{3,13}(h)\\
     \rho_{9,21}(-h) & \rho_{9,23}(-h) & \rho_{9,26}(-h)& \rho_{9,28}(h)& \rho_{6,28}(h) & \rho_{10,13}(h) & \rho_{10,10}(h)& \rho_{8,10}(h)& \rho_{5,10}(h)& \rho_{3,10}(h)\\
        \rho_{11,21}(-h) & \rho_{11,21}(-h) & \rho_{11,26}(h)& \rho_{9,26}(h)& \rho_{6,26}(h) & \rho_{8,13}(h) & \rho_{8,10}(h)& \rho_{8,8}(h)& \rho_{5,8}(h)& \rho_{3,8}(h)\\
          \rho_{14,21}(-h) & \rho_{14,23}(h) & \rho_{11,23}(h)& \rho_{9,23}(h)& \rho_{6,23}(h) & \rho_{5,13}(h) & \rho_{5,10}(h)& \rho_{5,8}(h)& \rho_{5,5}(h)& \rho_{3,5}(h)\\
            \rho_{16,21}(h) & \rho_{14,21}(h) & \rho_{11,21}(h)& \rho_{9,21}(h)& \rho_{6,21}(h) & \rho_{3,13}(h) & \rho_{3,10}(h)& \rho_{3,8}(h)& \rho_{3,5}(h)& \rho_{3,3}(h)\\
\end{array}\right).$}
\label{b1}
\end{align}
The specific elements $\rho_{i,j}(\pm h)$ are identical to those presented in Eqs.~\eqref{b6}-\eqref{b154}.   The corresponding eigenvalues $\lambda_i$ of all diagonal blocks are numerically computed using the MATLAB Symbolic Toolbox~\cite{matlab}. 

\section{\label{App C}Global bipartite entanglement ${\cal N}_{S_1|\mu_{1}\mu_{2}S_{2}}$  with a central spin $S_1$}

It should be mentioned that the system's symmetry is responsible for the identical magnitude of the global bipartite negativity ${\cal N}_{S_1|\mu_{1}\mu_{2}S_{2}}$ and ${\cal N}_{S_2|\mu_{1}\mu_{2}S_{1}}$. For this reason, the calculation details for determining the global bipartite negativity  ${\cal N}_{S_1|\mu_{1}\mu_{2}S_{2}}$ are sufficient for further analyses. The partially transposed density matrix  $\hat{\rho}_{S_{1}\mu_1\mu_{2}S_{2}}^{T_{S_{1}}}$ again has a block-diagonal structure, including two $1\times 1$ matrices $\mathbf{Q}_1^{S_{1}}(\pm h)$, two $4\times 4$ matrices $\mathbf{Q}_2^{S_1}(\pm h)$, two $8\times 8$ matrices $\mathbf{Q}_3^{S_1}(\pm h)$, and one $10\times 10$ matrix $\mathbf{Q}_4^{S_{1}}(h)$ with the following elements:
\begin{align}
\allowdisplaybreaks
&\mathbf{Q}_1^{S_{1}}(\pm h)\!=\!\left(
\begin{array}{c}
 \rho_{13,13}(\pm h)
\end{array}\right),\;\;
%\label{f24}
\mathbf{Q}_2^{S_{1}}(\pm h)\!=\!\left(
\begin{array}{cccc}
 \rho_{6,6}(\pm h) & \rho_{6,21}(\pm h) & \rho_{6,23}(\pm h)& \rho_{12,24}(\pm h)\\
 \rho_{6,21} (\pm h)& \rho_{21,21}(\pm h) & \rho_{21,23}(\pm h)& \rho_{24,27}(\pm h)\\
 \rho_{6,23}(\pm h) & \rho_{21,23}(\pm h) & \rho_{23,23}(\pm h)& \rho_{24,29}(\pm h)\\
 \rho_{12,24}(\pm h) & \rho_{24,27}(\pm h) & \rho_{24,29}(\pm h)& \rho_{30,30}(\pm h)\\
\end{array}\right),\;\;
\nonumber\\
&\mathbf{Q}_3^{S_{1}}(\pm h)\!=\!\left(
\begin{array}{cccccccc}
 \rho_{1,1}(\pm h) & \rho_{2,7}(\pm h) & \rho_{4,7}(\pm h)& \rho_{7,19}(\pm h) &\rho_{3,13}(\pm h) & \rho_{5,13}(\pm h) & \rho_{13,20}(\pm h)& \rho_{13,22}(\pm h)\\
\rho_{2,7}(\pm h) & \rho_{8,8}(\pm h) & \rho_{8,10}(\pm h)& \rho_{8,25}(\pm h)&\rho_{9,14}(\pm h) & \rho_{11,14}(\pm h) & \rho_{14,26}(\pm h)& \rho_{14,28}(\pm h)\\
\rho_{4,7}(\pm h) & \rho_{8,10}(\pm h) & \rho_{10,10}(\pm h)& \rho_{10,25}(\pm h) &\rho_{9,16}(\pm h) & \rho_{11,16}(\pm h) & \rho_{16,26}(\pm h)& \rho_{16,28}(\pm h)\\
\rho_{7,19}(\pm h) & \rho_{8,25}(\pm h) & \rho_{10,25}(\pm h)& \rho_{25,25}(\pm h) &\rho_{9,31}(\pm h) & \rho_{11,31}(\pm h) & \rho_{26,31}(\pm h)& \rho_{28,31}(\pm h)\\
\rho_{3,13}(\pm h) & \rho_{9,14}(\pm h) & \rho_{9,16}(\pm h)& \rho_{9,31}(\pm h) &\rho_{15,15}(\pm h) & \rho_{15,17}(\pm h) & \rho_{15,32}(\pm h)& \rho_{15,34}(\pm h)\\
\rho_{5,13}(\pm h) & \rho_{11,14}(\pm h) & \rho_{11,16}(\pm h)& \rho_{11,31}(\pm h) &\rho_{15,17}(\pm h) & \rho_{17,17}(\pm h) & \rho_{17,32}(\pm h)& \rho_{17,34}(\pm h)\\
\rho_{13,20}(\pm h) & \rho_{14,26}(\pm h) & \rho_{16,26}(\pm h)& \rho_{26,31}(\pm h) &\rho_{15,32}(\pm h) & \rho_{17,32}(\pm h) & \rho_{32,32}(\pm h)& \rho_{32,34}(\pm h)\\
\rho_{13,22}(\pm h) & \rho_{14,28}(\pm h) & \rho_{16,28}(\pm h)& \rho_{28,31}(\pm h) &\rho_{15,34}(\pm h) & \rho_{17,34}(\pm h) & \rho_{32,34}(\pm h)& \rho_{34,34}(\pm h)\\
\end{array}\right),
\nonumber\\
&\resizebox{1\textwidth}{!}{$
\mathbf{Q}_4^{S_{1}}(h)\!=\!\left(
\begin{array}{cccccccccc}
 \rho_{2,2}(h) & \rho_{2,4}(h) & \rho_{2,19}(h)& \rho_{3,8}(h) &\rho_{5,8}(h) & \rho_{8,20}(h) & \rho_{8,22}(h)& \rho_{6,14}(h)& \rho_{14,21}(h)& \rho_{14,23}(h)\\
 \rho_{2,4}(h) & \rho_{4,4}(h) & \rho_{4,19}(h)& \rho_{3,10}(h) &\rho_{5,10}(h) & \rho_{10,20}(h) & \rho_{10,22}(h)& \rho_{6,16}(h)& \rho_{16,21}(h)& \rho_{14,21}(-h)\\
  \rho_{2,19}(h) & \rho_{4,19}(h) & \rho_{19,19}(h)& \rho_{3,25}(h)& \rho_{5,25}(h)& \rho_{20,25}(h) & \rho_{22,25}(h)& \rho_{6,31}(h)& \rho_{6,16}(-h)& \rho_{6,14}(-h)\\
   \rho_{3,8}(h) & \rho_{3,10}(h) & \rho_{3,25}(h)& \rho_{9,9}(h)& \rho_{9,11}(h) & \rho_{9,26}(h) & \rho_{9,28}(h)& \rho_{22,25}(-h)& \rho_{10,22}(-h)& \rho_{8,22}(-h)\\
   \rho_{5,8}(h) & \rho_{5,10}(h) & \rho_{5,25}(h)& \rho_{9,11}(h)& \rho_{11,11}(h) & \rho_{11,26}(h) & \rho_{9,26}(-h)& \rho_{20,25}(-h)& \rho_{10,20}(-h)& \rho_{8,20}(-h)\\
    \rho_{8,20}(h) & \rho_{10,20}(h) & \rho_{20,25}(h)& \rho_{9,26}(h)& \rho_{11,26}(h) & \rho_{11,11}(-h) & \rho_{9,11}(-h)& \rho_{5,25}(-h)& \rho_{5,10}(-h)& \rho_{5,8}(-h)\\
     \rho_{8,22}(h) & \rho_{10,22}(h) & \rho_{22,25}(h)& \rho_{9,28}(h)& \rho_{9,26}(-h) & \rho_{9,11}(-h) & \rho_{9,9}(-h)& \rho_{3,25}(-h)& \rho_{3,10}(-h)& \rho_{3,8}(-h)\\
        \rho_{6,14}(h) & \rho_{6,16}(h) & \rho_{6,31}(h)& \rho_{22,25}(-h)& \rho_{20,25}(-h) & \rho_{5,25}(-h) & \rho_{3,25}(-h)& \rho_{19,19}(-h)& \rho_{4,19}(-h)& \rho_{2,19}(-h)\\
          \rho_{14,21}(h) & \rho_{16,21}(h) & \rho_{6,16}(-h)& \rho_{10,22}(-h)& \rho_{10,20}(-h) & \rho_{5,10}(-h) & \rho_{3,10}(-h)& \rho_{4,19}(-h)& \rho_{4,4}(-h)& \rho_{2,4}(-h)\\
            \rho_{14,23}(h) & \rho_{14,21}(-h) & \rho_{6,14}(-h)& \rho_{8,22}(-h)& \rho_{8,20}(-h) & \rho_{5,8}(-h) & \rho_{3,8}(-h)& \rho_{2,19}(-h)& \rho_{2,4}(-h)& \rho_{2,2}(-h)\\
\end{array}\right).$}
\label{c1}
\end{align}
The particular elements $\rho_{i,j}(\pm h)$ are the same as those presented in  Eqs.~\eqref{b6}-\eqref{b154}. The sum of the absolute values of all negative eigenvalues of $\mathbf{Q}_2^{S_1}(\pm h)$,  $\mathbf{Q}_3^{S_1}(\pm h)$, and  $\mathbf{Q}_4^{S_1}(h)$ determines the respective global bipartite negativity ${\cal N}_{S_1|\mu_1\mu_2S_2}$ between a single spin $S_1$ and spin cluster $\mu_1\mu_2S_2$.

\section{\label{App D}Global bipartite entanglement ${\cal N}_{\mu_{1}\mu_{2}|S_1S_2}$ with a central spin dimer $\mu_1\mu_2$}

It should be mentioned that system symmetry is responsible for the identical magnitude of the global bipartite negativity ${\cal N}_{\mu_{1}\mu_{2}|S_1S_{2}}$ and ${\cal N}_{S_1S_2|\mu_{1}\mu_{2}}$. Therefore, the calculation details for determining the global bipartite negativity  ${\cal N}_{\mu_{1}\mu_{2}|S_1S_{2}}$ are sufficient for further analyses.
The partially transposed density matrix  $\hat{\rho}_{\mu_1\mu_{2}|S_1S_{2}}^{T_{\mu_{1}\mu_2}}$ again has a block-diagonal structure, including two $1\times 1$ matrices $\mathbf{Q}_1^{\mu_{1}\mu_2}(\pm h)$, two $4\times 4$ matrices $\mathbf{Q}_2^{\mu_{1}\mu_2}(\pm h)$, two $8\times 8$ matrices $\mathbf{Q}_3^{\mu_{1}\mu_2}(\pm h)$, and one $10\times 10$ matrix $\mathbf{Q}_4^{\mu_{1}\mu_2}(h)$ with the following elements:
\begin{align}
\allowdisplaybreaks
&\mathbf{Q}_1^{\mu_{1}\mu_2}(\pm h)\!=\!\left(
\begin{array}{c}
 \rho_{15,15}(\pm h)
\end{array}\right),\;\;
%\label{f24}
\mathbf{Q}_2^{\mu_{1}\mu_2}(\pm h)\!=\!\left(
\begin{array}{cccc}
 \rho_{4,4}(\pm h) & \rho_{4,19}(\pm h) & \rho_{5,22}(\pm h)& \rho_{10,22}(\pm h)\\
 \rho_{4,19} (\pm h)& \rho_{19,19}(\pm h) & \rho_{20,22}(\pm h)& \rho_{22,25}(\pm h)\\
 \rho_{5,22}(\pm h) & \rho_{20,22}(\pm h) & \rho_{23,23}(\pm h)& \rho_{23,28}(\pm h)\\
 \rho_{10,22}(\pm h) & \rho_{22,25}(\pm h) & \rho_{23,28}(\pm h)& \rho_{28,28}(\pm h)\\
\end{array}\right),\;\;
\nonumber\\
&\mathbf{Q}_3^{\mu_{1}\mu_2}(\pm h)\!=\!\left(
\begin{array}{cccccccc}
 \rho_{1,1}(\pm h) & \rho_{2,4}(\pm h) & \rho_{4,7}(\pm h)& \rho_{2,19}(\pm h) &\rho_{3,22}(\pm h) & \rho_{7,19}(\pm h) & \rho_{8,22}(\pm h)& \rho_{13,22}(\pm h)\\
\rho_{2,4}(\pm h) & \rho_{5,5}(\pm h) & \rho_{5,10}(\pm h)& \rho_{5,20}(\pm h)&\rho_{6,23}(\pm h) & \rho_{10,20}(\pm h) & \rho_{11,23}(\pm h)& \rho_{16,23}(\pm h)\\
\rho_{4,7}(\pm h) & \rho_{5,10}(\pm h) & \rho_{10,10}(\pm h)& \rho_{5,25}(\pm h) &\rho_{6,28}(\pm h) & \rho_{10,25}(\pm h) & \rho_{11,28}(\pm h)& \rho_{16,28}(\pm h)\\
\rho_{2,19}(\pm h) & \rho_{5,20}(\pm h) & \rho_{5,25}(\pm h)& \rho_{20,20}(\pm h) &\rho_{21,23}(\pm h) & \rho_{20,25}(\pm h) & \rho_{23,26}(\pm h)& \rho_{23,31}(\pm h)\\
\rho_{3,22}(\pm h) & \rho_{6,23}(\pm h) & \rho_{6,28}(\pm h)& \rho_{21,23}(\pm h) &\rho_{24,24}(\pm h) & \rho_{21,28}(\pm h) & \rho_{24,29}(\pm h)& \rho_{24,34}(\pm h)\\
\rho_{7,19}(\pm h) & \rho_{10,20}(\pm h) & \rho_{10,25}(\pm h)& \rho_{20,25}(\pm h) &\rho_{21,28}(\pm h) & \rho_{25,25}(\pm h) & \rho_{26,28}(\pm h)& \rho_{28,31}(\pm h)\\
\rho_{8,22}(\pm h) & \rho_{11,23}(\pm h) & \rho_{11,28}(\pm h)& \rho_{23,26}(\pm h) &\rho_{24,29}(\pm h) & \rho_{26,28}(\pm h) & \rho_{29,29}(\pm h)& \rho_{29,34}(\pm h)\\
\rho_{13,22}(\pm h) & \rho_{16,23}(\pm h) & \rho_{16,28}(\pm h)& \rho_{23,31}(\pm h) &\rho_{24,34}(\pm h) & \rho_{28,31}(\pm h) & \rho_{29,34}(\pm h)& \rho_{34,34}(\pm h)\\
\end{array}\right),
\nonumber\\
&\resizebox{1\textwidth}{!}{$
\mathbf{Q}_4^{\mu_{1}\mu_2}(h)\!=\!\left(
\begin{array}{cccccccccc}
 \rho_{2,2}(h) & \rho_{3,5}(h) & \rho_{2,7}(h)& \rho_{5,8}(h) &\rho_{5,13}(h) & \rho_{3,20}(h) & \rho_{8,20}(h)& \rho_{9,23}(h)& \rho_{13,20}(h)& \rho_{14,23}(h)\\
 \rho_{3,5}(h) & \rho_{6,6}(h) & \rho_{3,10}(h)& \rho_{6,11}(h) &\rho_{6,16}(h) & \rho_{6,21}(h) & \rho_{11,21}(h)& \rho_{12,24}(h)& \rho_{16,21}(h)& \rho_{13,20}(-h)\\
  \rho_{2,7}(h) & \rho_{3,10}(h) & \rho_{7,7}(h)& \rho_{8,10}(h)& \rho_{10,13}(h)& \rho_{3,25}(h) & \rho_{8,25}(h)& \rho_{9,28}(h)& \rho_{12,24}(-h)& \rho_{9,23}(-h)\\
   \rho_{5,8}(h) & \rho_{6,11}(h) & \rho_{8,10}(h)& \rho_{11,11}(h)& \rho_{11,16}(h) & \rho_{6,26}(h) & \rho_{11,26}(h)& \rho_{8,25}(-h)& \rho_{11,21}(-h)& \rho_{8,20}(-h)\\
   \rho_{5,13}(h) & \rho_{6,16}(h) & \rho_{10,13}(h)& \rho_{11,16}(h)& \rho_{16,16}(h) & \rho_{6,31}(h) & \rho_{6,26}(-h)& \rho_{3,25}(-h)& \rho_{6,21}(-h)& \rho_{3,20}(-h)\\
    \rho_{3,20}(h) & \rho_{6,21}(h) & \rho_{3,25}(h)& \rho_{6,26}(h)& \rho_{6,31}(h) & \rho_{16,16}(-h) & \rho_{11,16}(-h)& \rho_{10,13}(-h)& \rho_{6,16}(-h)& \rho_{5,13}(-h)\\
     \rho_{8,20}(h) & \rho_{11,21}(h) & \rho_{8,25}(h)& \rho_{11,26}(h)& \rho_{6,26}(-h) & \rho_{11,16}(-h) & \rho_{11,11}(-h)& \rho_{8,10}(-h)& \rho_{6,11}(-h)& \rho_{5,8}(-h)\\
        \rho_{9,23}(h) & \rho_{12,24}(h) & \rho_{9,28}(h)& \rho_{8,25}(-h)& \rho_{3,25}(-h) & \rho_{10,13}(-h) & \rho_{8,10}(-h)& \rho_{7,7}(-h)& \rho_{3,10}(-h)& \rho_{2,7}(-h)\\
          \rho_{13,20}(h) & \rho_{16,21}(h) & \rho_{12,24}(-h)& \rho_{11,21}(-h)& \rho_{6,21}(-h) & \rho_{6,16}(-h) & \rho_{6,11}(-h)& \rho_{3,10}(-h)& \rho_{6,6}(-h)& \rho_{3,5}(-h)\\
            \rho_{14,23}(h) & \rho_{13,20}(-h) & \rho_{9,23}(-h)& \rho_{8,20}(-h)& \rho_{3,20}(-h) & \rho_{5,13}(-h) & \rho_{5,8}(-h)& \rho_{2,7}(-h)& \rho_{3,5}(-h)& \rho_{2,2}(-h)\\
\end{array}\right).$}
\label{d1}
\end{align}
The particular elements $\rho_{i,j}(\pm h)$ are identical to those presented in Eqs.~\eqref{b6}-\eqref{b154}.The sum of the absolute values of all negative eigenvalues of $\mathbf{Q}_2^{\mu_{1}\mu_2}(\pm h)$,  $\mathbf{Q}_3^{\mu_{1}\mu_2}(\pm h)$, and  $\mathbf{Q}_4^{\mu_{1}\mu_2}(h)$ determines the respective global bipartite negativity ${\cal N}_{\mu_1\mu_2|S_1S_2}$ between a single spin dimer $\mu_1\mu_2$ and spin dimer $S_1S_2$.

\section{\label{App E}Global bipartite entanglement ${\cal N}_{\mu_{1}S_1|\mu_2S_2}$ with a central spin dimer $\mu_1S_1$}

It should be noted that system symmetry is responsible for the identical magnitude of the global bipartite negativity ${\cal N}_{\mu_{1}S_1|\mu_{2}S_{2}}$ and ${\cal N}_{\mu_2S_2|\mu_{1}S_1}$. Therefore, the calculation details for determining the global bipartite negativity  ${\cal N}_{\mu_{1}S_1|\mu_2S_{2}}$ are sufficient for further analyses. The partially transposed density matrix  $\hat{\rho}_{\mu_1S_1|\mu_{2}S_{2}}^{T_{\mu_{1}S_1}}$ again has a block-diagonal structure, which includes two $1\times 1$ matrices $\mathbf{Q}_1^{\mu_{1}S_1}(\pm h)$, two $4\times 4$ matrices $\mathbf{Q}_2^{\mu_{1}S_1}(\pm h)$, two $8\times 8$ matrices $\mathbf{Q}_3^{\mu_{1}S_1}(\pm h)$, and one $10\times 10$ matrix $\mathbf{Q}_4^{\mu_{1}S_1}(h)$ with the following elements:
\begin{align}
\allowdisplaybreaks
&\mathbf{Q}_1^{\mu_{1}S_1}(\pm h)\!=\!\left(
\begin{array}{c}
 \rho_{6,6}(\pm h)
\end{array}\right),\;\;
%\label{f24}
\mathbf{Q}_2^{\mu_{1}S_1}(\pm h)\!=\!\left(
\begin{array}{cccc}
 \rho_{3,3}(\pm h) & \rho_{3,5}(\pm h) & \rho_{6,9}(\pm h)& \rho_{6,21}(\pm h)\\
 \rho_{3,5} (\pm h)& \rho_{5,5}(\pm h) & \rho_{6,11}(\pm h)& \rho_{6,23}(\pm h)\\
 \rho_{6,9}(\pm h) & \rho_{6,11}(\pm h) & \rho_{12,12}(\pm h)& \rho_{12,24}(\pm h)\\
 \rho_{6,21}(\pm h) & \rho_{6,23}(\pm h) & \rho_{12,24}(\pm h)& \rho_{24,24}(\pm h)\\
\end{array}\right),\;\;
\nonumber\\
&\mathbf{Q}_3^{\mu_{1}S_1}(\pm h)\!=\!\left(
\begin{array}{cccccccc}
 \rho_{2,2}(\pm h) & \rho_{2,4}(\pm h) & \rho_{3,8}(\pm h)& \rho_{5,8}(\pm h) &\rho_{6,14}(\pm h) & \rho_{3,20}(\pm h) & \rho_{5,20}(\pm h)& \rho_{6,26}(\pm h)\\
\rho_{2,4}(\pm h) & \rho_{4,4}(\pm h) & \rho_{3,10}(\pm h)& \rho_{5,10}(\pm h)&\rho_{6,16}(\pm h) & \rho_{3,22}(\pm h) & \rho_{5,22}(\pm h)& \rho_{6,28}(\pm h)\\
\rho_{3,8}(\pm h) & \rho_{3,10}(\pm h) & \rho_{9,9}(\pm h)& \rho_{9,11}(\pm h) &\rho_{12,15}(\pm h) & \rho_{9,21}(\pm h) & \rho_{11,21}(\pm h)& \rho_{12,27}(\pm h)\\
\rho_{5,8}(\pm h) & \rho_{5,10}(\pm h) & \rho_{9,11}(\pm h)& \rho_{11,11}(\pm h) &\rho_{12,17}(\pm h) & \rho_{9,23}(\pm h) & \rho_{11,23}(\pm h)& \rho_{12,29}(\pm h)\\
\rho_{6,14}(\pm h) & \rho_{6,16}(\pm h) & \rho_{12,15}(\pm h)& \rho_{12,17}(\pm h) &\rho_{18,18}(\pm h) & \rho_{15,24}(\pm h) & \rho_{17,24}(\pm h)& \rho_{18,30}(\pm h)\\
\rho_{3,20}(\pm h) & \rho_{3,22}(\pm h) & \rho_{9,21}(\pm h)& \rho_{9,23}(\pm h) &\rho_{15,24}(\pm h) & \rho_{21,21}(\pm h) & \rho_{21,23}(\pm h)& \rho_{24,27}(\pm h)\\
\rho_{5,20}(\pm h) & \rho_{5,22}(\pm h) & \rho_{11,21}(\pm h)& \rho_{11,23}(\pm h) &\rho_{17,24}(\pm h) & \rho_{21,23}(\pm h) & \rho_{23,23}(\pm h)& \rho_{24,29}(\pm h)\\
\rho_{6,26}(\pm h) & \rho_{6,28}(\pm h) & \rho_{12,27}(\pm h)& \rho_{12,29}(\pm h) &\rho_{18,30}(\pm h) & \rho_{24,27}(\pm h) & \rho_{24,29}(\pm h)& \rho_{30,30}(\pm h)\\
\end{array}\right),
\nonumber\\
&\resizebox{1\textwidth}{!}{$
\mathbf{Q}_4^{\mu_{1}S_1}(h)\!=\!\left(
\begin{array}{cccccccccc}
 \rho_{1,1}(h) & \rho_{2,7}(h) & \rho_{4,7}(h)& \rho_{3,13}(h) &\rho_{5,13}(h) & \rho_{2,19}(h) & \rho_{4,19}(h)& \rho_{3,25}(h)& \rho_{5,25}(h)& \rho_{6,31}(h)\\
 \rho_{2,7}(h) & \rho_{8,8}(h) & \rho_{8,10}(h)& \rho_{9,14}(h) &\rho_{11,14}(h) & \rho_{8,20}(h) & \rho_{10,20}(h)& \rho_{9,26}(h)& \rho_{11,26}(h)& \rho_{5,25}(-h)\\
  \rho_{4,7}(h) & \rho_{8,10}(h) & \rho_{10,10}(h)& \rho_{9,16}(h)& \rho_{11,16}(h)& \rho_{8,22}(h) & \rho_{10,22}(h)& \rho_{9,28}(h)& \rho_{9,26}(-h)& \rho_{3,25}(-h)\\
   \rho_{3,13}(h) & \rho_{9,14}(h) & \rho_{9,16}(h)& \rho_{15,15}(h)& \rho_{15,17}(h) & \rho_{14,21}(h) & \rho_{16,21}(h)& \rho_{10,22}(-h)& \rho_{10,20}(-h)& \rho_{4,19}(-h)\\
   \rho_{5,13}(h) & \rho_{11,14}(h) & \rho_{11,16}(h)& \rho_{15,17}(h)& \rho_{17,17}(h) & \rho_{14,23}(h) & \rho_{14,21}(-h)& \rho_{8,22}(-h)& \rho_{8,20}(-h)& \rho_{2,19}(-h)\\
    \rho_{2,19}(h) & \rho_{8,20}(h) & \rho_{8,22}(h)& \rho_{14,21}(h)& \rho_{14,23}(h) & \rho_{17,17}(-h) & \rho_{15,17}(-h)& \rho_{11,16}(-h)& \rho_{11,14}(-h)& \rho_{5,13}(-h)\\
     \rho_{4,19}(h) & \rho_{10,20}(h) & \rho_{10,22}(h)& \rho_{16,21}(h)& \rho_{14,21}(-h) & \rho_{15,17}(-h) & \rho_{15,15}(-h)& \rho_{9,16}(-h)& \rho_{9,14}(-h)& \rho_{3,13}(-h)\\
        \rho_{3,25}(h) & \rho_{9,26}(h) & \rho_{9,28}(h)& \rho_{10,22}(-h)& \rho_{8,22}(-h) & \rho_{11,16}(-h) & \rho_{9,16}(-h)& \rho_{10,10}(-h)& \rho_{8,10}(-h)& \rho_{4,7}(-h)\\
          \rho_{5,25}(h) & \rho_{11,26}(h) & \rho_{9,26}(-h)& \rho_{10,20}(-h)& \rho_{8,20}(-h) & \rho_{11,14}(-h) & \rho_{9,14}(-h)& \rho_{8,10}(-h)& \rho_{8,8}(-h)& \rho_{2,7}(-h)\\
            \rho_{6,31}(h) & \rho_{5,25}(-h) & \rho_{3,25}(-h)& \rho_{4,19}(-h)& \rho_{2,19}(-h) & \rho_{5,13}(-h) & \rho_{3,13}(-h)& \rho_{4,7}(-h)& \rho_{2,7}(-h)& \rho_{1,1}(-h)\\
\end{array}\right).$}
\label{e1}
\end{align}
The particular elements $\rho_{i,j}(\pm h)$ are identical to those presented in Eqs.~\eqref{b6}-\eqref{b154}. The sum of the absolute values of all negative eigenvalues of  $\mathbf{Q}_2^{\mu_{1}S_1}(\pm h)$,  $\mathbf{Q}_3^{\mu_{1}S_1}(\pm h)$, and  $\mathbf{Q}_4^{\mu_{1}S_1}(h)$ determines the respective global bipartite negativity ${\cal N}_{\mu_1S_1|\mu_2S_2}$ between a single spin dimer $\mu_1S_1$ and spin dimer $\mu_2S_2$.

\section{\label{App F}Global bipartite entanglement ${\cal N}_{\mu_{1}S_2|\mu_2S_1}$ with a central spin dimer $\mu_1S_2$}

It should be noted that the system's symmetry results in the identical magnitude of a global bipartite negativity  ${\cal N}_{\mu_{1}S_2|\mu_{2}S_{1}}$ and ${\cal N}_{\mu_2S_1|\mu_{1}S_2}$. Therefore, the calculation details for determining the global bipartite negativity  ${\cal N}_{\mu_{1}S_2|\mu_2S_{1}}$ are sufficient for further analyses. The partially transposed density matrix  $\hat{\rho}_{\mu_1S_2|\mu_{2}S_{1}}^{T_{\mu_{1}S_2}}$ again has a block-diagonal structure, including two  $1\times 1$ matrices $\mathbf{Q}_1^{\mu_{1}S_2}(\pm h)$, two $4\times 4$ matrices $\mathbf{Q}_2^{\mu_{1}S_2}(\pm h)$, two $8\times 8$ matrices $\mathbf{Q}_3^{\mu_{1}S_2}(\pm h)$, and one $10\times 10$ matrix $\mathbf{Q}_4^{\mu_{1}S_2}(h)$ with the following elements:
\begin{align}
\allowdisplaybreaks
&\mathbf{Q}_1^{\mu_{1}S_2}(\pm h)\!=\!\left(
\begin{array}{c}
 \rho_{16,16}(\pm h)
\end{array}\right),\;\;
%\label{f24}
\mathbf{Q}_2^{\mu_{1}S_2}(\pm h)\!=\!\left(
\begin{array}{cccc}
 \rho_{3,3}(\pm h) & \rho_{3,20}(\pm h) & \rho_{6,21}(\pm h)& \rho_{9,21}(\pm h)\\
 \rho_{3,20} (\pm h)& \rho_{20,20}(\pm h) & \rho_{21,23}(\pm h)& \rho_{21,26}(\pm h)\\
 \rho_{6,21}(\pm h) & \rho_{21,23}(\pm h) & \rho_{24,24}(\pm h)& \rho_{24,27}(\pm h)\\
 \rho_{9,21}(\pm h) & \rho_{21,26}(\pm h) & \rho_{24,27}(\pm h)& \rho_{27,27}(\pm h)\\
\end{array}\right),\;\;
\nonumber\\
&\mathbf{Q}_3^{\mu_{1}S_2}(\pm h)\!=\!\left(
\begin{array}{cccccccc}
 \rho_{2,2}(\pm h) & \rho_{3,5}(\pm h) & \rho_{3,8}(\pm h)& \rho_{2,19}(\pm h) &\rho_{5,20}(\pm h) & \rho_{8,20}(\pm h) & \rho_{11,21}(\pm h)& \rho_{14,21}(\pm h)\\
\rho_{3,5}(\pm h) & \rho_{6,6}(\pm h) & \rho_{6,9}(\pm h)& \rho_{3,22}(\pm h)&\rho_{6,23}(\pm h) & \rho_{9,23}(\pm h) & \rho_{12,24}(\pm h)& \rho_{15,24}(\pm h)\\
\rho_{3,8}(\pm h) & \rho_{6,9}(\pm h) & \rho_{9,9}(\pm h)& \rho_{3,25}(\pm h) &\rho_{6,26}(\pm h) & \rho_{9,26}(\pm h) & \rho_{12,27}(\pm h)& \rho_{15,27}(\pm h)\\
\rho_{2,19}(\pm h) & \rho_{3,22}(\pm h) & \rho_{3,25}(\pm h)& \rho_{19,19}(\pm h) &\rho_{20,22}(\pm h) & \rho_{20,25}(\pm h) & \rho_{21,28}(\pm h)& \rho_{21,31}(\pm h)\\
\rho_{5,20}(\pm h) & \rho_{6,23}(\pm h) & \rho_{6,26}(\pm h)& \rho_{20,22}(\pm h) &\rho_{23,23}(\pm h) & \rho_{23,26}(\pm h) & \rho_{24,29}(\pm h)& \rho_{24,32}(\pm h)\\
\rho_{8,20}(\pm h) & \rho_{9,23}(\pm h) & \rho_{9,26}(\pm h)& \rho_{20,25}(\pm h) &\rho_{23,26}(\pm h) & \rho_{26,26}(\pm h) & \rho_{27,29}(\pm h)& \rho_{27,32}(\pm h)\\
\rho_{11,21}(\pm h) & \rho_{12,24}(\pm h) & \rho_{12,27}(\pm h)& \rho_{21,28}(\pm h) &\rho_{24,29}(\pm h) & \rho_{27,29}(\pm h) & \rho_{30,30}(\pm h)& \rho_{30,33}(\pm h)\\
\rho_{14,21}(\pm h) & \rho_{15,24}(\pm h) & \rho_{15,27}(\pm h)& \rho_{21,31}(\pm h) &\rho_{24,32}(\pm h) & \rho_{27,32}(\pm h) & \rho_{30,33}(\pm h)& \rho_{33,33}(\pm h)\\
\end{array}\right),
\nonumber\\
&\resizebox{1\textwidth}{!}{$
\mathbf{Q}_4^{\mu_{1}S_1}(h)\!=\!\left(
\begin{array}{cccccccccc}
 \rho_{1,1}(h) & \rho_{2,4}(h) & \rho_{2,7}(h)& \rho_{3,10}(h) &\rho_{3,13}(h) & \rho_{4,19}(h) & \rho_{7,19}(h)& \rho_{10,20}(h)& \rho_{13,20}(h)& \rho_{16,21}(h)\\
 \rho_{2,4}(h) & \rho_{5,5}(h) & \rho_{5,8}(h)& \rho_{6,11}(h) &\rho_{6,14}(h) & \rho_{5,22}(h) & \rho_{8,22}(h)& \rho_{11,23}(h)& \rho_{14,23}(h)& \rho_{13,20}(-h)\\
  \rho_{2,7}(h) & \rho_{5,8}(h) & \rho_{8,8}(h)& \rho_{9,11}(h)& \rho_{9,14}(h)& \rho_{5,25}(h) & \rho_{8,25}(h)& \rho_{11,26}(h)& \rho_{11,23}(-h)& \rho_{10,20}(-h)\\
   \rho_{3,10}(h) & \rho_{6,11}(h) & \rho_{9,11}(h)& \rho_{12,12}(h)& \rho_{12,15}(h) & \rho_{6,28}(h) & \rho_{9,28}(h)& \rho_{8,25}(-h)& \rho_{8,22}(-h)& \rho_{7,19}(-h)\\
   \rho_{3,13}(h) & \rho_{6,14}(h) & \rho_{9,14}(h)& \rho_{12,15}(h)& \rho_{15,15}(h) & \rho_{6,31}(h) & \rho_{6,28}(-h)& \rho_{5,25}(-h)& \rho_{5,22}(-h)& \rho_{4,19}(-h)\\
    \rho_{4,19}(h) & \rho_{5,22}(h) & \rho_{5,25}(h)& \rho_{6,28}(h)& \rho_{6,31}(h) & \rho_{15,15}(-h) & \rho_{12,15}(-h)& \rho_{9,14}(-h)& \rho_{6,14}(-h)& \rho_{3,13}(-h)\\
     \rho_{7,19}(h) & \rho_{8,22}(h) & \rho_{8,25}(h)& \rho_{9,28}(h)& \rho_{6,28}(-h) & \rho_{12,15}(-h) & \rho_{12,12}(-h)& \rho_{9,11}(-h)& \rho_{6,11}(-h)& \rho_{3,10}(-h)\\
        \rho_{10,20}(h) & \rho_{11,23}(h) & \rho_{11,26}(h)& \rho_{8,25}(-h)& \rho_{5,25}(-h) & \rho_{9,14}(-h) & \rho_{9,11}(-h)& \rho_{8,8}(-h)& \rho_{5,8}(-h)& \rho_{2,7}(-h)\\
          \rho_{13,20}(h) & \rho_{14,23}(h) & \rho_{11,23}(-h)& \rho_{8,22}(-h)& \rho_{5,22}(-h) & \rho_{6,14}(-h) & \rho_{6,11}(-h)& \rho_{5,8}(-h)& \rho_{5,5}(-h)& \rho_{2,4}(-h)\\
            \rho_{16,21}(h) & \rho_{13,20}(-h) & \rho_{10,20}(-h)& \rho_{7,19}(-h)& \rho_{4,19}(-h) & \rho_{3,13}(-h) & \rho_{3,10}(-h)& \rho_{2,7}(-h)& \rho_{2,4}(-h)& \rho_{1,1}(-h)\\
\end{array}\right).$}
\label{d1}
\end{align}
The particular elements $\rho_{i,j}(\pm h)$ are identical to those presented in Eqs.~\eqref{b6}-\eqref{b154}.The sum of the absolute values of all negative eigenvalues of $\mathbf{Q}_2^{\mu_{1}S_2}(\pm h)$,  $\mathbf{Q}_3^{\mu_{1}S_2}(\pm h)$, and  $\mathbf{Q}_4^{\mu_{1}S_2}(h)$ determines the respective global bipartite negativity ${\cal N}_{\mu_1S_2|\mu_2S_1}$ between a single spin dimer $\mu_1S_2$ and spin dimer $\mu_2S_1$.

\printcredits

%% Loading bibliography style file
%\bibliographystyle{model1-num-names}
%\bibliographystyle{cas-model2-names}

% Loading bibliography database
%\bibliography{my-refs}

\end{document}